# ARSSEM

## Active Radiation Shield for Space Exploration Missions

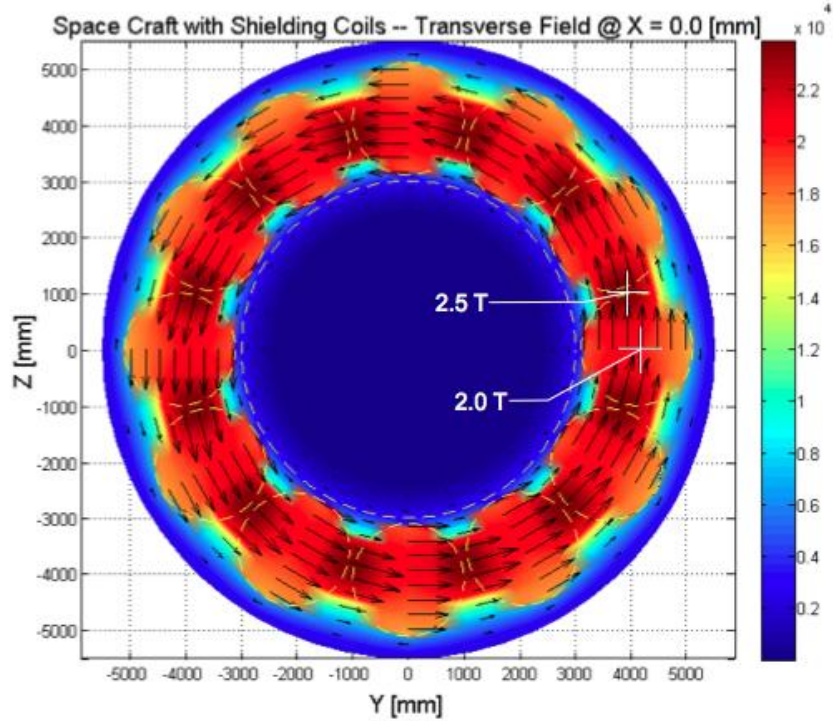

R. Battiston[1], W.J. Burger[1], V. Calvelli[2], R. Musenich[2], V. Choutko[3], V.I. Datskov[4], A. Della Torre[5], F. Venditti[5], C. Gargiulo[6], G. Laurenti[7], S. Lucidi[8], S. Harrison[9] and R. Meinke[10]

[1]*INFN-Perugia and Università degli Studi di Perugia, I-06123 Perugia (PG), Italy*
[2]*INFN-Genova, via Dodecaneso 33, I-16146 Genova (GE), Italy*
[3]*Massachusetts Institute of Technology, Cambridge MA 02139, USA*
[4]*2INFN-Milano Bicocca, Piazza della Scienza 3, I-20126 Milano, Italy*
[5]*CGS S.p.A, via Gallarate 150, I-20151 Milan (MI), Italy*
[6]*INFN-Roma, P.le Aldo Moro 2, I-00185 Roma (RM), Italy*
[7]*INFN-Bologna, Viale B. Pichat 6/2, I-40127 Bologna (BO), Italy*
[8]*SERMS S.r.l., Viale Pentima Bassa 21, I-05100 Terni (TR), Italy*
[9]*Scientific Magnetics,7 Suffolk Way, Abingdon, OX14 5JX United Kingdom*
[10]*Advanced Magnet Lab, 1720 Main St, NE Bldg. 4, Palm Bay FL 32905, USA*

# Active Radiation Shield for Space Exploration Missions

# Contents

ACTIVE RADIATION SHIELD FOR SPACE EXPLORATION MISSIONS

# List of Acronyms

ALARA As Low as Reasonably Achievable
AMS     Alpha Magnetic Spectrometer
AML     Advanced Magnet Laboratory
APM     AMS Program Manager
AR      Accepted Risk
BFO     Blood Forming Organs
CERN    Centre Europeen Pour la Recherche Nucleaire
ECAL    Electric CALorimeter
ELIPS   European Programme for Life and Physical Sciences and
ESA     European Space Agency
ESTEC   European Space Research and Technology Centre
EVA     Extra-Vehicular Activity
GCR     Galactic Cosmic Rays
GSI     Gesellschaft für Schwerionenforschung (Darmstadt, Germany)
Gy      Gray, unit of absorbed dose
HZE     High Z Events
INFN    Istituto Nazionale di Fisica Nucleare
ISS     International Space Station
LEO     Low Earth Orbit
LET     Linear Energy Transfer
LLO     Low Lunar Orbit
LSS     Life Support System
NEA     Near Earth Asteroids
KSC     Kennedy Space Center
RICH    Ring Imaging Cherenkov
SCL     Space Cryomagnetics Ltd
SEE     Solar Energetic Events
SEP     Solar Energetic Particles
SPE     Solar Particle Event
STS     Space Transportation System
TBD     To Be Defined
TOF     Time of Flight
TRD     Transition Radiation detector

# Foreword

Human exploration of space is among most ambitious goal of mankind. However, this strong ambition is not supported by our evolutionary DNA code. Space, unfortunately, is a very hostile environment for man. Long duration permanence in space on nearby LEO already requires the development of a technology marvel like the ISS. The future of human exploration of the solar system (Moon, Mars, NEA) would pose much more difficult challenges, requiring the best of our technology and ingenuity to overcome them.

One of the major issues to be solved is the protection from the effects of ionizing radiation. On Earth, exposure to ionizing radiation is seldom an issue: the effects of nuclear plant accidents or usage of nuclear weapons are the most extreme, but unlikely cases. The atmosphere, the geomagnetic field and the Earth shadow, all three contributes to almost eliminate the effect due to the cosmic radiation. The situation changes substantially already on LEO: the protection of the atmosphere is lost and the dose absorbed by the astronauts increase nearly 2 orders of magnitude. But permanence on LEO rarely exceeds 6 months, then, by simply *"returning home"*, the absorbed dose can be limited below the professional exposure limits. The Apollo missions to the Moon were lasting only about ten days and were not an issue from the point of view of the absorbed dose. But an exploration mission, lasting two to three years in space, represents a very significant step from the point of view of radiation protection: both the duration (up to 5 times) and the intensity (up to 5 times) of the exposure to radiation are increased *at the same time*, reaching and sometime exceeding professional career limits.

This issue is known since the time of Werner von Braun: several studies attempted during the last 40 years to find practical ways to protect the astronauts from the sudden, very intense, low energy, Solar Particle Events and, at the same time, from the dose due to the continuous flux of penetrating, high energy Galactic Cosmic Rays.

The usage of intense magnetic fields, enveloping the spacecraft and deflecting the various forms of charged cosmic radiation, has been considered by various authors. Magnetic shielding, indeed, appears likely to be the most effective, although technologically challenging, active protection method. Permanent magnets in space, however, can only be based on superconducting materials, due to basic power and mass considerations. So far the use of superconductivity in space has been almost non existing.

During the last 10 years, the AMS Collaboration was involved in the single largest effort to design and build a large superconducting magnet to be deployed in space. During a decade, an international team of experts based in Europe, developed a 3 ton, 1 Tesla, superfluid Helium cooled, superconducting magnet which was designed, built and fully space qualified, although it was eventually not used on the AMS-02 experiment deployed ont the ISS in May 2011.

The knowledge accumulated by this team during the AMS project, the analysis tools which have been developed, the available instrumentation and test data, represent a uniquely qualified starting point to further study a realistic active magnetic shield for space exploration and for the definition of a development path towards its realization. This was the reason why ESA Human Spaceflight Directorate supported this one year study devoted to the analysis of the Active Radiation Shields for Space Exploration Missions (ARSSEM).

The team involved in this study comprised:

<u>Università di Perugia</u> (prime contractor) and Sezione INFN of Perugia, Italy, led by Professor Roberto Battiston associated with Dr. William J. Burger;
<u>SERMS srl</u>, Terni, Italy, represented by Ing. Stefano Lucidi;
<u>Sezione INFN di Genova,</u> Italy, represented by Dr. Riccardo Musenich;
<u>Sezione INFN di Roma</u>, Italy, represented by Dr. Ing. Corrado Gargiulo;
<u>Sezione INFN di Bologna</u>, Italy, represented by Dr. Ing. Giuliano Laurenti;
<u>Sezione INFN di Milano Bicocca</u>, Italy, represented by Dr. Vladimir I. Datskov;
<u>Scientific Magnetic</u>, Abingdon, UK, represented by Mr. Steve Harrison;
<u>Compagnia Generale dello Spazio</u>, Milano, Italy represented by Dr. Floriano Venditti associated with Dr. Alberto Della Torre;
<u>AGS</u>, Genova, Italy, represented by Dr. Valerio Calvelli;
<u>Advanced Magnet Laboratory</u>, Melbourne, Florida, US, represented by Dr. Reinar Meinke
<u>MIT</u>, Boston, US, represented by Dr. Vitaly Choutko

After reviewing the physics basis of the issue of radiation protection in space, the study uses full physics simulation with code built by the AMS collaboration to understand the interplay among the the various factors determining the dose absorbed by the astronauts during a long duration mission: radiation composition and energy spectrum, 3D particle propagation through the magnetic field, secondary production on the spacecraft structural materia, dose sensitivity of the various parts of the human body.

With the goal of defining what would be achievable using current state of the art technology and then identify a technology development roadmap, the study analyze and compares the structural and magnetic issue related to two toroid shield configuration, one based on the traditional *"racetrack"* coil solution and another based on an innovative solution using *Double Helix* coils. The study shows the *Double Helix* solution exhibits a number of interesting features which are suited for the active shield application, in particular considering future developments of various related technologies. The study then proposes an technology R&D roadmap for active radiation shield development which would match ESA decadal development strategy for human exploration of space.

*Perugia, November 2011*                                       *Roberto Battiston*

# Executive summary

**ACTIVE RADIATION SHIELDING FOR SPACE EXPLORATION MISSIONS**

**(ARSSEM)**

After the completion of the International Space Station *(ISS)*, missions beyond low Earth orbit *(LEO)*, to Moon, Mars or Near Earth Asteroids *(NEA)* are considered as the next step of manned, peaceful, space exploration. These missions would inevitably last much longer that the long duration missions on the *ISS* (Expeditions), posing severe challenges in several areas of life sciences like, in particular, radiation health.

Space radiation environment represents a serious challenge for long duration mission. On *LEO* the shadow of the Earth and the effect of the magnetosphere, reduces by a factor *4/5* the dose absorbed by astronauts. During a long duration exploration mission, the dose could easily reach and exceed the current dose yearly limits.

For this reason, during at last four decades, means to actively shield the Galactic Cosmic Ray *(GCR)* component have been considered in particular using superconducting magnets creating a toroidal field around the habitable module. Theses studies indicated that magnetic shielding with up to a factor of *10* of reduction of the *GCR* dose could, in principle, be developed.

Europe has a significant amount of experience in this area thanks previous human missions in *LEO* (*Spacelab, MIR* and *ISS*), related programs (e.g. *HUMEX, AMS-02*) and *Topical Teams* activities: this background represent a solid starting point for further progress in this area.

Due to recent and significant progress in superconducting magnet technology both in ground laboratory (*ITS* – Intermediate Temperature Superconductors - and *HTS* - High Temperature Superconductors) and on the preparation of space experiments (development of the space qualified *AMS-02* superconducting magnet), it is interesting to re-evaluate active shielding concepts as potentially viable solutions to crew protection from exposure to high energy cosmic radiation.

Past active radiation shielding concepts yielded architectures that are significantly massive and too costly to be launched and assembled in space. This is largely due to the magnet size and field strength required to shield Galactic Cosmic Radiation (*GCR*) and Solar Proton Events (*SPE*) for meaningful level of crew protection from radiation in space.

Since then, state-of-the-art superconducting magnet technology has made significant progress in performance including higher temperature superconductivity (*ITS* and *HTS*) and new mechanical solutions better suited to deal with the Lorentz forces created by the strong magnetic fields. In addition, ten years of design, research and development, construction and testing of the *AMS-02* magnet, the only space qualified superconducting magnet built so far, provides an heritage of European based experience which motivates further developments of this technology for space applications.

Use of *ITS* or *HTS* allows for simpler magnet cooling systems using liquid hydrogen, gaseous helium or even liquid nitrogen as cryogens instead of the more complex liquid helium systems required for Low Temperature (*LTS*) superconducting magnets like *AMS-02*. Due to the large energy margin of *ITS* and *HTS* conductors, significantly lighter support structures are needed for the superconducting coils providing for more creative ways to shield from radiation a crew in space.

We compared *15* different magnets geometries producing toroidal field, both from the point of view of their mechanical properties (coil mass, structural mass) as well as from the point of view of their shielding properties. For the purpose of the study we analyzed magnetic configurations which would be possible to build *"today"*, using state of the art technologies: the identification of the most promising solutions would then provide the starting basis for *R&D* and future design studies. For this we have considered relatively low bending power/shielding power configurations, assuming that future improvements in the technology would allow to increase their shielding capabilities. The structural design of two coil configurations having magnetic bending power of *4-5 Tm* were analyzed more in detail, to determine their structural mass and material budget: the classical *"racetrack"* toroid magnet design and a new multi-coil geometry based on a technology called *"Double Helix" (DH)*. The radiation shielding efficiency of these two solutions have been evaluated and compared using an advanced *3D* simulation tool developed in the frame of the AMS Collaboration, able to simulate Cosmic Rays interaction with the magnet structure, including secondary particle production, and their *3D* propagation in the magnetic field.

The accuracy of the simulation tools we developed and used for this study allow for a better understanding of the interplay among different elements concurring to the definition of the dose absorbed by the different parts of the body of the astronauts during an long duration mission like: shielding effect of the structural mass versus shielding effect of the magnetic field, the dependence of these effects on the ion charge $Z$ and on their energy.

The result of the study shows that the two coil configurations analyzed have similar performances in terms of shielding capability, in particular for the low BdL configurations which we have studied in details. At *4-5 Tm* the combined effect of passive and active shielding would provide a *40%* reduction of the dose due to *GCR* with respect to empty space, reducing it to a value *30%* below the currently recommended yearly dose limit of *50 rem*. These configuration would also ensure a good shielding against the sudden radiation spikes due to *SPE*. Increasing the magnetic bending power (higher field and/or larger magnetized volumes) would increase the active shielding efficiency, provided that the amount of the material crossed by the *GCR* is kept to a minimum, which one of the key technological challenge.

The *DH* design looks promising from the point of view of the weight and complexity of the supporting structure, transportation to orbit and modularity of the multi coil designs, since a good fraction of the resulting Lorentz forces are distributed within the coils winding, requiring less additional supporting structures. Extrapolation to high value of *BdL* should make the situation even more favorable for the *DH* design. In addition *DH* design has more flexibility both in terms of magnetic as well as of mechanic design, including the possibility of designing coils with complex field configurations matching the needs of the shield application or of *"game changing"* new technologies based on *"deployable"* coils.

Following the main idea behind this study, namely to identify designs based on state of the art but existing technologies, which could be developed within a decade or so with an high degree of confidence, we defined a *Radiation Protection System Roadmap*, identifying *Ten Critical Technologies* needed to develop within the next ten years a realistic *DH* active magnetic shield, identifying the corresponding Technology Tree as well as the technological *R&D* and development of demonstrators which are needed over a period of ten years to bring these technologies from the current status to the *TRL* needed to test such a system in space.

The ten critical technologies which have been identified are:

#1  High performances *ITS* and *HTS* cables (*MgB$_2$*, *YBCCO*)

#2  *Double Helix* coil

#3  Cryogenic stable, light mechanics

#4  Gas based recirculating cryogenic systems

#5  Cryocoolers operating at low temperature

#6  Magnetic field flux charging devices

#7  Quench protection for *ITS* and *HTC* coils

#8  Large cryogenic cases for space operation

#9  Superinsulation, Radiation Shielding, Heat Removal

#10 Deployable *SC* Coils

Most of these critical technologies are within the background of the European industry and top research laboratories. As a recommendation of this study, an European led effort to develop them to the level of ground based and space based active shield demonstrators should be supported. We present in the study a *Radiation Protection System Planning*, taking as reference the latest version (2011) of the *"The Global Exploration Roadmap"* developed by *ISECG* (International Space Exploration Coordination Group). A first step toward a final objective of human Mars exploration could a mission to *NEA* (Near Earth Asteroids), the so-called *"deep space first"* as an alternative to a *"moon-first"* approach. Such a mission would require as well a Radiation Protection System for the crew as the mission duration could be in the range of *400* days, so that radiation protection becomes an issue. A reference mission with *4* astronauts has been considered, as well as an extension to an *8* crew mission.

The development plan is in principle divided in four main sequential phases:

1. Technology development, in which the critical and/or not yet mature technologies are brought to *TRL5*.

2. Ground Demonstrators, aiming at integrating, with different steps, the required different technologies in one system, allowing to validate the capability of the system in providing the expected functions and performance.

3. In-space Demonstration, where a representative Radiation Shield model is flown in *LEO* (maybe as a free-flying item in proximity of the *ISS*).

4. Deep Space Habitat (*DSH*) Radiation Shielding System; this is the operational system for the *DSH*.

This report summarizes - as the Final Report of the "*Active Radiation Shielding for Space Exploration Missions" (ARSSEM)*) - the findings presented in details in four Technical Notes (ESTEC/Contract N° 4200023087/10/NL/AF):

*TN1* - Data Collection and Requirements Generation

*TN2* - Radiation Protection System Concept Report

*TN3* - Radiation Protection System Preliminary System Engineering Plan

*TN4* - Results of Active Shielding Round Table Discussions

The study has been led by University of Perugia (I) with the cooperation of INFN (I), CGS (I), SCL (UK), AML (USA).

# 1. The Reference Scenario in Space

## 1.1 The Interplanetary Travel Case

After the completion of the International Space Station (*ISS*), exploration missions beyond low Earth orbit (*LEO*), to Moon, Mars or Near Earth Asteroids (*NEA*) are considered as the next step of manned, peaceful, space exploration. These missions would inevitably last much longer that the long duration missions on the *ISS* (Expeditions), traveling on deep space orbits, posing severe challenges in several areas of life sciences like, in particular, radiation health, as discussed in the 2003 *HUMEX* study by *ESA*[1].

The most challenging of these exploration mission will certainly be a manned Mars mission and for this reason it will be taken as reference scenario on this study, assuming as guideline the mission described in the *ESA Exploration Reference Architecture Document*[2].

The overall mission to Mars would foresee several launches from the Earth and the capability to assembly modules in earth orbit. At first, two cargo transportation missions will bring on Mars surface the habitation module and other surface elements (rovers, power plant, etc.); also the manned lander/ascent vehicle will be transported to Mars orbit, where it will wait for the crew arrival.

Then, the human mission foresees the assembly in Earth orbit of the transfer vehicle, with the propulsion stage and the crewed elements.

The mission timing foresees a transfer phase from Earth to Mars orbit of about of *200* days, a stay on Mars surface of about *400-500* days and a return trip of about *200* days (so-called '*conjunction mission*'). An alternative mission (the '*opposition mission*'), would need about *300* days each way and a stay on Mars of about *40* days. The whole crew is transferred to the Mars surface, leaving the transfer vehicle in dormant mode until the crew return.

It follows that the total duration of a crewed Mars mission would be between *2* and *3* years.

## 1.2 Ionizing radiation in space

In space the main contribution to the absorbed dose of ionizing radiation are

1) Solar Particle Events *(SPE)*
2) Galactic Cosmic Rays *(GCR)*

These two contributions differ very significantly in rate, chemical composition and energy spectrum as it can be seen in *Figure 1.1* and *1.2*.

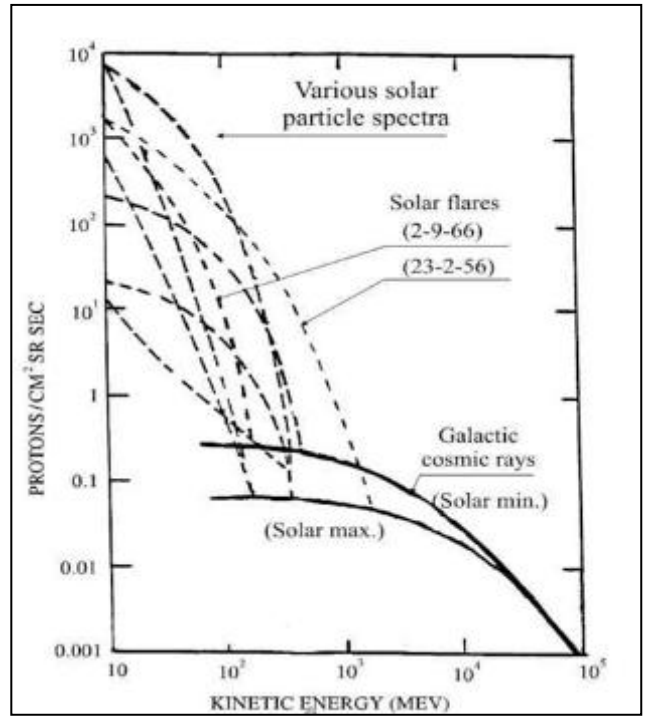

*Figure 1.1 Particle spectra observed in SPE compared with the GCR spectrum (Physics Today, Oct. 1974, p. 25)*

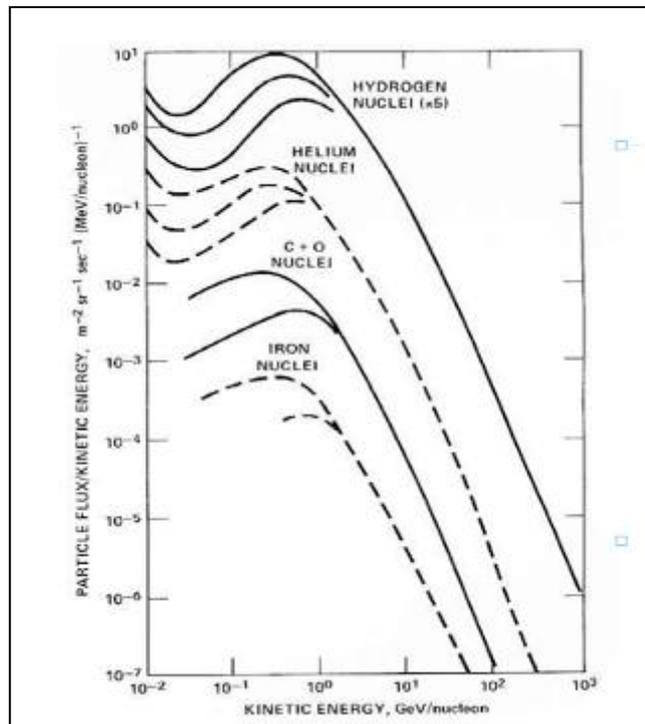

*Figure 1.2 Distribution of energies of GCR. This is a graph of the more abundant nuclear species in CR as measured near Earth. Below a few GeV/nucleon these spectra are strongly influenced by the Sun. The different curves for the same species represent measurement extremes resulting from varying solar activity (Physics Today, Oct. 1974, p. 25)*

We note in these figures

- the large flux increase during a *SPE*: although these events have a typical rate of *1* to a few/year they are very dangerous for the astronauts as well as for the instrumentation
- the significant fraction of the *GCR* flux at energies below *1 GeV* (the flux is dominated by protons, although not the radiation dose, as seen in the following): without adequate shielding the astronauts would be exposed to dangerous radiation doses
- the anti-correlation of the *GCR* flux with the solar activity, which could be considered as one of the factors in choosing the timing of a interplanetary mission

In the following we will concentrate our attention on the case of *GCR*, since, the requirements of a shield for the continuous flux of these high energy, penetrating ionizing particles bring to shielding techniques (active or passive) which would also effectively block the intense, low energy, low penetration *SPE* ionizing radiation.

On average astronauts and cosmonauts on *ISS* receive $0,6$ $mSv\ d^{-1}$ (**229 mSv y$^{-1}$**), with ~75 % coming from *GCR* and *25 %* coming from protons encountered in passages through the *South Atlantic Anomaly* (*SAA*) region of the Van Allen belts: we recall that a long duration stay on the *ISS* typically does not exceed 6 months. About a factor *5,5* exists from *ISS* to deep space (**1.100 mSv y$^{-1}$**) where no protection from the magnetosphere or planetary shadow exists: this factor reduces to *3,7* namely *2,2* $mSv\ d^{-1}$ (**740 mSv y$^{-1}$**) inside a Mars habitable module with *1,5 cm Al* thick walls ($4\ g\ cm^{-2}$).

## 1.3 Galactic Cosmic Rays (GCR)

Exposure to *GCR* could pose a serious hazard for long-duration space missions. *GCR* radiation consists of particles of charge from hydrogen to uranium arriving from outside the heliosphere. These particles range in energy from ~$10\ MeV\ n^{-1}$ to ~$10^{12}\ MeV\ n^{-1}$, with fluence - rate peaks around *300* to *700 MeV n$^{-1}$*. Because of the vast energy range, it is difficult to provide adequate shielding, thus these particles provide a steady source of low dose-rate radiation.

Integrations of energy spectra show that ~75 % of the particles have energies below ~$3\ GeV\ n^{-1}$. Under modest aluminum shielding, nearly *75 %* of the dose equivalent is due to particles with energies <$2\ GeV\ n^{-1}$. Thus, the most important energy range for risk estimation is from particles with energies below ~$2\ GeV\ n^{-1}$, and nearly all of the risk is due to particles with energies <$10\ GeV\ n^{-1}$. The local interstellar energy spectrum (outside the heliosphere) is a constant, but inside the heliosphere the spectrum and fluence of particles below ~$10\ GeV\ n^{-1}$ is modified by solar activity. The assessment of radiation risk requires a detailed knowledge of the composition and energy spectra of *GCR* in interplanetary space, and their spatial and temporal variation.

In the case of deep space interplanetary mission, neglecting the contribution of *SPE*, different *GCR* species contribute to the absorbed radiations with fractions of the total dose depending on the electric charge, due to the energy deposition mechanism based on ionization, which follows a $Z^2$ law. Due to that reason, *Fe* nuclei, although much less abundant than protons, provide the most important contribution to the radiation dose of all *GCR* nuclear species (*Figure 1.3*).

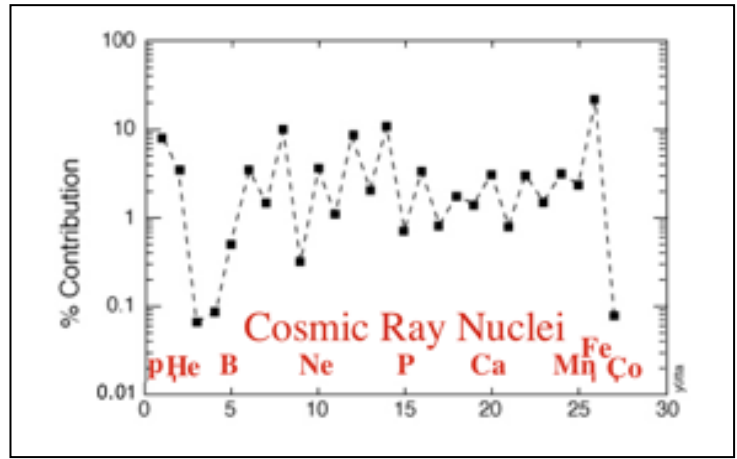

*Figure 1.3 Dose contribution of different CR species*

Considering the effect the *11 years* solar modulation, the fluxes of *GCR* are about *40-60%* lower during solar maximum (*Figure 1.4*): the corresponding doses are *60%* lower during this part of the solar cycle. This is a major difference in dose, which could be used as factor when planning for an exploratory mission.

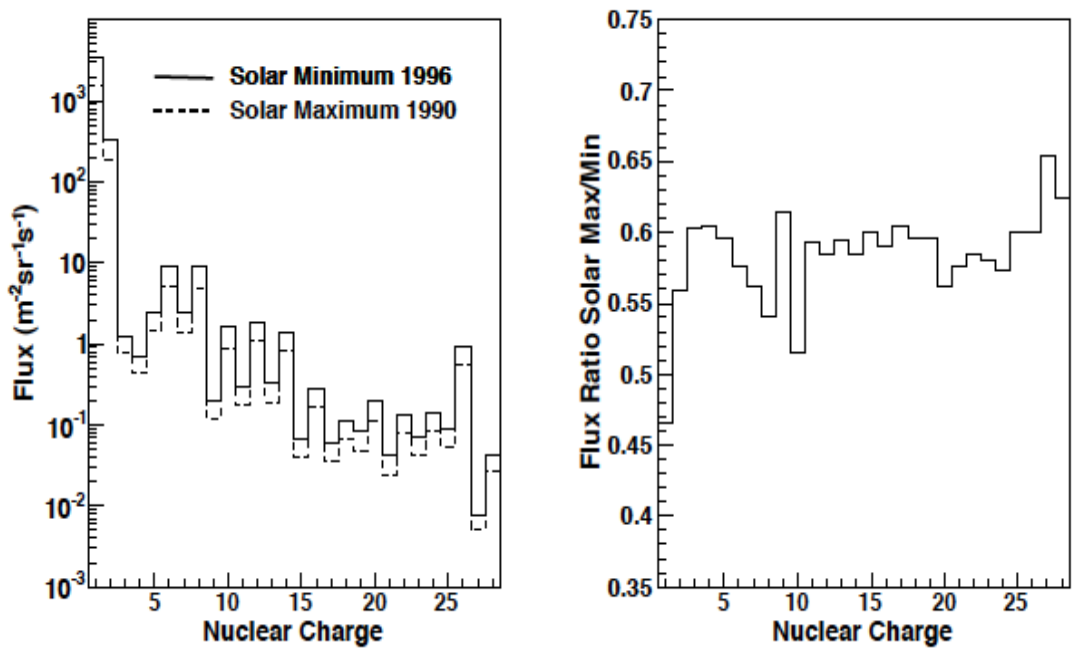

*Figure 1.4 The energy-integrated cosmic-ray nuclei fluxes for the solar minimum and maximum pe-*

*riods (left) and the flux ratio (right). The smaller reduction of the fluorine flux is explained by the absence of an anomalous component.*

Given the wide range of energies, flux intensities and amount of energy deposited by the different nuclear species, it is fundamentally important, for the scope of this work, to be able to reliably compute the absorbed doses for various shielding conditions by means of a full *Monte Carlo* (*MC*) simulation, including

- accurate composition and spectrum of the *GCR*
- detailed modeling of the spacecraft material, including the shielding and the related structural parts
- up to date interaction and fragmentation cross sections for *GCR*-matter interactions
- *3D* propagation of primary and secondary particles through the magnetic field

In this study we used a physics simulation code derived from the *AMS* experiment simulation code [3] and specifically optimized for the analysis of active magnetic shields: the physics simulation used in this work is expected to be significantly more accurate than other studies existing in the literature [4][5].

## 1.4 The AMS Physics Simulation code

The performance of the different shielding configurations has been studied using a full physics simulation based on *GEANT3* [6], which performs particle propagation in magnet fields and materials with a detailed treatment of electromagnetic interactions. Hadron interactions of protons, helium nuclei, the generated secondary singly-charged meson and baryons, and secondary deuterons and triton are simulated with *GEANT-FLUKA*[7]. The Relativistic Quantum Molecular Dynamics model (*RQMD*)[8] is used for the hadron interactions of higher charge nuclei.

The *GEANT3* simulation code is a modified version of the simulation/reconstruction program of the *AMS* experiment, which was used for a previous magnetic screen study. In terms of the simulated physics, *GEANT3* remains a reference for the electromagnetic interactions, the *FLUKA* hadron transport code incorporated in *GEANT3*, designated *GEANT-FLUKA,* remains a good reference for hadron interactions. The *RQMD* model continues to be used for hadron interactions in *GEANT4* applications.

The reliability of models depends on the availability of experimental data. The lack of data for the high charge nuclei, in particular concerning the biological effects, may be considered to be a more important uncertainty, than the relative merit of the different hadronic models quoted above.

The *AMS* software includes both *GEANT3* and *GEANT4* versions of the simulation. In the context of the *1* year *ESA* study, it was decided to update the *GEANT3* version in order to minimize the time devoted to program development.

The particles are traced through magnetic field as well as structural materials where they can release energy by ionization processes or by nuclear interactions: secondary particles are also traced through the volume of interest.

The ionization losses recored during the track propagation, $dE_i$, are converted to an equivalent dose $\varepsilon_i (Sv)$ by multiplying the absorbed dose $dE_i/m$ *(Gy)*, where *m* is the mass of the volume considered (*skin, Blood Forming Organ - BFO, whole body*) by the quality factor $Q(L)$ defined by the unrestricted linear energy transfer in water *L (keV/mu)*:

$$L\ (keV/\mu m):\qquad \varepsilon_i = Q(L) \cdot \frac{dE_i}{m}$$

$$\text{with}\quad L = \frac{dE_i}{dx}\qquad Q(L) = \begin{cases} 1 & \text{for} \quad L \leq 10 \\ 0.32\,L - 22 & \text{for} \quad 10 < L < 100 \\ 300/\sqrt{L} & \text{for} \quad L \geq 100 \end{cases}$$

The total equivalent dose $d_z(E_j)(Sv)$ for an exposure time *t*, released by *GCR* of charge *z* and kinetic energy $E_j$, is the sum of the recorded equivalent doses produced by the $N_j$ particles generated with the flux $f_z(E_j)$ ($cm^2\ sr^{-1} s^{-1}\ MeV^{-1}$) over the acceptance *A* ($cm^2\ sr$):

$$d_z(E_j) = \sum_i \varepsilon_i \cdot \frac{A}{N_j} \cdot f_z(E_j) \cdot t$$

The *GCR* equivalent dose *D (Sv)* is obtained by extending the event generation over suitable ranges in energy and charge. The contribution from charge *Z* is given by

$$d_z = t \cdot \sum_j \left[ \sum_i \varepsilon_i \right]_j \cdot \frac{A}{N} \cdot \sum_j f_z(E_j)$$

and the equivalent dose for *Z* up to *28 (Ni)*,

$$D = \sum_{z=1}^{z=28} d_z$$

The *GCR* kinetic energy spectra $f_z(E)$, *1* to $10^5$ *MeV/n*, are taken from *CREME 2009* database [9] (Figure 1.5). They include the contribution of anomalous cosmic rays which contribute at low energies. Depending on the needs, equivalent doses are evaluated at solar minimum or at solar maximum.

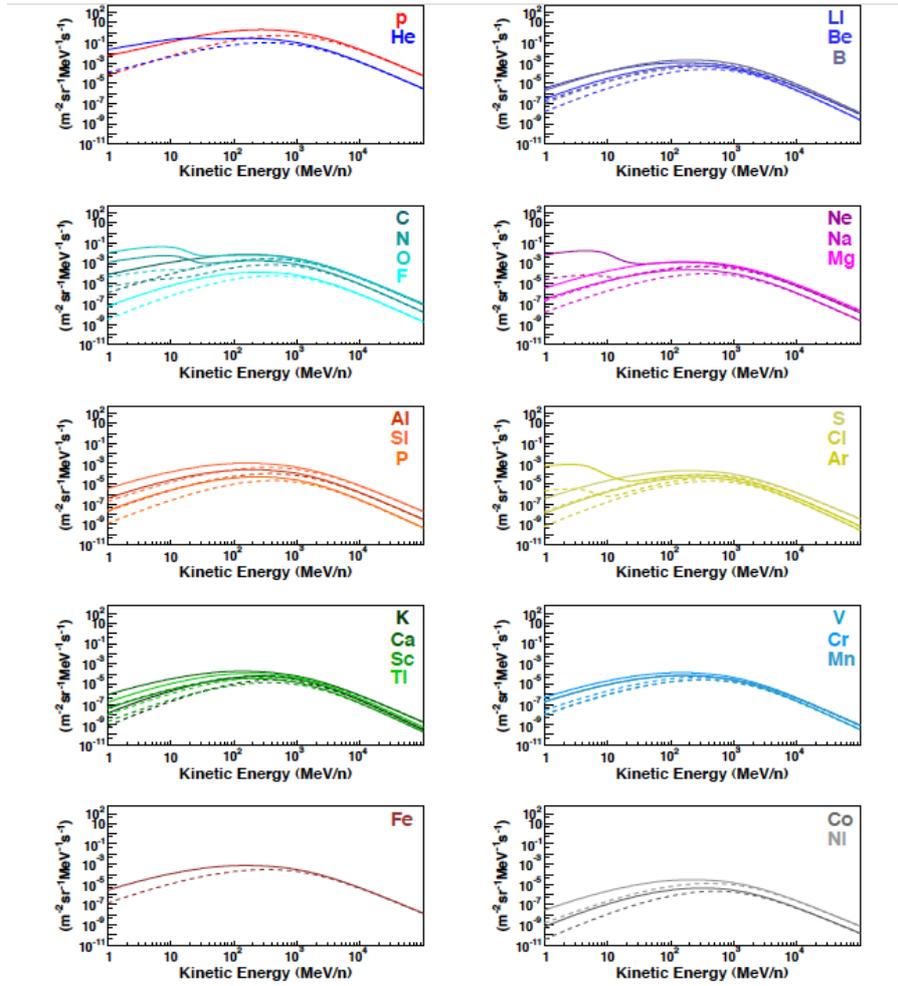

*Figure 1.5 The cosmic-ray energy spectra from the CREME 2009[9] data base for a solar maximum in 1990 (dashed) and a solar minimum in 1996 (solid).*

Using the formulas above is is possible to transform the *GCR* fluxes into doses as a function of the energy and *CR* species (charge), once we define volumes corresponding to the different parts of a human body: *skin, whole body* and *BFO (Blood Forming Organs)*. The integrated fluxes for the two periods are presented in *Figure 1.4*. A reduction of *45-55 %* is observed in the cosmic-ray nuclei fluxes between the solar maximum and minimum. Interestingly, the reduction at solar maximum is highest for the *proton* flux, *GCR protons* being one of the species causing the largest dose deposition and at the same time the component most difficult to stop using both passive and active means.

For the purpose of this study "*astronauts*" have also been modeled. From the absorbed dose point of view his simulation is much simpler than more detailed descriptions of the human body used elsewhere [10], but it is adequate for the scope of this work. The human body is represented in the simulation as a *24 cm* diameter, *180 cm* long, water filled cylinder. The cylinder is sub-divided in cylindrical volumes to define the regions used to compute the dose associated with the *skin* and *Blood-Forming Organs (BFO),* respectively the first *2 mm* at the surface of the cylinder and a *2 mm* thick layer located a depth of *5 cm* from the cylinder surface (*Figure 1.6*). The *whole body* dose re-fers to the ionization losses recorded in the full volume of the cylinder.

To test the AMS simulation code, these *"astronauts"* have been placed in different locations within the shielded volume (*Figure 1.7*) to measure the doses absorbed by the various part of their bodies, during different solar period and as a function of the electric charge of the incoming particles and of their energy *(Figures 1.8a, b)*. The results are summarized in *Table 1.1*.

A *< 40 %* reduction of the annual doses is observed between the solar maximum and minimum periods. The contribution of the higher charge nuclei ($Z > 2$) approximately *85, 75* and *70%* of the annual *skin, BFO* and *whole body* doses at solar minimum, respectively *90, 80* and *73%* at solar maximum. *GCR* protons, which are the most difficult to shield, are indeed quite effectively reduced by the solar activity. These values are consistent with previously reported results [11].

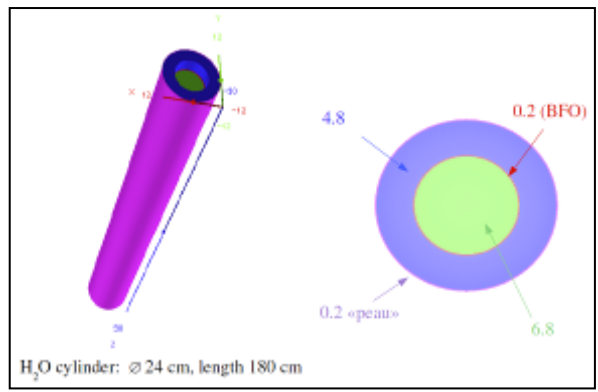

*Figure 1.6 The cylindrical water volumes used to compute the dose of the skin and blood forming organs (BFO). The total dose refers to the full volume of the 24 cm diameter, 180 cm long cylinder.*

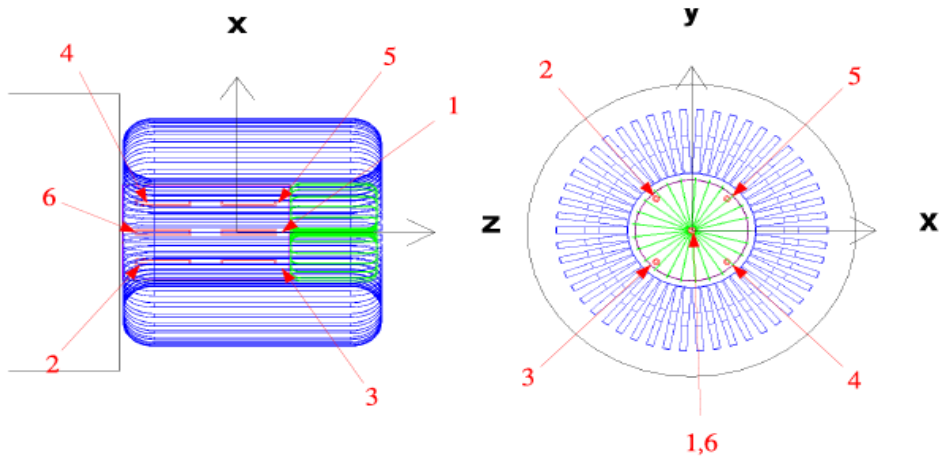

*Figure 1.7 The location of the water cylinders in the habitat.*

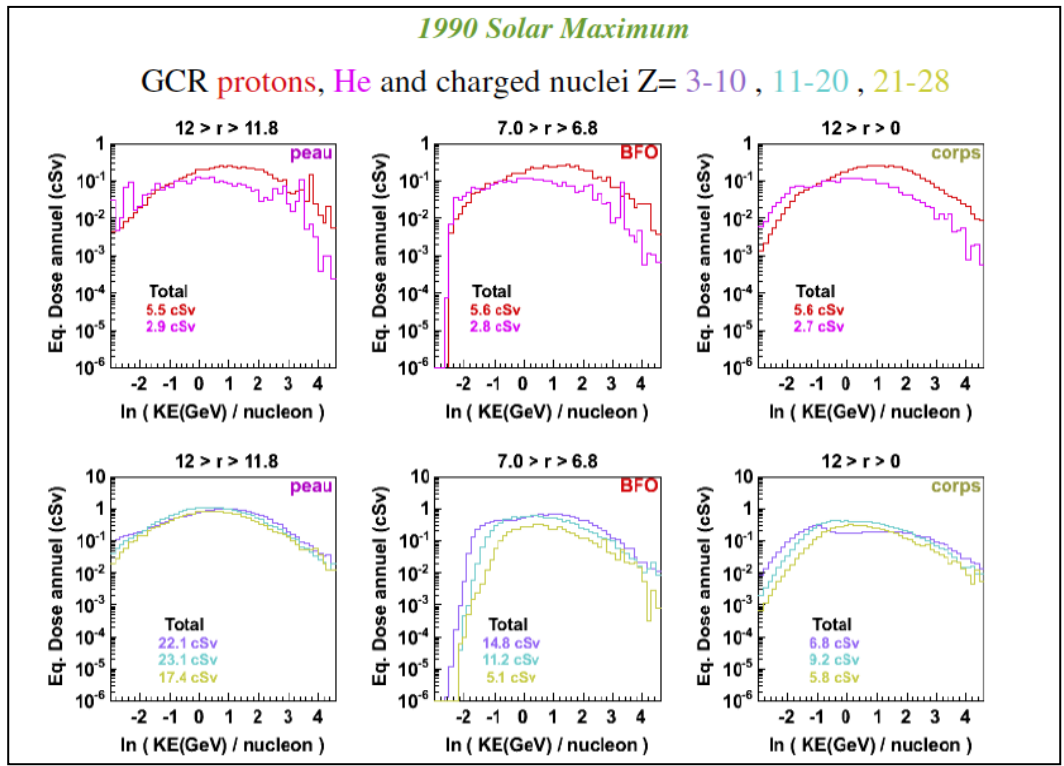

*Figure 1.8a Annual Equivalent Doses vs Kinetic Energy in Free Space (solar maximum)*

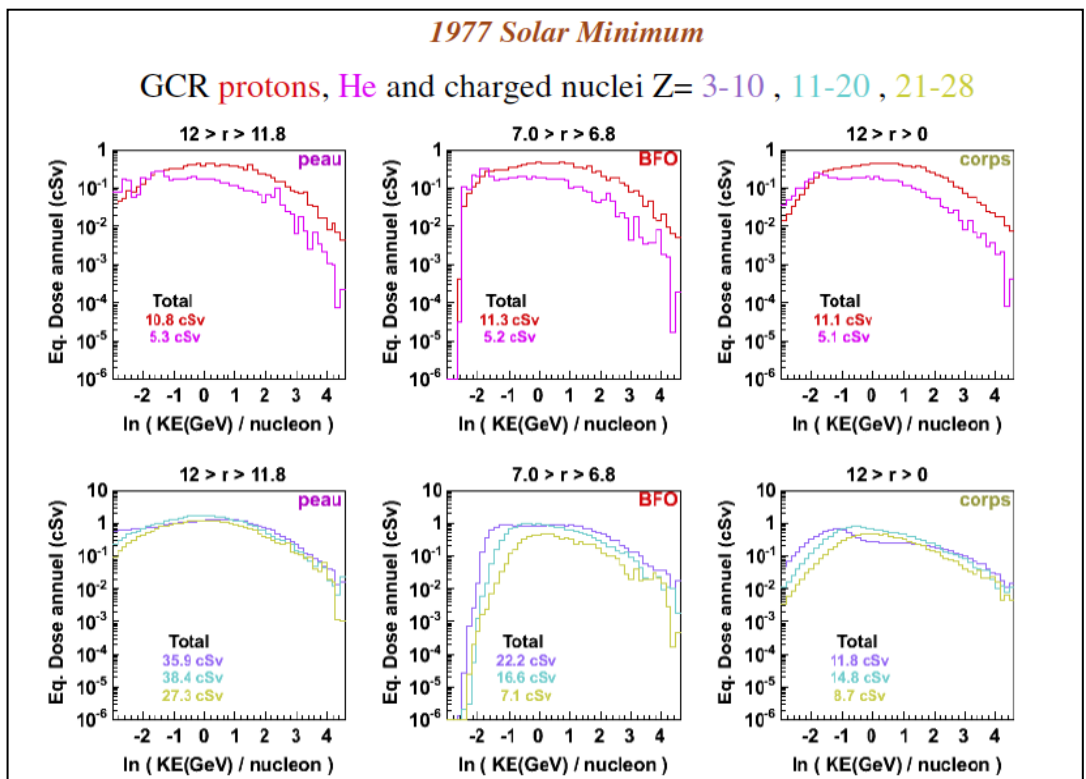

*Figure 1.8b Annual Equivalent Doses vs Kinetic Energy in Free Space (solar minimum)*

| Z | Solar Minimum | | | Solar Maximum | | |
|---|---|---|---|---|---|---|
| | skin | BFO | body | skin | BFO | body |
| 1 | 10.8 | 11.3 | 11.1 | 5.5 | 5.6 | 5.6 |
| 2 | 5.3 | 5.2 | 5.1 | 2.9 | 2.8 | 2.7 |
| 3-10 | 35.9 | 22.2 | 11.8 | 22.1 | 14.8 | 6.8 |
| 11-20 | 38.4 | 16.6 | 14.8 | 23.1 | 11.2 | 9.2 |
| 21-28 | 27.3 | 7.1 | 8.7 | 17.4 | 5.1 | 5.8 |
| total | 117.7 | 62.4 | 51.5 | 71.0 | 39.5 | 30.1 |

*Table 1.1  Annual Skin, BFO and Whole Body Equivalent Doses in Free Space (cSv/rem)*

*Table 1.1* is important, since it provides the reference yearly doses to be compared with the various cases when the astronaut will be placed behind an active/passive radiation shield. It also helps us to understand the following basic facts related to the radiation doses which could be absorbed in space.

· **Doses vary significantly with the phase of the solar cycle**. There is a *60%* difference between *solar maximum* and *solar minimum* total doses (factor *1,6*): however this difference exceeds *100%* (factor of *2,0*) in case of protons, *86%* for *He* (factor *1,86*), *66-71%* (factors *1,4* to *1,5*) for heavier ions *(Z>2)*.

· **Dose sensitivity is different on various parts of the body**. *Skin* is exposed to the highest dose, *90%* higher (factor *1,9*) than *BFO* and *130%* higher (factor *2,3*) than *whole body*. The *skin* is the most external part of the body and absorbs the highest dose. *Whole body* are shielding *BFO* as well as self-shielding: these parts get a factor *1,6* to *3,8* (on average a factor *2*) less dose than *skin*, the largest reduction taking place at higher Z since this shielding effect is due to ionization energy losses which are proportional to $Z^2$. It is then not possible to provide a single number to define the total absorbed dose: the most radiation sensitive parts of the body (typically *BFO*) will set the dose limits.

· **Doses strongly depend on *CR* species**. Although protons are by far most abundant (*85%* of the total flux), their dose is only a factor *2* higher than He (*14%* of the total flux), but it is *10* time lower (*skin*) than the contributions of ions having Z>2 (1% of the total flux). This factor of *10* becomes factor of *4* and of *3* for *BFO* and *whole body*, respectively, mostly because of self-shielding effects.

It follows that it is basically impossible to obtain reliable dose estimates without an accurate *MC* simulation. In additions to the trends described above here are additional subtle effects which influence doses as a function of solar cycles, *CR* species and body parts, causing *10-30%* fluctuations on the determination of doses. In order to obtain accurate prediction details on the shielding materials and active shield geometry surrounding the astronauts, modeling of their body as well of the mission and orbit parameters should be carefully studied and analyzed through a reference, detailed *MC* simulations involving all known effects.

## 1.5 Doses on exploration missions

*IN DEEP SPACE*

For transits to Mars the main concerns are exposures from large *SPE*s and chronic exposures from both *SPE*s and the background *GCR* environment. Since transit times of approximately six months are thought to be necessary, **effective doses in excess of *1.100 mSv /y (>110 rem/y)* in deep space** have been estimated from the *GCR* environment (see also *Table 1.1*). Much of this effective dose comes from *high-LET* components of the spectrum, such as high-energy heavy ions (the so-called *HZE* particles). Because of weight reasons, typical shielding thicknesses for interplanetary spacecraft are likely not to exceed ~ *8 cm* of aluminum (~*20 g cm$^{-2}$*) or other structural materials. Doses from large *SPEs*, mainly from energetic protons with energies as large as several hundreds of mega electron volts and higher, are likely to be well below any acute radiation syndrome response levels for spacecraft with ~*15-20 g cm$^{-2}$* of shielding.

*ON THE MARS SURFACE*

For operations on the surface of Mars, the main sources of concern are chronic exposures to *SPEs* and the *GCR* environment. Acute exposures to *SPE* protons are unlikely because the overlying atmosphere of Mars ( ~*16* to *20 g cm$^{-2}$* carbon dioxide) provides substantial shielding for all surface operations, except those that might take place at high mountainous altitudes. The overlying atmosphere on Mars will also provide some shielding against incident *GCR* particles. Especially important will be secondary neutrons, which come from nuclear fragmentation interactions between the incident protons and heavy ions and the atmosphere, and from albedo neutrons emanating from the Martian soil[12]. These neutron energies range from thermal up to hundreds of mega electron volts or more**. Then, the GCR dose on Mars is expected to be between *100* and *200 mSv/y (10-20 rem/y)*, depending on the location** (*Figure 4.10*).

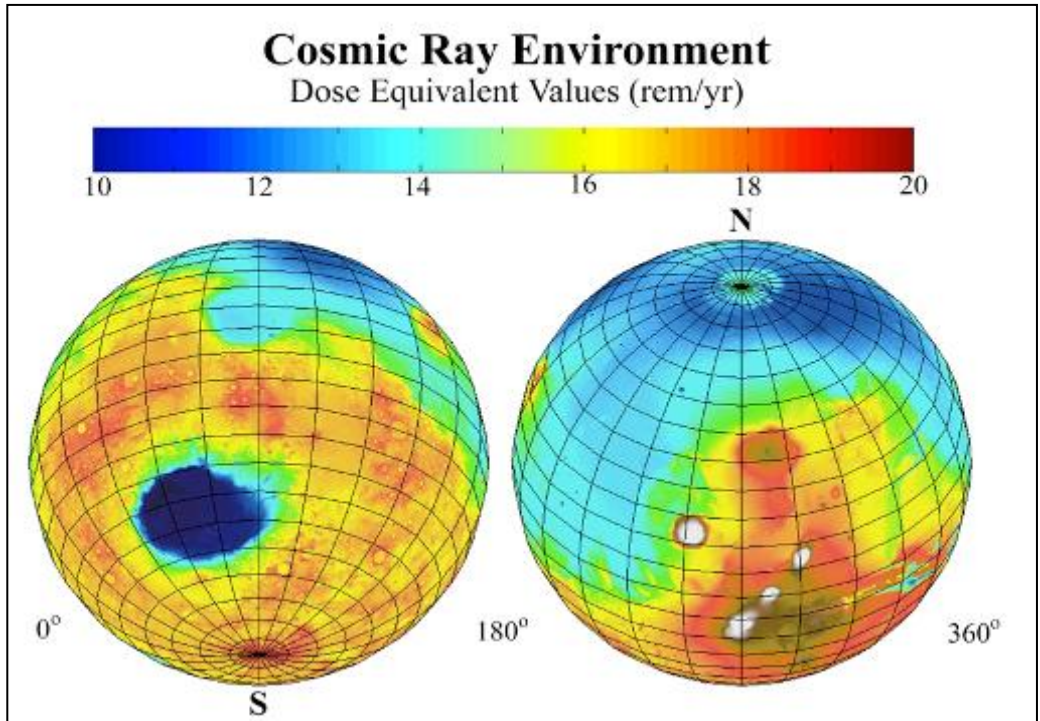

*Figure 1.9 Dose map on the surface of Mars* [13] *.*

*ON THE MOON SURFACE*

The ambient dose equivalent on the lunar surface is dominated by the contribution from galactic cosmic rays, secondary neutrons and γ rays from the lunar surface. **Total annual dose equivalent of about *223 mSv/y (22,3 rem/y)* on the lunar surface is expected during the quiet time of solar activity, with oscillations of ± *100 mSv/y* as a function of solar activity** (see *Table 1.2* [14]). The dose of neutrons and γ rays reaches to *50 mSv/y* and *20 mSv/yr*, respectively. In particular, fast neutrons have the largest contribution to the total neutron dose. The dependency of total neutron dose for mare and highland region of the Moon is small on the lunar surface if compared with that for solar activity

| particle | H*(10) [mSv/yr] in highland (A16) | | | in mare (A11) |
|---|---|---|---|---|
| | Solar Min. | Solar Ave. | Solar Max. | Solar Ave. |
| Fast neutron | 52.9 | 37.8 | 18.5 | 40.2 |
| Epithermal neutron | 19.6 | 13.9 | 6.8 | 16.0 |
| Thermal neutron | 0.35 | 0.24 | 0.13 | 0.15 |
| Neutron total | 72.9 | 51.9 | 25.4 | 56.3 |
| GCR | 233.8 | 168.6 | 65.8 | 168.6 |
| Gamma ray | 3.3 | 2.5 | 1.6 | 2.5 |
| total | 310.0 | 223.0 | 92.8 | 227.4 |

*Table 1.2 Yearly doses on the Moon surface in different solar conditions (mSv/y)* [14]

## 1.6 Radiation Exposure Limits

No radiation exposure limits have been established yet for exploration missions above LEO, to the Moon, Mars or other Near Earth Asteroids (*NEA*). As guideline the limits recommended for Low Earth Orbit (*LEO*) operations are used [1] [15] – see *Tables 1.3a,b*: they corresponds to the level of radiation that would cause an excess risk of *3%* for fatal cancer. These values are the ones that *ESA* currently uses for mission planning and operations [15].

| Exposure interval | Dose Equivalent (cSv) |
|---|---|
| 30 days | 25 |
| Annual | 50 |
| Career | See table below |

Table 1.3a  Recommended organ dose equivalent limits for all ages – from NCRP 132 (2001) (cSv/rem)

| Age | 25 yrs | 35 yrs | 45 yrs | 55 yrs |
|---|---|---|---|---|
| Male – Dose eq. (cSv) | 70 | 100 | 150 | 190 |
| Female – Dose eq. (cSv) | 40 | 60 | 90 | 160 |

Table 1.3b  LEO career Whole Body effective dose limits – from NCRP 132 (2001) (cSv/rem)

Since radiation doses absorbed during a space mission might be reduced, not eliminated, a strategy called *As Low As Reasonably Achieavable (ALARA)* is used, with the goal to define the smallest possible amount of dose which can be acceptable for a given mission scenario. It is obvious that both the reduction of the carcinogenic risks mentioned above as well as the *ALARA* strategy call for the most efficient, although practical, shielding techniques.

However, the uncertainties in biological effects of space radiation are still large and, when considering a *95%* confidence level of remaining below the *3%* risk of fatal cancer, a factor of *3* of reduction of the dose could be applied to increase the safety margin. These uncertainties could only be reduced by improving the knowledge of *GCR* and *SPE* biological effects [1]. This is illustrated in *Figure 1.10*[16], where the cancer risk estimates and the *95%* upper limit after a travel to Mars, are plotted vs. shielding thickness. Dashed and continuous lines represent water and aluminum shields, respectively. The median estimate and the upper *95%* confidence level are shown. The calculation was performed using the *HZETRN* deterministic code used by *NASA* [16].

*Table 1.4* shows the result of dose calculations performed using the *AMS* code for different scenarios: we see that in most cases the dose are below the yearly limit provided by *NCRP 132 (Table 1.3a)* even in the unrealistic case (*Deep Space*) where there is no shield at all around the astronaut. The most exposed body parts are *BFO* and *Whole Body (WB)* during *solar minimum*, when the *50* *cSv/y* limits are exceeded. However, without an efficient shielding, during a *2-3* years mission taking place during *solar minimum* career dose limits would be exceeded for most cases listed in *Table 1.3b*.

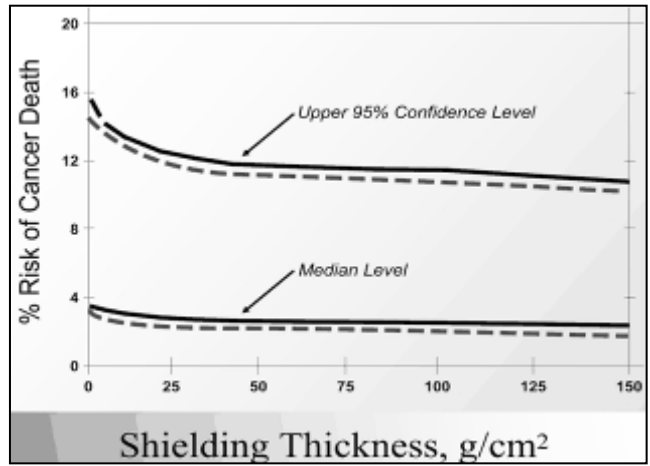

*Figure 1.10 How the cancer risk for a mission to Mars varies for increasing amounts of (passive) shielding materials after considering the tissue shielding of the human body* [16].

| Dose | Skin (cSv/y) | Skin (cSv/y) limit | BFO (cSv/y) | BFO (cSv/y) limit | WB (cSv/y) | WB (cSv/y) limit |
|---|---|---|---|---|---|---|
| Deep Space (solar maximum, this work) | 71 | 300 | 39,5 | 50 | 30,1 | 50 |
| Deep Space (solar minimum, this work) | 117 | 300 | 62,4 | 50 | 51,5 | 50 |
| Moon surface (solar maximum, this work) | 36 | 300 | 19,5 | 50 | 15 | 50 |
| Moon surface (solar minimum, this work) | 58 | 300 | 31 | 50 | 26 | 50 |
| Deep Space (solar minimum, HZETRN*) | 107 | 300 |  | 50 |  | 50 |
| Inside ISS (average, measured) | 20 | 300 |  | 50 |  | 50 |

*Table 1.4 GCR yearly doses and limits for various mission scenarios. In most cases the calculated doses are lower than the current professional yearly limits (which are defined for LEO missions). In some cases (red) they exceed these limits. The 95%/3% confidence level/risk criterion discussed in the text would require a shielding capability for an interplanetary mission matching the reduction of a factor of about three of the BFO yearly limits, namely to about 170 mSv/y.*

## 1.7 References

1) SP-1264: The HUMEX Report
2) Exploration Reference Architecture Document, ESA doc. HSF-EA/HSP/DOC/OM/2009-05002 iss.1/0
3) V. Choutko, H. Hofer and S.C.C. Ting, "The AMS Experiment and Magnet Faraday Cage for Human Space Exploration", presented at the NASA Active Radiation Shielding Workshop, Ann Arbor, MI, August 17-18, 2004.
4) J. Hoffman, P. Fisher and O. Batishchev, Use of superconducting magnet technology for astronauts radiation protection", NASA Institute for Advanced Concepts, Phase 1 Report Final (2005), American Physical Society Bulletin 49 (2004) 261.
5) P. Spillantini, Advances in Space Research 43 (2010) 900-916; P. Spillantini, Acta Astronautica 68 (2011) 1430–1439
6) GEANT - Detector description and simulator tool, CERN Program Library Long Write-up W5013, CERN, Geneva (1993).
7) A. Fassò, A. Ferrari, J. Ranft and P.R. Sala, FLUKA: present status and future developments, Proceedings of 4th International Conference on Calorimetry in High Energy Physics, La Bidola (Italy) 21-26 September 1993, World Scientific, pp. 493-502.
8) H. Sorge, Physical Review C, volume 52, number 6 (1995) 3291.
9) Cosmic Rays Effects on Micro Electronics (CREME), https://creme.isde.vanderbilt.edu/CREME-MC
10) Cucinotta FA, Kim MY, Willingham V, George KA. (2008) Physical and biological dosimetry analysis from International Space Station astronauts. Radiat. Res., 170:127–138.
11) Radiation Effects and Shielding Requirements in Human Missions to the Moon and Mars, D. Rapp, Mars Volume 2, 46-71 (2006); doi:10.1555/mars.2006.0004.
12) Martian Radiation Environment Models (MarsREM) Project: Contract Final Report.
13) http://mars.jpl.nasa.gov/odyssey/gallery/latestimages/latest2002/march/Fig.4_marie.jpg
14) K. Hayatsu, et al. Environmental radiation dose on the moon, Proceedings of ICAPP 2007.
15) Report of the ESA Topical Team in Life & Physical Sciences, "Shielding from Cosmic Radiation for Interplanetary Missions: Active & Passive Methods".
16) Durante M. and Cucinotta FA, Heavy ion carcinogenesis and human space exploration. Nat. Rev. Cancer 8 (2008) 465-472.

# 2 Previous studies on radiation shielding in space

## 2.1 Passive Shielding

Protons and nuclei are subject to electromagnetic and strong nuclear interactions. The latter produce lower energy secondary charged particles and neutrons, which if not contained in the absorbing material represent a higher radiation risk than that of the ionization produced by the initial particle. Since the discovery of radioactivity, there is a vast experience related to radiation shielding on ground. By using a sufficiently **thick** slab of suitably chosen material any flux of charged particles or neutrons can be exponentially reduced.

The situation in space is, however, quite different, since mass is very costly and it is very inefficient to put an heavy shield in orbit. It follows that, in the case of an interplanetary mission, the available shield around an *ISS*-size habitable module will necessarily be **thin** : considering that *5 g/cm$^2$* is the typical surface density of the walls of an *ISS* module (about *2 cm* of *Al*), a passive thickness of *20 g/cm$^2$* (as we will see, a modest value for *GCR* shielding purposes) would increase by a factor of *4* the module weight.

The energy lost in a given material via electromagnetic interactions is proportional to $Z^2/\beta^2$ where Z is the charge of the particle and $\beta=v/c$ its velocity expressed in terms of the speed of light. The critical parameter is the *maximum particle energy* to be stopped, since it defines the required thickness and consequently the mass of the passive absorber. The use of proton-rich materials, such as water or polyethylene, reduces the production of fragment nuclei, protons and neutrons by spallation, and lowers the radiation risk for a given amount of shielding weight. Practical consideration related to the mass to be transported in space, however, will limit the usage of shielding materials with surface density of the order of *20 g/cm$^2$* ( *8 cm* of *Al* or *20 cm* of *polyethylene*), corresponding already to about *40 t* of shielding structure for a module having *3 m* of radius and *10 m* of length. This level of shielding would be adequate for shielding from *SPE* but it would not significantly reduce the dose due to *GCR*. In addition it would create secondary particles which tends to increase the radiation dose to which the astronauts are exposed, partially destroying the shielding effect of a *thin* absorber.

In general, a passive shield of a given thickness would stop ionizing radiation up to a fixed energy, depending on the particle mass and charge. For a given material thickness, above this given energy, there will be leakage of particles through the shield. *Figure 2.1* shows the stopping power of an *Al* slab for low energy *protons* and as a function of the *proton* energy: from this curve we can derive that *2-4-10 cm* (*5-10-25 g/cm$^2$*) of *Al* can stop *protons* having energies up to *50-70-125 MeV*, reducing quite effectively the flux due to a *SPE*, but having a very little effect on *GCR*.

Spacecraft structures have traditionally been made of aluminum, and many studies and measurements have been done to show cosmic ray shielding by different thicknesses of aluminum. Passive shielding can be quite effective for particles with high Z and/or energies *< 1 GeV/nucleon*.

However, for higher energy or low *Z* particles, nuclear interactions with the shielding material produce secondary particles which can actually be more numerous and detrimental to astronaut health than the incident radiation.

Figure 2.2[1] shows that the first ~*10 cm.* of aluminum partially reduces the radiation, in particular in the case of *solar minimum* where the low energy part of the *GCR* spectrum still increases: however doubling the thickness produces almost no more net reduction.

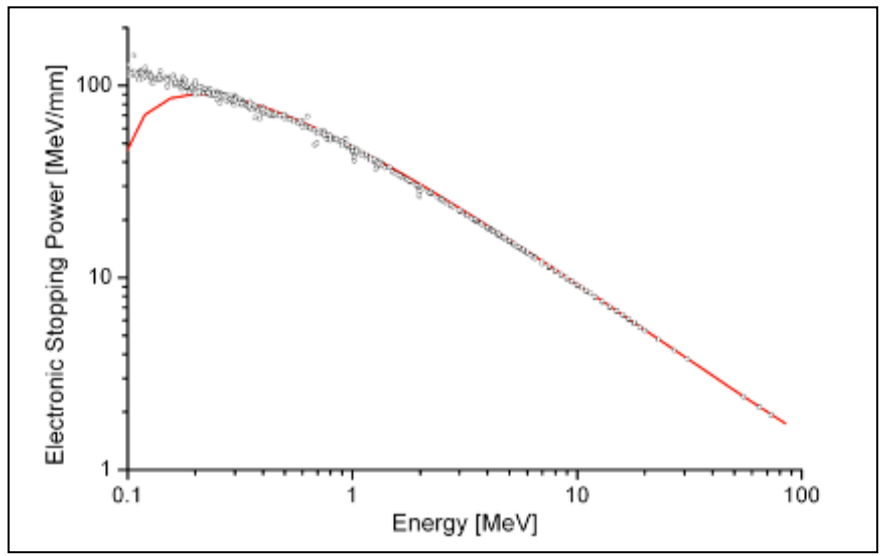

*Figure 2.1  Aluminum stopping power*

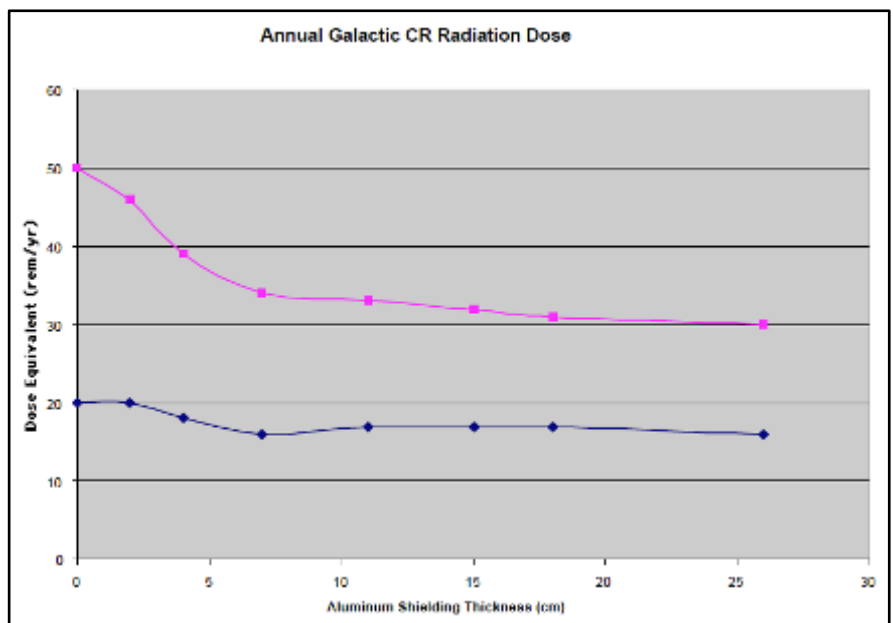

*Figure 2.2  Radiation doses received from galactic cosmic radiation with different thicknesses of passive aluminum shielding. (Top curve: solar minimum; Bottom curve:solar maximum)* [1]

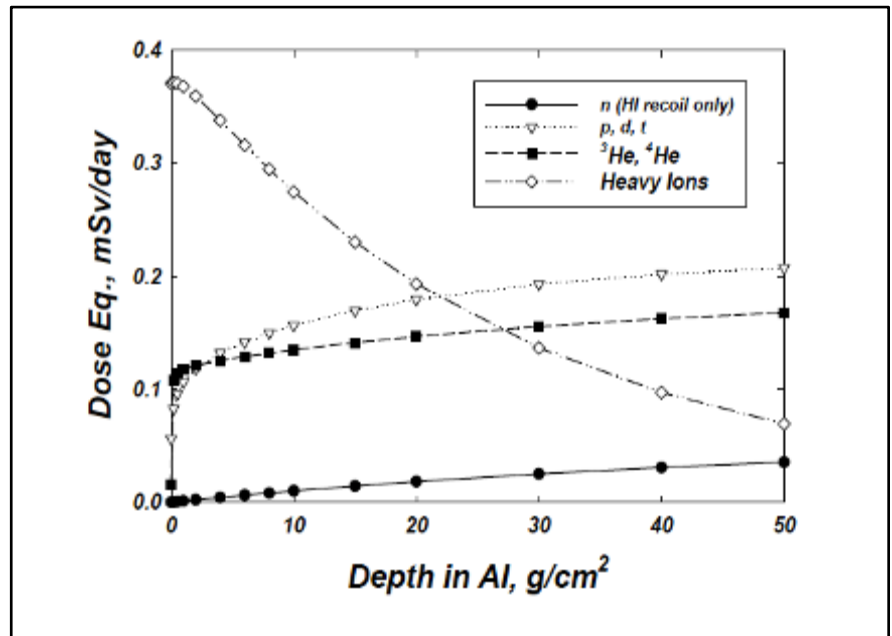

*Figure 2.3 Effect of passive absorber on the doses of different GCR species.*

In addition, looking more carefully to the dependence of the *GCR* dose on the passive shield thickness, we note that for a given *thin Al* thickness, as the one available in space, the proton component actually increases, as well as the neutron component, while only the dose due to *High Z GCR* is reduced (*Figure 2.3*). This effect is quite subtle, and underline the importance of full physics simulations in order to understand quantitatively the shielding effectiveness of an active (or passive) system.

The production of secondary particles by *GCR* shows some dependence on the shield atomic number: *hydrogen* is a better shielding than *Al* and *polyethylene* it is probably a more practical material to be used for shielding purposes (*Figure 2.4*). However for *thin* shields these effects are *small*.

In conclusion there is consensus on the fact that realistic passive shields would not be able to significantly reduce the dose due to *GCR*, although would be useful for the construction of small shelters to temporarily protect the astronauts from the occurrence of large *SPE*. The usefulness of an active radiation shield during an interplanetary mission is then evident.

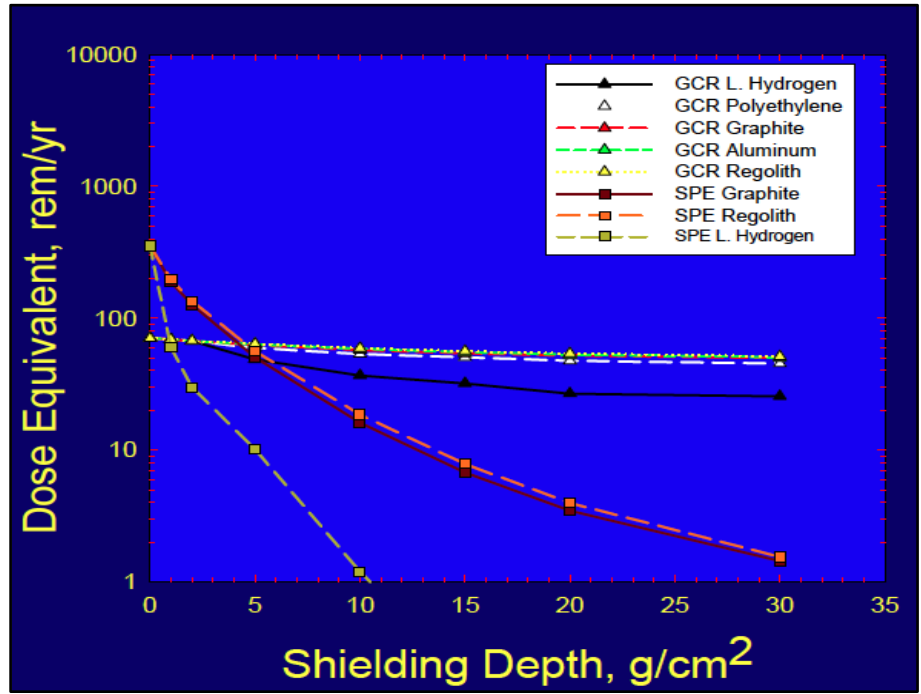

*Figure 2.4 Effective doses vs depth in several materials for GCR at solar minimum and the 1972 SPE* [2]

## 2.2 Active Magnetic Shielding using Superconducting Magnets

The particle *"containment"* by a magnetic field *B* consists on the Lorentz deflection of the particle trajectory, by an amount proportional to *BL/R*, where *B* is the field flux density, *L* the particle path length and *R* the particle rigidity *(R=pc/Ze)*. In principle, the value of *BL* may be chosen to protect efficiently a geometric volume up to the rigidity where the residual radiation dose represented by the penetrating particles is acceptable. In practice, the specific values of *B* and *L* have important consequences for the engineering design; consequently both *confined* and *unconfined* magnetic shield configurations have been considered. Here we will only consider *confined* magnetic solutions.

Since the magnetic field does not degrade the particle energy and secondary particle production is limited to the coils and supporting structures, the radiation risk represented by the "*secondaries*" is expected to be considerably less than for the *"punch-through"* particles created by a passive absorber. The principal uncertainty for a magnetic shield design is the radiation risk associated with high-charge-high-energy (*HZE*) *Cosmic Rays*, due to the lack of knowledge of the biological effects of the high-charge nuclei[3]

The most promising configuration considered consists on a toroidal field around the habitable part of the spacecraft. The principle of the toroidal shield is shown in *Figure 2.5a,b*. Particles entering the magnetic volume defined by the coils (red), will be subjected to the Lorenz force ***F*** *= q **v*** × ***B** /c = q v B sinθ/c* where *θ* is the angle between the particle's velocity and the magnetic field.

If they travel in a radial direction, orthogonal to the toroidal axis (*Figure 2.5a*), they are def-lected back, with a radius of curvature defined by the formula $\rho = m\gamma v / \kappa B$ where $\kappa = 0.3 GeV/T\ m$. When they reach a kinetic energy corresponding to a gyro radius which is <u>equal to the coil thickness L</u>, they are able to penetrate in the habitable module.

If they travel in a radial direction, at an angle φ with the toroidal axis (*Figure 2.5 b*), the maximum energy at which will be deflected back is a function of φ. In the most unfavorable conditions, particle having a kinetic energy corresponding to a gyroradius which is equal to <u>half of coil thickness</u> L, are able to penetrate in the habitable module. This energy is defined as **cutoff radial energy** for a given magnetic field configuration since all particle with energy <u>below the cutoff</u> cannot enter the habitable module. For example *5 Tm* would correspond to $E_{cutoff}$ = *250 MeV* for *GCR* protons.

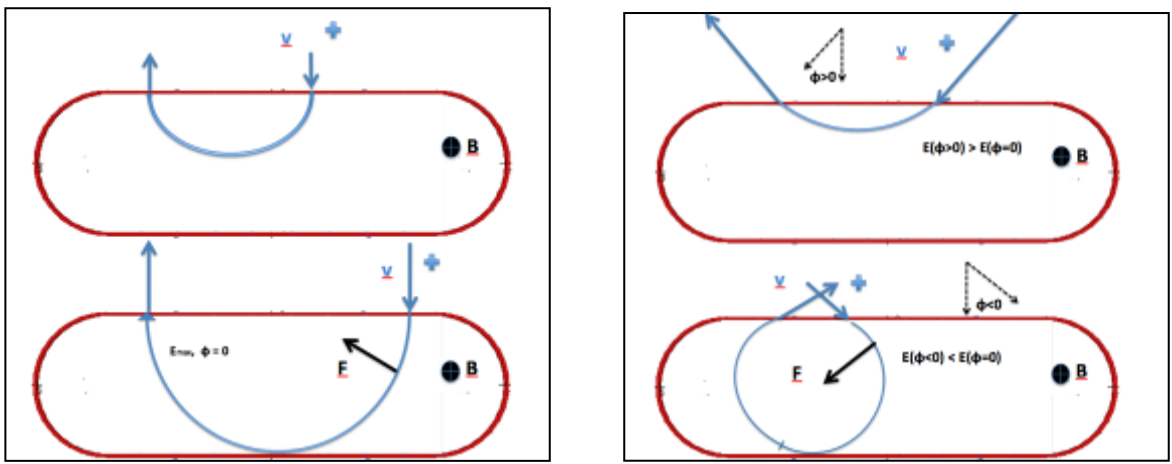

*Figure 2.5 Principle of the toroidal magnetic shield : a) magnetic field repulsion for particle traveling in a radial direction, orthogonal to the toroidal axis; b) magnetic field repulsion for particle traveling in a radial direction, at an angle φ with to the toroidal axis*

There are three studies of confined magnet shield configurations which present quantitative estimates of the expected dose rates.

**Hoffman et al.**[1] present results for a double-toroidal-solenoid design which shields a *7 m* long, *7 m* diameter field-free cylindrical volume in the center of the *endcap-barrel-endcap* configuration (*Figure 2.6*). The objective was to provide complete shielding in all directions for protons and nuclei up to *2-4 GeV/n*. Results are reported in terms of the expected reduction of the dose due to the galactic Cosmic Ray radiation. With a *9,2 T* field and a *1,7 m* path length, the simulation results suggest the annual dose received in interplanetary space would be decreased to *10%* of the total expected dose (*90 rem/y*). The total dose received during a *3 year* mission would be less than 10% of a the career dose level for a senior astronaut (*300 rem*). The authors also present very rough estimates for resulting the magnet mass *(400-1600 t)*, the liquid helium mass *(52-75 t)*, as well as of the cryocooler power (*117-169 kW*) and of the stored energy (*16 GJ*). These estimates are based on extrapolations from the parameters of the *AMS-02* superconducting magnet *(0.8 T)*.
This study is based on a number of simplifying assumptions, which would definitely affect its con-

clusions:
- the particle magnetic transport is based on a *2D* calculation
- there is no attempt to describe the mechanical structure of the magnetic system, even if it is stated that its mass will be very large
- the *MC* does not simulate the interaction of the particles, in particular *HZE*, with the material of the spacecraft or of the magnetic system and its structure

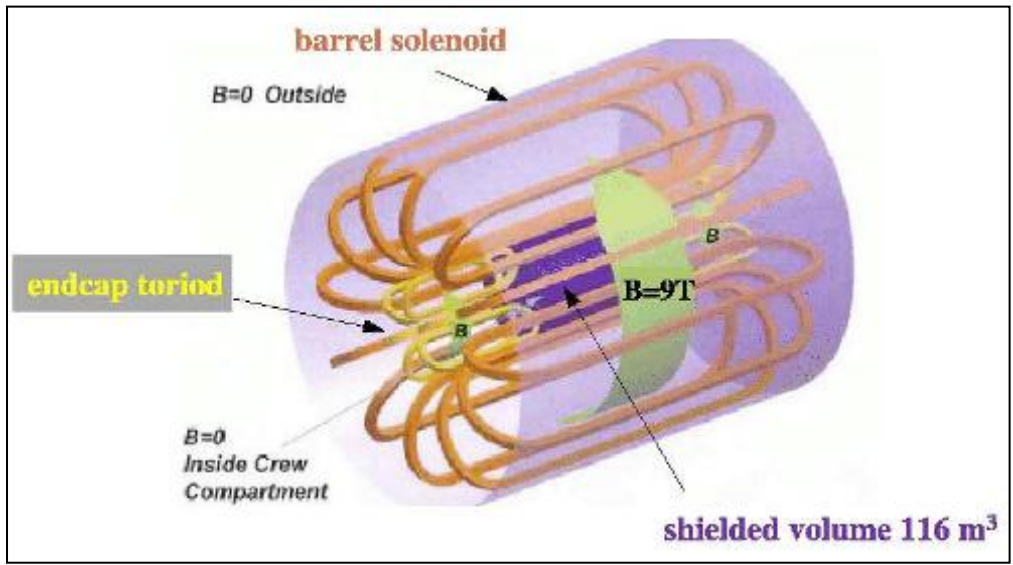

*Figure 2.6 Active shielding: toroidal field configuration considered in Ref. 4.*

**Choutko et al.**[4], from the *AMS* Collaboration, have presented results of a study of a magnet shield configuration consisting of a *single toroid endcap and toroid barrel*. The *"free"* end of the barrel is shielded by the propulsion system of the spacecraft. The largest shielded volume considered is a *5,5 m* long, *4 m* cylinder (Fig. 2.7). Both *"worst case" SEP* proton fluxes and the galactic cosmic ray protons and nuclei are included in the *GEANT3*[5] simulation, which incorporates *FLUKA*[6] for nuclear interactions. The results are reported in terms of the annual dose for *Blood Forming Organs* (*BFO*) in the human body, represented in the simulation by a *1,8 m* long, *0,12 m* diameter water cylinder. For an average *BL* of *17 Tm*, the annual *BFO* dose is estimated to be *13-24 rem*. The magnetic system weight is estimated at *31 t*, however no details are provided about its design; the mass of an equivalent performance aluminum absorber is estimated at *800 t*. The study is based on a state of the art *MC* simulation code, which can not only transport all kind of cosmic ray particles in a *3D* magnetic field but also simulate their interactions with the spacecraft or with the magnetic system material, including the effects of the secondaries particles produced in these interactions. There is no attempt, however, to realistically consider the magnetic system mass and supporting structure of the large magnetic system which is proposed and this would affect the conclusions of the study.

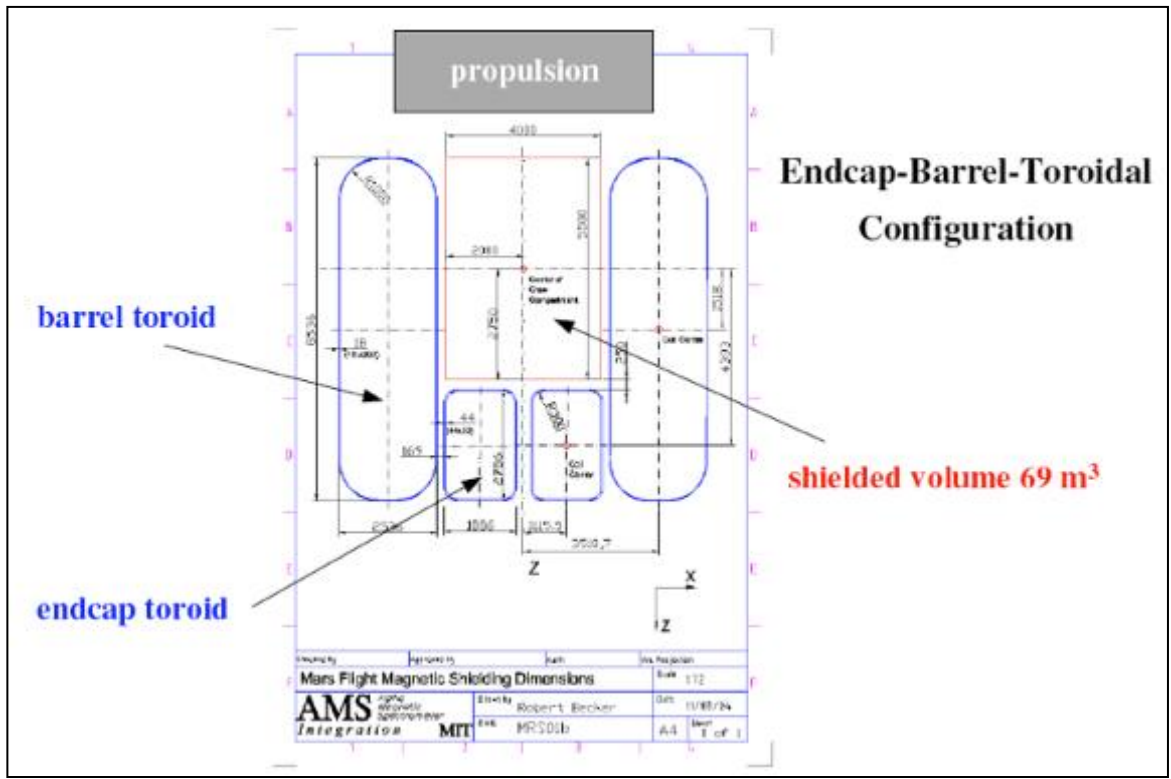

*Figure 2.7 Active shielding: one of several toroidal field configurations considered in Ref. 4*

**Spillantini et al.** [7] studied different scenarios considered by the Topical Team "*Shielding from cosmic radiation for interplanetary missions: active and passive methods*", under the auspices of the European Space Agency (*ESA*). The toroid barrel shield considered protect a *10 m* long, *6 m* ▢ cylindrical habitat (*Figure 2.8*). With a *2 m* path length, the distance between $R_1$ and $R_2$, the galactic cosmic ray proton flux is expected to be reduced by *34%, 59%, 75%, 80%,* for field strengths at $R_1$ of *2, 4, 6* and *8 T* (*Figure 2.9*). The doses are assumed to scale in the same way. However there is no attempt to simulate by *MC* the interaction of protons and *HZE* with the spacecraft and magnet materials. In order to estimate the complexity of the magnetic system, the author present a rough estimate of its mass as a function of the external coil radius (*Figure 2.10*). For a given value of the bending power *BL*, the larger is the external radius the lower is the *B* field and correspondingly ligh-ter are both the coils and the structural parts. However exceeding an external radius of 5 *m* would require the development of a new class of launchers. For example, a *5 m* radius, *12,5 Tm* magnet would, according to this estimate, weigh *62 t*.

The study does not make use of *MC* simulation while uses simple scaling laws to estimate the magnetic system mass and supporting structure of the large magnetic system which is being consi-dered proposed. These simplifications would definitely affect the accuracy of the conclusions of the analysis presented.

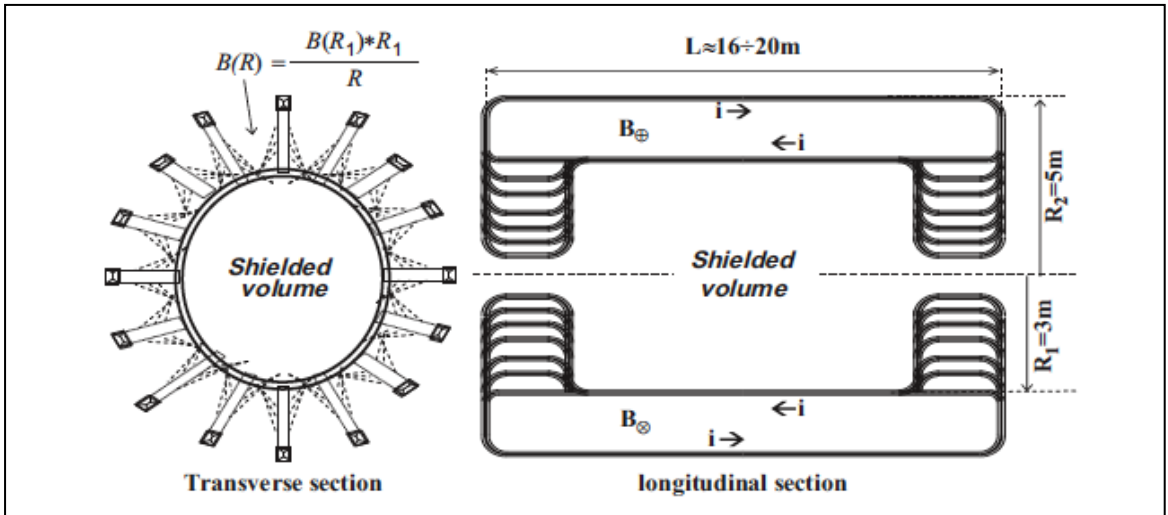

*Fig. 2.8. Configuration assumed to evaluate the protection of a 6 m diameter cylindrical habitat* [7]

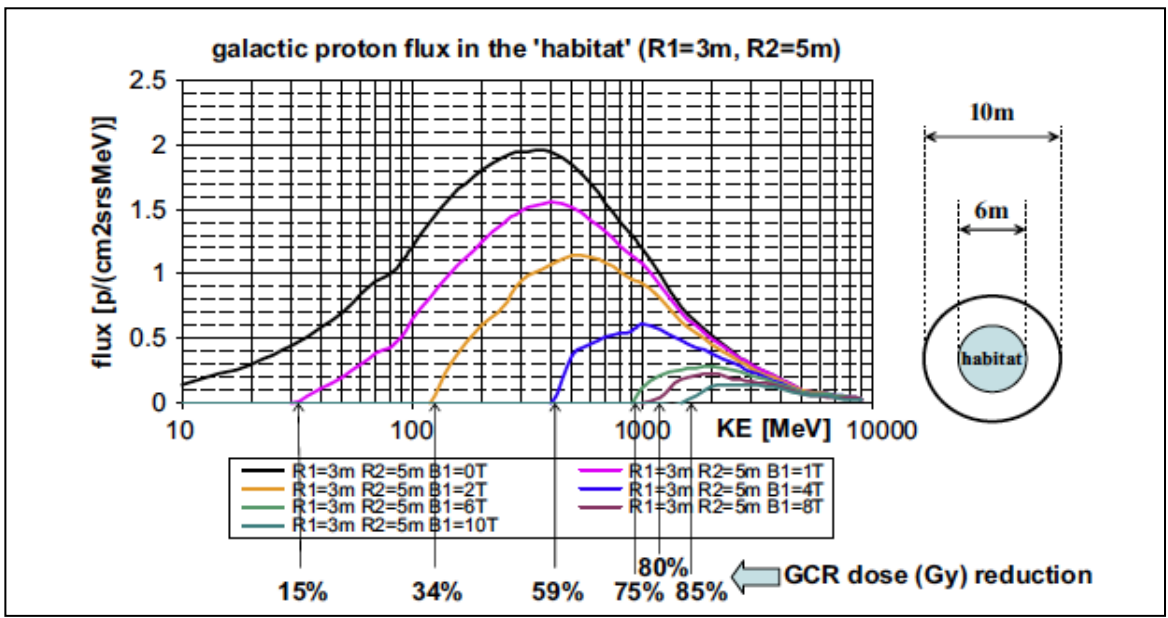

*Fig. 2.9. Reduction of the galactic proton flux inside the habitat. The corresponding reduction of the dose due to GCR flux is reported at the bottom of the figure for different values of the maximum magnetic field (1,2,4,6,8, and 10 T) of the system* [7]

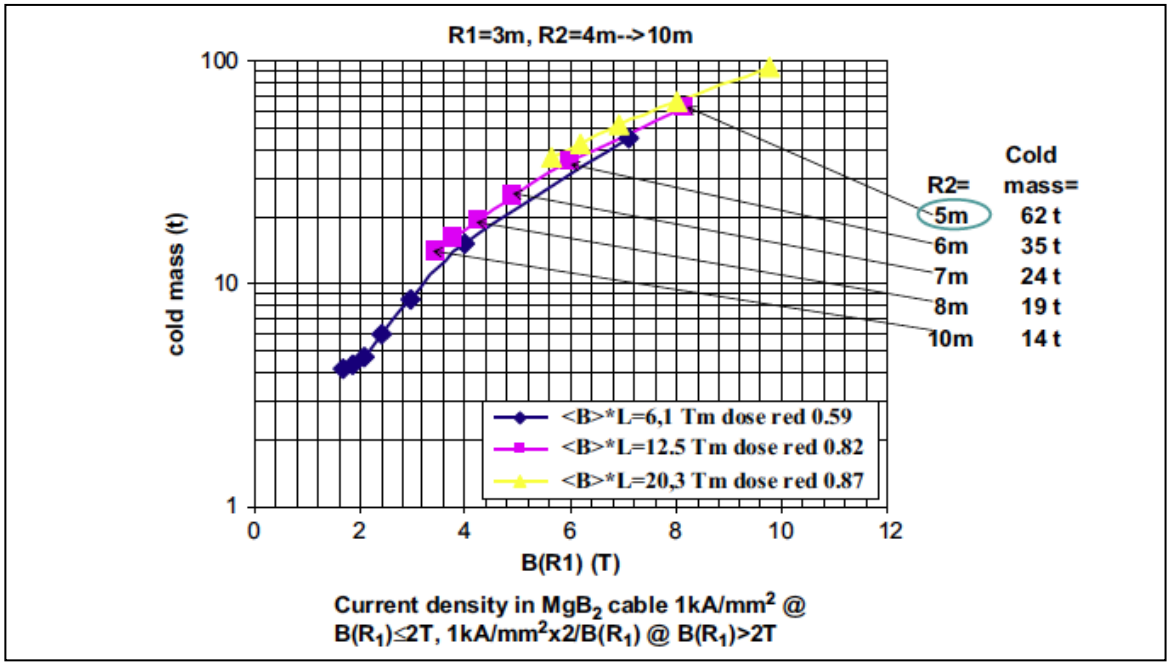

*Fig. 2.10 Superconductor mass of the system realized by $MgB_2$ SC cable, for the values 6.1, 12.5 and 20.3 Tm of the bending power <B>(R2 -R1) (corresponding to 0.59, 0.82 and 0.85 reduction of the GCR dose) and several values of the outer diameter as a function of the maximum magnetic field intensity* [7]

*Table 2.1* provides a comparison among the main parameters of the active toroidal shields studied in the literature. From the comparison of these results we can derive some considerations which are relevant for the study presented here:

1) the magnet weight estimates provided in all studies, although very approximate, results in very large weight budgets for a space project, in particular in the first and the third study;
2) it is necessary to perform a more accurate overall weight determination in order to establish which shielding efficiency with this kind of toroidal shield can be realistically proposed;
3) the two studies which did not perform a *MC* simulation of the interactions of the *CR* with the material of the magnet/spacecraft as well as the corresponding secondary particle productions, computed *Flux* or *Dose* reduction factors of *10*, while the study where the full *MC* simulation has been done only reached a *1,5-2* less effective shielding factor.

For these reasons in the the study presented here:

- first, we considered the issue of the magnetic system weight by designing two different types of toroidal magnetic systems and considering both the superconducting cable used in the coils as well as the structural supporting material;
- second, we implemented the material budget derived by these designs into a full *MC* simula-

tion able to take into account both the interactions of the particles with the material, the secondary particles production as well as the transport of the particles in the resulting magnetic field.

- third, we computed the radiation doses using the energy deposition by the different components of the radiation in the various body parts.

| Configuration | 1 Hoffman et al. | 2 Choutko et al. | 3 Spillantini et al. |
|---|---|---|---|
| Magnet Mass (t) | 400-1600$^{(1)}$ | 31$^{(2)}$ | 90$^{(3)}$ |
| BL (Tm) | 15,6 | 17 | 20,3 |
| Flux reduction factor | 10 | 4-7 | 10 |
| Dose (rem/y) | 9 | 13-24 | - |
| Diameter/Length (m) | 10/10 | 4/5,5 | 6/10 |
| Shielded Volume (m$^3$) | 269 | 69 | 282 |
| 3D Magnetic Transport | No | Yes | No |
| Full MC CR Simulation | No | Yes | No |
| Structural mass in MC | No | No | No |

*(1)   total mass including coil, mechanical structure, cryocooler, liquid helium*
*(2)   quoted as "magnet system weight"*
*(3)   cold mass x 1,5*

*Table 2.1  Summary of previous studies on toroidal magnetic shield systems*

## 2.3 References

1) J. Hoffman, P. Fisher and O. Batishchev, Use of superconducting magnet technology for astronauts radiation protection", NASA Institute for Advanced Concepts, Phase 1 Report Final (2005), American Physical Society Bulletin 49 (2004) 261.
2) Cucinotta FA, Kim MY, Willingham V, George KA. (2008) Physical and biological dosimetry analysis from International Space Station astronauts. Radiat. Res., 170:127–138.
3) R.B. Setlow, Mutation Research 430 (1999) 169-175.
4) V. Choutko, H. Hofer and S.C.C. Ting, "The AMS Experiment and Magnet Faraday Cage for Human Space Exploration", presented at the NASA Active Radiation Shielding Workshop, Ann Arbor, MI, August 17-18, 2004.
5) GEANT - Detector description and simulator tool, CERN Program Library Long Write-up W5013, CERN, Geneva (1993).
6) A. Fassò, A. Ferrari, J. Ranft and P.R. Sala, FLUKA: present status and future developments, Proceedings of 4th International Conference on Calorimetry in High Energy Physics, La Bidola (Italy) 21-26 September 1993, World Scientific, pp. 493-502.
7) P. Spillantini, Advances in Space Research 43 (2010) 900-916; P. Spillantini, Acta Astronautica 68 (2011) 1430–1439

# 3 Technology background for the design of an active magnetic shield

## 3.1 The AMS-02 Superconducting Magnet

*AMS-02* is a cosmic ray experiment which has been installed on the International Space Station (*ISS*) in May 2011. In its final configuration, the experiment consists of a system of permanent magnets and a suite of particle detectors. The function of the magnets is to provide a field which will alter the trajectories of incoming charged particles to assist in their identification: in particular, to distinguish between matter and antimatter.

In its original conception, the experiment used a superconducting magnet to generate the field. The development of the superconducting magnet took around *12* years from the first concepts to the final commissioning. Although not used in the final experiment configuration because of the limited lifetime of the liquid helium, the magnet was successfully commissioned and operated with the particle detectors both at *CERN* in Geneva and at the European Space Agency facility in Noordwijk. It is the first and only large superconducting magnet system designed, built and qualified to be operated in space.

The superconducting magnet system *AMS-02* consists of a pair of large dipole coils together with two series of six racetrack coils, radially distributed between them (*Figure 2.11*). The magnet is thermally insulated and mechanically suspended inside a vacuum case (*Figure 2.12*).

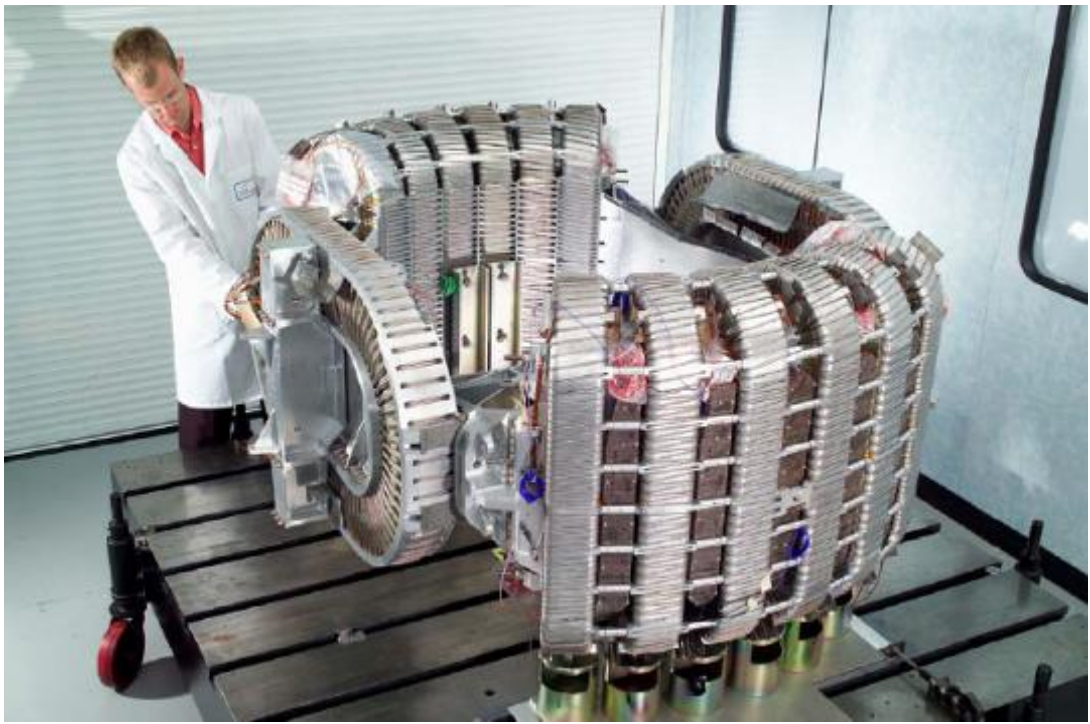

*Fig. 2.11 The coils of the AMS-02 Superconducting Magnet*

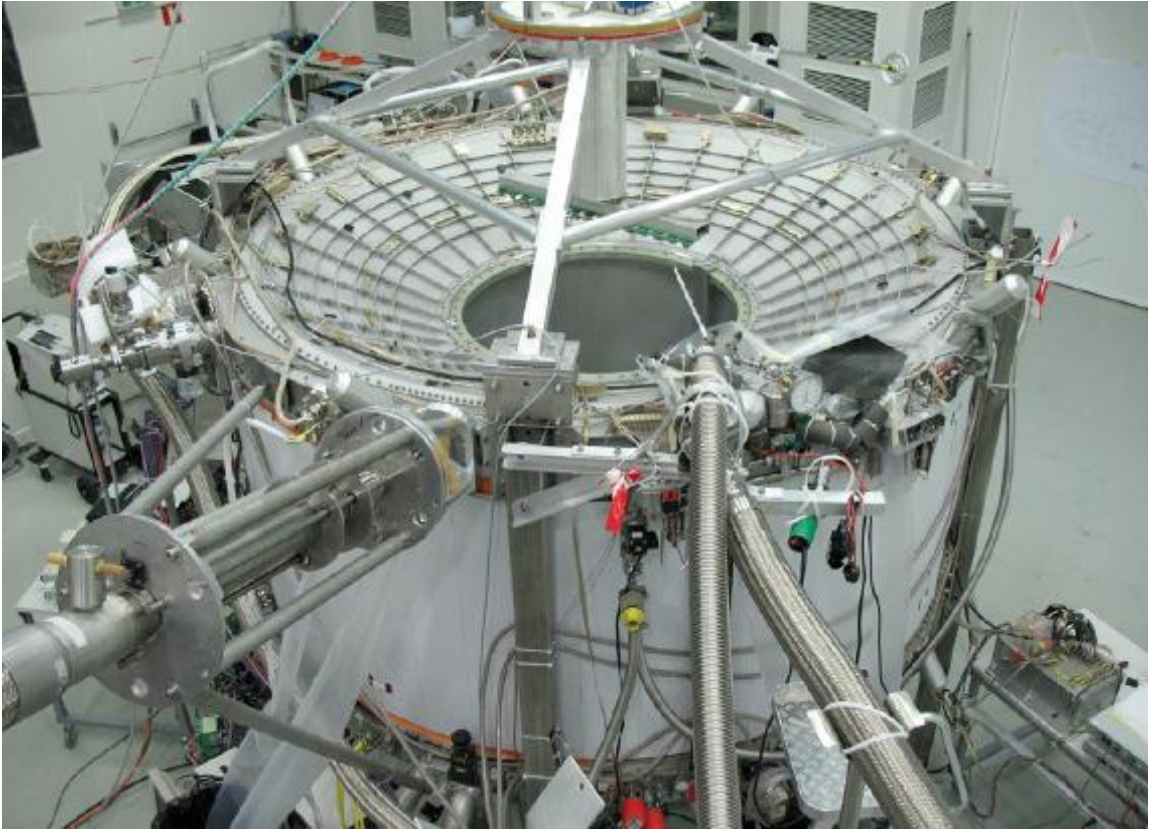

*Fig. 2.12 The AMS-02 superconducting magnet during the tests at CERN*

The principle of operation of the *AMS-02* superconducting magnet was based on the evaporative cooling of *superfluid He* of a low temperature *Al-stabilized SC NbTi* wire operating at *1.7 K* *(Fig. 2.13)*.

The choice of *superfluid He* cooling was based on the fact that:

- the extraordinary thermal properties of *superfluid He* allows for very efficient and uniform *"dry cooling by contact"* of the magnet wire
- the variation in *enthalpy* from *superfluid He* to *room temperature He* vapor maximize the amount of cryogenic power for unit of weight, thus the endurance, given that *LiHe* refilling was not an option for *AMS* on the *ISS*
- this large amount of *enthalpy* for unit of weight was used to maintain the temperature gradient between the cold mass and the external side of the vacuum vessel at room temperature
- the large variation of thermal conductivity from *SC* to normal *LiHe* was instrumental to allow for a design of the cooling pipe carrying the *SF He* to the *SC* coils limiting to about *30%* the amount of *LiHe* lost in case of a quench.

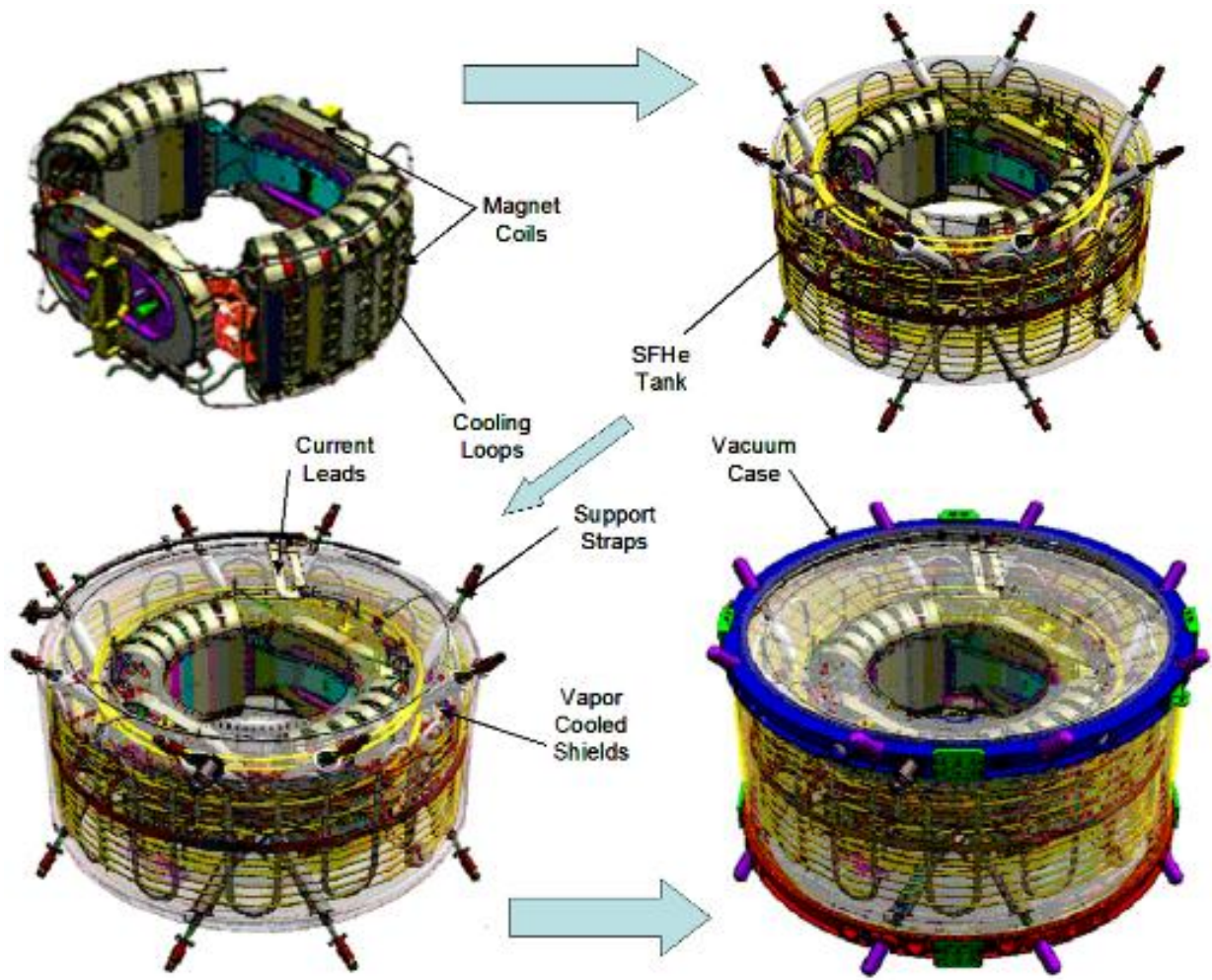

*Fig. 2.13 The principle of operation of the thermal system of the AMS-02 superconducting magnet: a) the magnet 14 coils operate at 1,7 K generating a dipole field of 0,8 T; b) the coils are maintained at low temperature by contact cooling with a pipe filled with superfluid He, going through the Liquid He vessel surrounding the coils; c) the magnet cold mass is thermally insulated by multilayers super-insulating sheets, thermal shields actively cooled by 4 cryocoolers and by the enthalpy of the evaporating He liquid; d) the super-insulated cold mass is suspended with 16 thermally insulated straps to a vacuum case.*

However, the choice of evaporative cooling of *SF Helium* in turn dictated a number of technical choices used in the *AMS-02* magnet

- the need to operate the cold mass of the magnet and the stored *SC He* at $T < 1.7\ K$
- the use of low temperature *SC* wire, *NbTi*. Operating at *1.7 K* (vs *9.2 K* critical temperature) this wire ensures high operational stability (insensitiveness to quenches) at remarkably high current in strong magnetic field
- the use of *SF* based thermo-mechanical pumps to remove the heat load from the *SF He* bath
- the design of a *SC* coil quenching system, based on a quench alarm signal triggering a set of heaters allowing fast transition of the whole volume of *SC* cable

*38*

- the need to use low temperature cryogenic valves operated by *He* gas
- the need for a extremely efficient super-insulation system separating the cold mass at *1.7 K* from the nearly *50 sqm* of vacuum case surface at *300 K*. *AMS-02* is in fact designed to be installed on the *ISS*, where it is impossible to be shielded from the Earth albedo and from the Sun direct radiation. The ratio between the radiative load at the vacuum case *(20 kW)* and the *He* vessel load (*O(100) mW* for a three years operation in space) is a staggering factor of $2 \cdot 10^5$, partly obtained by the evaporative *SC He* cooling and cryocoolers, but mostly obtained through the use of a large number of super-insulation layers.

## 3.2 AMS-02 heritage for future SC magnet for active radiation shielding

The experience gained on the *AMS-02* development is of basic importance when considering the design of future generation, space borne, large magnets to be used as an active shield against the ionizing radiation. This heritage suggested some fundamental decision about the design of future active SC radiation shields as those discussed in this study.

Long operation in space requires in practice indefinite endurance of the *SC* shield. It would be unrealistic to use evaporative cooling for a large *SC* magnet shielding to be operated in space during a long mission, for the following reasons:

- the amount of *He* to be used to cool a system having a size an order of magnitude larger than *AMS-02* (that is an exposed area two orders of magnitudes larger than *AMS-02*, and a cold mass volume/thermal inertia three orders of magnitude larger than *AMS-02*) would require an unrealistic *O(30) t* of mass of He to operate *O(1) year* in space;
- the magnet endurance depends linearly on the value of the thermal leak into the cold mass: this parameter must be the smallest possible number, depending on the manufacturing of the insulation system. If this *small* number turns out to be larger by, let us say, a factor of two, the endurance would be reduced by *50%* or the amount of *He* should be increased proportionally;
- over long period of time (three years), the cryogenic system should allow for a number of *SC* magnet quenches, without loosing the cooling fluid.

In addition, although *NbTi* cables currently have good current-density/mass characteristics their low operation temperature (< *4 K*) makes them not particularly attractive for a space application, both for the requirements on the cryogenic system and for the exposure to quenching instabilities. The continuous progress on *ITS* (Intermediate Temperature Superconductor) and *HTS* (High Temperature Superconductor) materials and cables properties, make them very attractive for a long term program like the one discussed in this project.

| SERVICES/HARDWARE | DEVELOPER | USEFUL TECHNOLOGY | COMMENT |
|---|---|---|---|
| UPS System Design | MIT (USA) | yes | Non-critical technology |

| SERVICES/HARDWARE | DEVELOPER | USEFUL TECHNOLOGY | COMMENT |
|---|---|---|---|
| Flight Cryocooler Electronics | INFN (I) | yes | Critical technology |
| Magnet | | no | New geometry, structure and superconducting cable |
| SFHe Tank Design | | no | Operation at higher temperature and with cold gas would not require this critical technology for storing SF Helium |
| VCS & VCS Supports | | yes | Non-critical technology |
| Non-linear Strap Design | | yes | Non-critical technology |
| Cryosystem | | yes | To be scaled up to larger volume |
| Cold Mass Replica | SCL (UK) | yes | Non-critical technology |
| Current Leads | | yes | Connection from higher to SC temperature will always be needed |
| Flight Barometric Switch | | yes | Non-critical technology |
| Vent System | | no | Space vacuum will be used, cooling in space will be by shadowing and cryocooling |
| Vent Pump | | yes | Non-critical technology |
| Burst Disks (Complete System) | | yes | For He gas vessel |
| Warm Helium Supply | | yes | Gas only |
| SFHe Tank Manufacturing | HBE (CH) | yes | Light and large metal structures |
| Cryovalves | WEKA (CH) | yes | Critical technology |
| Warm Cryosystem Valves | | yes | Non critical technology |
| Cryocoolers | GSFC (US) | yes | Critical technology, the operating temperature should be reduced |
| Porous Plug | | no | For SF Helium only |
| Mass Gauges | Linde (D) | no | No Li He anymore |
| Relief Valves | | yes | Non critical technology |
| Thermo-Mechanical Pumps | SCL (UK) | no | For SF Helium only |
| Cryomagnet Avionics Box | CRISA (S) | yes | To control the operation of the SC magnet |
| Dump Rectifiers | JS (US) | yes | Non critical technology, to be developed for HTC SC |
| Vacuum Cases | | yes | Manly for the super insulation |

| SERVICES/HARDWARE | DEVELOPER | USEFUL TECHNOLOGY | COMMENT |
|---|---|---|---|
| Non-linear Strap Manufacturing | SCL(UK), CTG (UK) | yes | Non critical technology |
| SFHe Tank Thru Tube Machining | Turku/Artekno (FI) | no | For SF Helium only |
| UPS Batteries | Yardney-Lithion (UK) | yes | Non critical technology |
| UPS Battery Box | CSIST (Taiwan) | yes | Non critical technology |

*Table 2.2 Technology heritage of the AMS-02 superconducting magnet development*

For these reasons future superconducting active shield system would be rather
- based on *ITS* or *HTS* operating between *20* and *70 K;*
- operated in a cyro-free mode, that is without evaporation of the cooling fluid; the energy for the cryogenic recirculation system would be provided by solar panels. Cryo-free systems are routinely available on ground, not yet for a space application: on ground their efficiency in term of ratio of input power/cooling power currently ranges from $10^{-2}$ to $10^{-3}$;
- operated at low ambient temperature *(< 100 K)*, namely behind a solar shield.

There are however a number of space qualified technologies which are heritage of the development of the *AMS-02* superconducting magnet. They are listed in *Table 2.2* together with the Institute/Company who leaded their development: some of these technologies will be useful/critical for the development of future *ITS/HTS* based superconducting radiation shields.

## 3.3 Technology background: ITS/HTS cables

The concept of active radiation shielding is based on the availability of high performance, Super Conducting (*SC*) cables, able to carry very large currents at relatively high temperature.

One hundred years after the discovery of superconductivity, the availability of *SC* wires which can be used to build powerful magnets is limited to about half a dozen of materials. A superconducting material for a magnetic application should have some specific characteristics, namely it should:

- be chemically stable
- have critical temperature $Tc >> 4\ K$
- be able to carry high current densities at high magnetic field
- be easy to draw in long wires or tapes with uniform properties

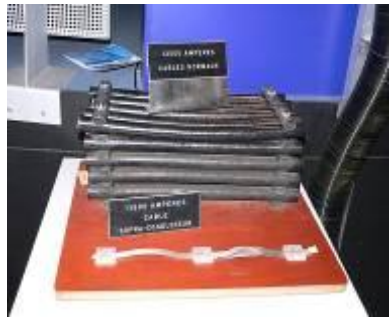

*Figure 2.14. The current which can be transported by SC cables without any loss is huge if compared to the best high power cables.*

At present, the industrially used materials (98% of the world market) are the

**Low Tc Superconductors, (LTS):**

- **Niobium-Titanium alloy (NbTi)**            Tc= 9.2 K
         $J_c$(4.2K, 5 T) ~ 3000 (5.500) A/mm$^2$, easy coil manufacturing
- **Niobium Tin (Nb3Sn)**            Tc= 18 K
         $J_c$(4.2K, 12 T) ~ 3000 A/mm2,
         wind-and-react technique for coil manufacturing

For special applications (O(2%) of the world market) there are the

**ITS (Intermediate Tc superconductors):**

- **Magnesium Diboride (MgB$_2$)**            Tc= 39 K
         $J_e$(21.5K, 4T) ~ 200 A/mm2, $I_e$(21.5K, 4T) ~ 30 A
         0.8-mm round wire
         isotropic, wind-and-react technique for coil manufacturing

**HTS (High Tc superconductors):**

- **YBCO and BSCCO**            Tc~ 90 – 110 K
    **BSCCO-2223** $J_e$(77K, 0T) < 160 A/mm2, $J_e$(20K, 5T) < 1.5 - 2.5*160 A/mm2
         Magnet-grade wire available, tape, anisotropic
    **BSCCO-2223** $J_e$(20K, 4T) ~ 200 A/mm$^2$, $B_{max}$(4.2K) > 45 tesla
         0.8-mm round wire, isotropic, wind-and-react
    **YBCO**      $J_e$(77K, 0T) > 300 A/1-cm tape, $B_{c2}$ higher than BSCCO
         2-G wire under development, operation ≥ 30 K possible

*Figure 2.14* and *2.15* show the *"critical current density"*, $J_c$, and the *"engineering critical current density"*, $J_e$, of practical superconductors[1]. As can easily be seen the *"engineering currents"* are one to two orders of magnitude lower than the *"critical currents"*, since the manufacturing of the cables requires a number of additional layers to create a matrix which will ensure the thermal and mechanical stability of the wires.

One important issue in designing a *SC* magnet is the protection from thermal instabilities which can cause a transition of the wire from the *SC* to the normal state: in this case, the huge amount of energy accumulated in the magnet, would tend to dissipate, through ohmic heating, within a small cable volume, destroying the cable. In a superconducting magnet a disturbance (like a resin crack or a wire displacement) can occur, generating a heat release and, in turn, a temperature increasing $(\Delta Q \rightarrow \Delta T)$. If the final temperature is higher than the critical temperature (at the magnet field and current), i.e. $T_0 + \Delta T > Tc(B, J)$, dissipation begin in the conductor.

To increase the conductor stability we can either choose high conductivity materials (*Cu* or *Al*) for the matrix or to use superconductors with higher critical temperature. High critical temperature means also high enthalpy margin that leads to higher stability (*Figure 2.17*).[1]

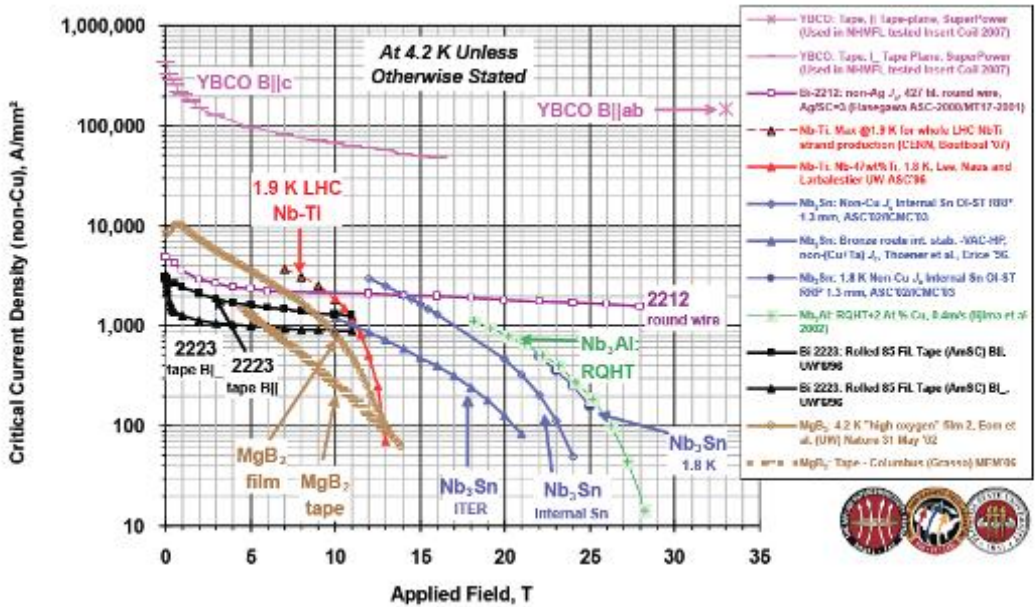

*Figure 2.15   $J_c$ for existing SC materials*

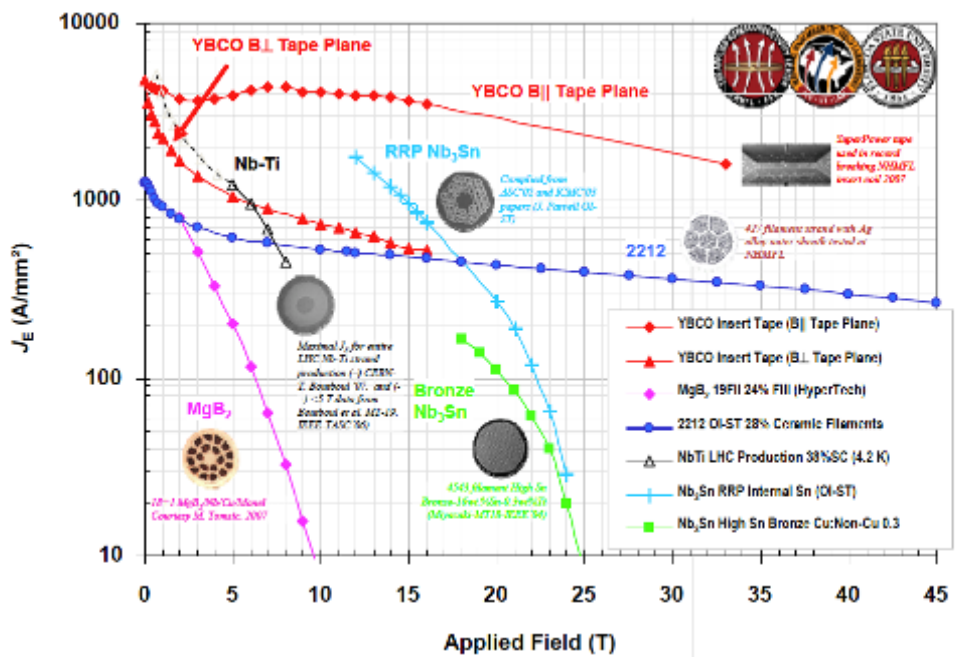

*Figure 2.16   $J_e$ for existing SC wires*

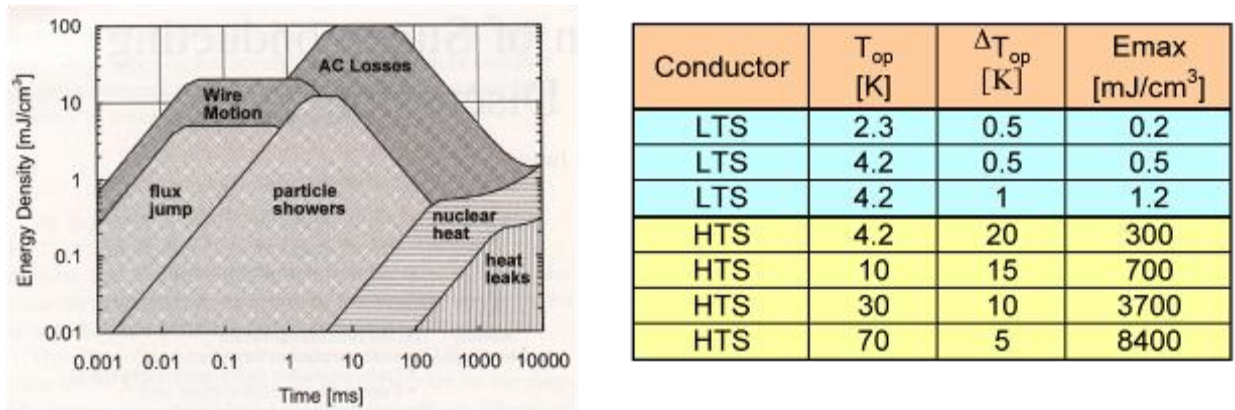

*Figure 2.17   Stability of SC wires. Left: Disturbance Energy Spectrum. Right: Temperature and Energy margins for LTS and ITS/HTS.*

The high thermal stability of *ITS/HTS* conductors, however, makes it more difficult the design of an active *"quench protection system"*, since such systems are based on the fact that a relatively small amount of thermal energy is sufficient to make the whole coil transit to the normal state, thus distributing the ohmic dissipation to the whole volume. Typical solutions to this issue are based on the segmentation of the coil in several smaller circuits which, in case of a quench, can be controlled more easily from an electrical point of view, while mechanical stability, in case of an asymmetric quench of the magnet, should be ensured by the supporting structure.

The cables considered in this study are two *ITS/HTS* wires: **$MgB_2$** (in the design of the coils of the *"racetrack"* toroid) and **YBCO** (in the design of the coils of the *"Double Helix"* toroid).

They have been chosen because of

  a) operability at high temperature (above Liquid He temperatures)
  b) high $J_e$
  c) good $J_e / \rangle$ ratio
  d) good mechanical properties
  e) intense R&D ongoing worldwide to improve the characteristics of these cables

**$MgB_2$** : amongst all superconducting materials, $MgB_2$ might represent the best choice for space applications, thanks to its low mass density and to its broad chemical compatibility with many elements, although not with copper. $MgB_2$ is about *2-3* times lighter than any other industrially produced superconducting compound. Also, looking at *Table 2.3*, it appears clear that $MgB_2$ represents a compromise between *Nb-alloy* low temperature superconductors, showing low critical temperatures, but large coherence length, relatively simple manufacturing into wires and useful critical fields, and *Cu-oxide* high temperature superconductors, capable of carrying supercurrents in liquid nitrogen, but afflicted by a very low coherence length and complexity in wire manufacturing. Being indeed the coherence length $\xi$ the key parameter that determines the maximum distance that guarantees superconducting coupling between adjacent particles (the longer the better), $MgB_2$ manufactur-

*44*

ing methods significantly benefits from a long $\xi$ as the low temperature superconductors. Research on $MgB_2$ based cables and related manufacturing technologies is proceeding worldwide in view of a number of industrial applications: it follows that the properties of $MgB_2$ based cables have a potential for a significant improvement in the coming years (at least a factor of 3 in $J_e$). Companies pursuing the development of $MgB_2$ based cables include Columbus Superconductors (I)[2].

| Superconducting Compound | $T_c$ in Kelvin | $H_{c_2}$ at 4.2 K in Tesla | $\xi$ (nm) | Mass Density (g/cm³) |
|---|---|---|---|---|
| Nb-Ti | 9 | 10 | 5 | 6.0 |
| Nb$_3$Sn | 18 | 28 | 5 | 7.8 |
| MgB$_2$ | 39 | up to 70 | 5 | 2.5 |
| YBCO-123 | 90 | > 50 | << 1 c | 5.4 |
| BSCCO-2223 | 108 | > 50 | << 1 c | 6.3 |

*Table 9.1 Properties of SC compounds used for SC magnet wires*

**YBCO :** *Yttrium Barium Copper Oxide*, is a crystalline chemical compound. This material achieved prominence because it was the first material to achieve superconductivity above the boiling point (*77 K*) of liquid nitrogen. Several commercial applications of this type of high temperature superconducting materials have been realized. For example, superconducting materials are finding use as magnets in magnetic resonance imaging, magnetic levitation, and Josephson junctions. *YBCO* has yet to be used in many applications involving superconductors due to its intrinsic granularity and small coherence length which limits it manufacturability in the form of cables.

The most promising method developed to utilize this material for cables involves deposition of *YBCO* on flexible metal tapes coated with buffering metal oxides. This is known as coated conductor. Texture (crystal plane alignment) can be introduced into the metal tape itself (the *RABiTS* process) or a textured ceramic buffer layer can be deposited, with the aid of an ion beam, on an untextured alloy substrate (the *IBAD* process). Subsequent oxide layers prevent diffusion of the metal from the tape into the superconductor while transferring the template for texturing the superconducting layer. Novel variants on *CVD, PVD,* and solution deposition techniques are used to produce long lengths of the final *YBCO* layer at high rates. Companies pursuing these processes include American Superconductor, Superpower (a division of Intermagnetics General Corp), Sumitomo, Fujikura, Nexans Superconductors, and European Advanced Superconductors. A much larger number of research institutes have also produced YBCO tape by these methods.

A second generation *(2G) of YBCO* tape conductors is now on the market[3], in the form of *4 cm* wide, *<0.23 mm* thick, tapes (Figure 2.18) carrying an outstanding *1000 A @ 75 K* on *0,8 ⌈m* of *YBCO* (*Figure 2.19*).

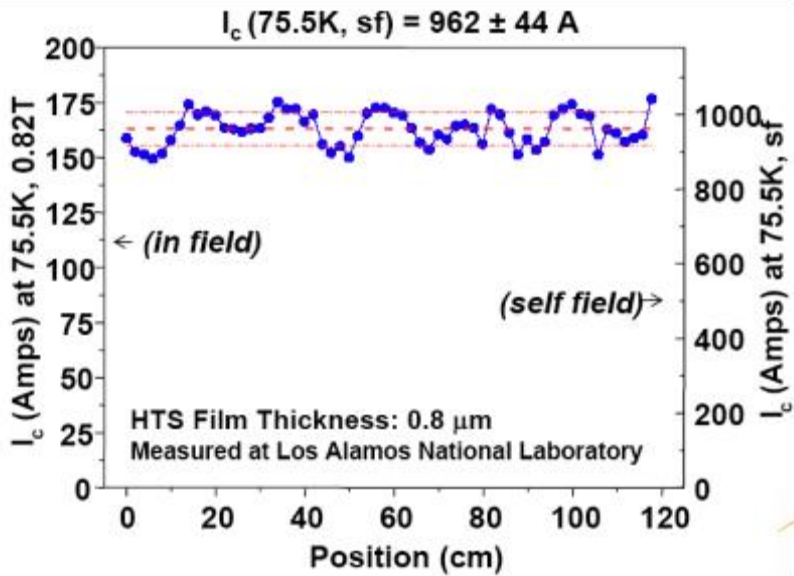
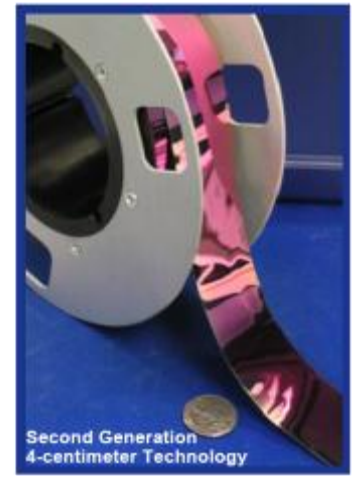

*Figure 2.18 Properties of 2G YBCO SC tapes*

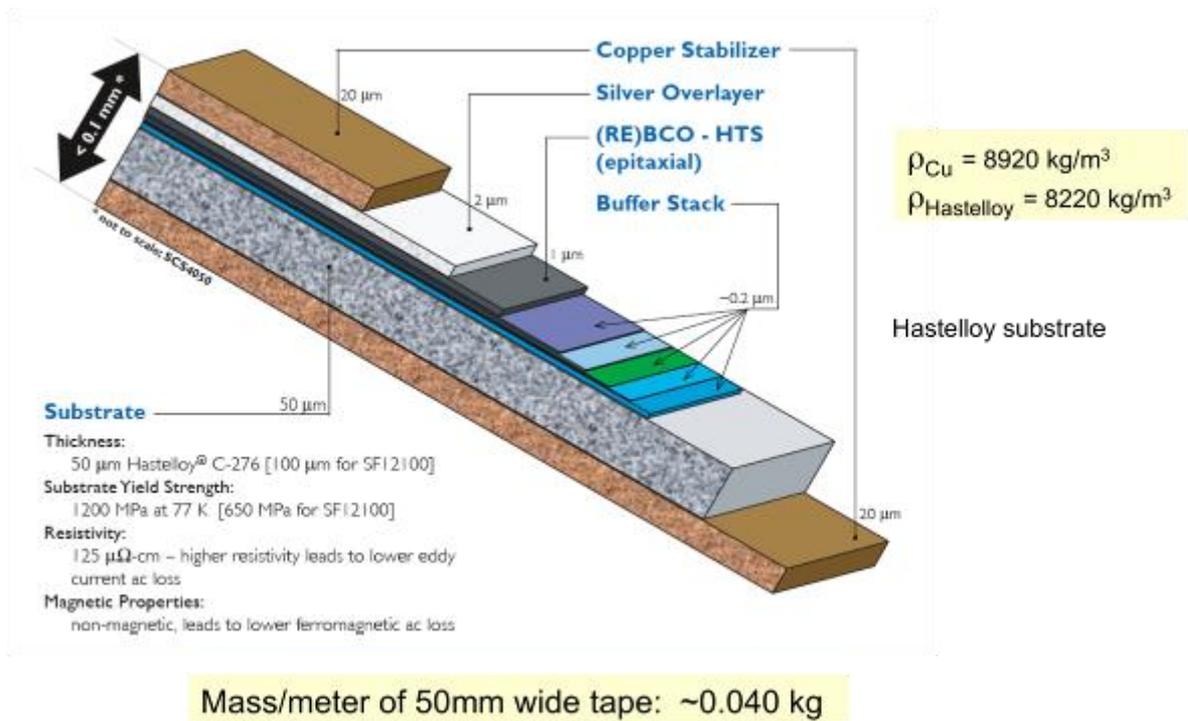

*Figure 2.19 Structure of 2G YBCO SC tapes : note the thickness of the YBCO film (1 μm)*

Scaling the properties of these *2G* tapes to lower temperatures and to larger thickness (*Figure 2.20*) it would be possible to develop *YBCO* conductors carrying *11.000 A* on a *5 cm* wide tape operated at *25 K*, values which would be suitable for the development of the *Double Helix* coils discussed in this report. Development of multi-km long tapes with these characteristics will nevertheless take a few more years of R&D.

*46*

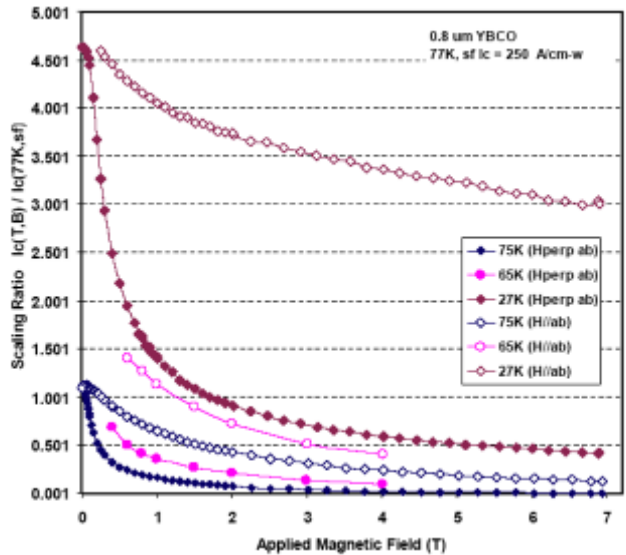

*Figure 2.20 Properties of 2G YBCO SC tapes*

Research on *YBCO* as well as $MgB_2$ based cables and related manufacturing technologies is proceeding worldwide in view of the large number of possible industrial applications.

In conclusion, the potential of *ITS/HTS* as building block of the coils for space radiation shields is very significant. Improvement of the properties of the *ITS/HTS* wires or of the construction processes will be driven by the huge potential application market. The specific application for space shields will benefit from the results of the ongoing world *R&D* on these materials. However, as we will show in the following, it is already possible, using today state of the art technology, to design shields capable to reduce significantly the *GCR* dose.

## 3.4 References

1) http://www.magnet.fsu.edu
2) Columbus Superconductors SpA, Via delle Terre Rosse, 30 - 16133 Genova, Italy, http://www.columbussuperconductors.com
3) SuperPower Inc., 450 Duane Avenue, Schenectady , NY 12304 USA, http://www.superpower-inc.com

# 4  Design of Active Toroid Shield

## 4.1 Introduction

Past studies of confined magnetic shields have considered multi coil ("*racetrack*") toroid systems. These analyses were based on different scaling assumptions on the mechanical or magnetic subsystems which were studied (cold mass versus cryogenics, structural mass, etc.). For these reasons the resulting weights estimates (showing a spread of more than one order of magnitude), cannot reliably be compared to each other.

In this study we used a different approach. Thanks to the expertise of engineers who designed and built large superconducting magnetic systems for *CERN* and other laboratories we made a preliminary design of the active shields configurations. We followed the approach of designing magnets which could be built using state-of-the-art but existing technology, which through a suitable *R&D* effort could be qualified for space utilization. This approach lead us to the development of relatively low efficiency shields (*4-5 Tm*): through the understanding the merits and limits of the designs which have been studied, however, we identified a roadmap for technology developments needed to reach for magnetic shields with higher shielding efficiency. The roadmap is discussed in Chapter 7.

We studied two types of coil configurations.

First of we studied the "*racetrack*" toroid system as a function of the effective magnetic cutoff, to first approximation the integral of *BdL* provided by the magnet, in the range from *4* to *10 (15) Tm*. The choice of the superconducting cable, the design of the coils, the analysis of the induced forces and torques, both in case of static operation as well as of quenching conditions, have been considered in this analysis as a function of the shielding effectiveness.

This analysis provides a set of mass estimates which can be extrapolated over a range of parameters and a number of assumptions related to mechanical and magnetic properties of the "*racetrack*" design, which would be a reference for future studies. It also helps us to identify some critical issues concerning the use of "*racetrack*" toroidal designs as an active radiation shield for interplanetary space mission.

We also analyzed an innovative toroid magnetic design based on an array of suitably arranged magnetic dipoles based on the "*Double Helix*" coil technology. Also in this case the magnet design has been developed with the help of professional engineers who are world experts in designing these type of coils: the choice of the suitable superconducting cable as well as of the power and thermal systems needed for the *"Double Helix"* solution have also been considered.

This second round of analysis provided us a new set of mass estimates and extrapolations related to the new *"Double Helix"* design. From these data it turns out that the new configuration exhibits some interesting properties, being more effective in dealing with the Lorenz forces created by the strong magnetic field and being intrinsically modular from the point of view of the assembly in space or on the surface of a planet. After a suitable amount of *R&D*, this solution could potentially provide a lighter shield solution than the *"racetrack"* shield for similar (low) *BdL* values and open the way to potentially interesting developments towards medium to high *BdL* values.

For the scope of this study each shield configuration was studied using a different *SC* cable: $MgB_2$ for the *"racetrack"* shield and *YBCO* for the *Double Helix* shield.

## 4.2 Study of a shield based on a "racetrack" toroid magnet

The design of a *"racetracks"* toroidal magnets requires a detailed structural analysis in order to establish forces and torques acting on the coils, defining a possible mechanical structure and estimating a realistic mass of the magnetic systems. The analysis performed, although preliminary and developed with conservative assumptions, provides a better estimates than what is available in the literature. This analysis has been done by engineers of *ASG Superconductors s.p.a.* (I), Genova, the company who has designed several of the superconducting magnet for *CERN*, *DESY* and *SLAC*. The *SC* cable has been studied at *Columbus Superconductors*, Genova (I) a company designing and building advanced $MgB_2$ cables for industrial and scientific applications.

In designing this magnet, we set the bending strength (integral of *B*dL*) at *5 Tm*, corresponding to an omni-directional threshold of about *250 MeV* of kinetic energy for protons, for an expected dose reduction of about a factor of *2*. Two additional key parameters of the design are the radial dimensions of the habitable module *(3 m)* and the radial extension of the coils *(3,1 m)*: in this way the total radial dimension would be close to *6 m*, compatible with the dimensions of future exploration-class fairings.

The analysis of this magnet configuration has been performed in a parametric way: in this way the extension of the analysis to lower *(4 Tm)* or higher *(6 Tm)* bending power can be easily calculated. Extensions to higher bending strength (*10 Tm* and *15 Tm*) have also been analyzed in term of mass, although the very significant increase in magnetic forces and torques, poses, in the last case, engineering challenges which should be carefully addressed.

In this way we are able to provide an estimate of the weight dependence of a toroidal magnet system including its structural mass in the interval *4 Tm* to *15 Tm*: these values correspond, according to the literature, to the interval in expected dose reduction ranging between a factor *2* and an asymptotical factor close to *10*.

The *5 Tm* toroid magnet studied here has the characteristics described in the following *Tables 4.1* and *4.2*. The model of the toroidal magnet results as in *Figure 4.1*.

| Toroid (12 racetrack coils) | |
|---|---|
| Inner radius | 3000 $mm$ |
| Outer radius | 6100 $mm$ |
| Axial length | 10000 $mm$ |
| **Racetrack coil** | |
| Width | 50 $mm$ |
| Length | 500 $mm$ |
| Section | 0.025 $m^2$ |
| Inner diameter | 3000 $mm$ |

Table 4.1 - Dimensions of the magnet cold mass (without the mechanical structure)

| Toroid | |
|---|---|
| Overall density current | 117 $A/mm^2$ |
| Overall current per coil | $2,92 \cdot 10^6$ $A$ |
| Average integral of field | 4.899 $Tm$ |
| Maximum integral of field | 4,902 $Tm$ |
| Minimum integral of field | 4,897 $Tm$ |
| Ripple Parameter on the minor axis | 0.071 |

Table 4.2 - Current density, magnetic field and integral of field

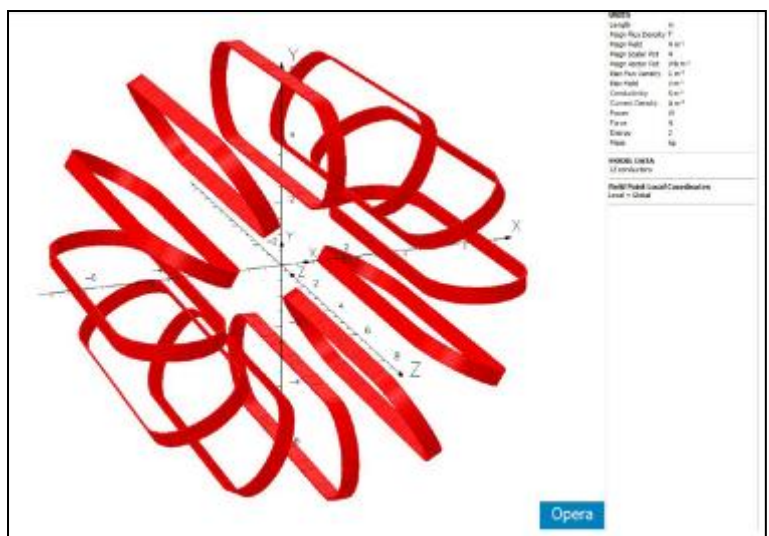

Figure 4.1 - Racetrack Toroid magnet configuration

Forces and torques over racetracks have been calculated through the *FEM* simulation software *OPERA 3D – Cobham*, subdividing every racetrack in different conductors of different sizes. This approach allows to increase the precision of the computation and it offers a detailed distribution of forces and torques.

Calculations have been performed not only in the standard operative condition, but also for single racetrack displacements along radius, angles, axes and current variations.

In case of displacements or inhomogeneous current variations due to a quench (conditions which can happened if coils are not connected in series), the symmetry is broken: therefore different distributions of forces and torques could develop.

The details of the analysis show that in case of displacements or inhomogeneous current variations forces don't increase dramatically and the order of magnitude remains the same. In case of torques, however, the breaking of the symmetry induces high torques, in particular in the case of a quench.

These values have been used as an input to the design of a possible mechanical structure.

Radial forces and forces generated by torques have to be supported by a structure placed around the habitable module. Such structure has to be as thin and light as possible in order to save space and weight for the astronauts living quarters.

A possible solution, already developed by *Hexcel*, is to use honeycomb panels that provide high stiffness and low density. Moreover, they could be assembled in orbit since they are modular.

*Honeycomb* disposition can follow a dodecagonal form. This kind of structure has been designed for the *Atlas Barrel Pixel Detector* (here with 8 sides) and it could easily be cleared for windows with a rendering study [4].

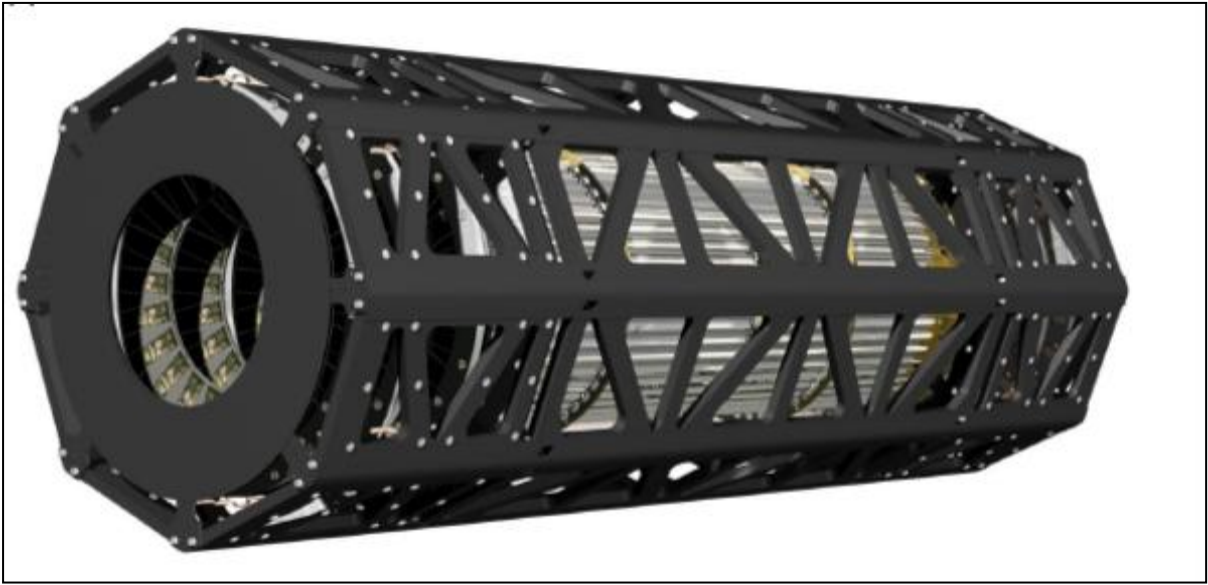

*Figure 4.2 - ATLAS Barrel Pixel Detector*

*Hexcel* produces *honeycomb* panels for space applications made in *Epoxy UD Carbon* tape and *Al 5052*[3].

Since torques represent the main problem for a toroidal magnet has in case of quench, the mechanical structure must provide the necessary mechanical stability. A standard solution is to use beams across the magnet for counter the forces induced by the torques: we assumed to use carbon fibers to counter these forces.

The overall weight resulting from this analysis of a *5 Tm* racetrack magnet is given in *Table 4.3*. This design exercise represents a preliminary but sufficiently detailed study to provide an estimate of the weight of a *SC* toroidal magnetic shielding system based on advanced, light *SC* cable based on $MgB_2$ wire, as a function of its radiation shielding efficiency (bending power). The study has been repeated/extrapolated for different values of the toroid bending power, as shown in the following *Table 4.4*. It should be stressed that the structural mass estimate is based on the assumption on the mechanical deviation from perfect symmetry: a lower error budget on the mechanics would decrease the structural weight of the toroid magnet configuration.

| TOROIDAL MAGNET | | |
|---|---|---|
| **dimension and number of racetracks** | | |
| Number of racetracks | 12 | |
| Axial height | 10.0 | m |
| Inner radius | 3.0 | m |
| Outer radius | 6.1 | m |
| Volume | 7.075 | m³ |
| Volume free for the cockpit | 246 | m³ |
| **Supply** | | |
| Current per cable | 1168 | A |
| Maximum voltage | 3000 | V |
| **Energy and inductance** | | |
| Total energy | 9.85E+08 | J |
| Inductance L | 1444 | H |
| **Mass** | | |
| Coils | 2.62E+04 | kg |
| Tie-rods | 3.93E+03 | Kg |
| Mechanical structure | 6.98E+03 | Kg |
| Bearing structure | 4.44E+03 | Kg |
| Torque retaining beams | 4.66E+03 | Kg |
| **Total mass** | **4,64E+04** | **kg** |

*Table 4.3 – Racetrack toroid magnet specifications and total mass (5 Tm)*

| | BdL (Tm) | Habitable volume (m^2) | Weight (t) | Coils weight (t) | Structural weight (t) |
|---|---|---|---|---|---|
| 12 coils single toroid | 4 | 286 | 40,0 | 21,5 | 18,6 |
| 12 coils single toroid | 5 | 286 | 46,4 | 26,4 | 20 |
| 12 coils single toroid | 6 | 286 | 58,0 | 27,5 | 30,5 |
| 12 coils double toroid | 10 | 286 | 149,0 | 62,9 | 85,6 |
| 12 coils double toroid | 15 | 286 | NA | NA | NA |

*Table 4.4 – Extrapolation of the weight of the toroidal racetrack magnet to different bending power*

From this analysis we can derive some preliminary conclusions on the *"racetrack"* shield

configuration:

- the weight of a racetrack toroidal configuration is rather large, even for moderate bending power (i.e. *5 Tm*) corresponding to a factor of about *2* expected shielding effectiveness, we are in the range of *40* to *45 t* for a shield covering the barrel of habitable module;
- this technology would be quite effective in designing an active shelter: with a weight of about *7 t (3 m □ x 3 m* length) it would provide a better shield from *SPE* than an equivalent weight of passive shelter (*15 g/cm$^2$*) providing also factor two reduction in *GCR* dose during sleep time (*1/3* of the time)
- the weight of the magnet, at least for moderate bending power, is due to *50%* to the coil and *50%* to the supporting structure
- approaching high bending power (*10 - 15 Tm*) the structural weight grows with the square power of the bending power, quickly reaching unrealistic weight values *>> 50 t* (*Figure 4.3*). In order to deal with the large forces present in these large toroidal systems, on could develop multi toroidal system used in ground based application ( e.g. double toroidal system for *BdL = 10 Tm*, see *Figure 4.4*). However, an attempt to develop a mechanical structure for a *15 Tm* magnet based on the extrapolation of the concepts used for the lower *BdL* configurations, was not successful since the assumption used for the structural strength on the cable were not anymore applicable due to the large induced loads.

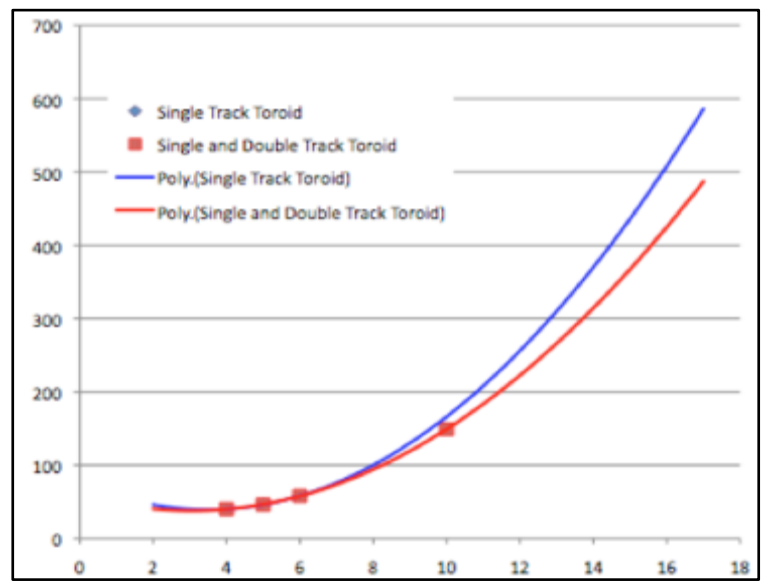

*Figure 4.3 Extrapolations of the magnetic system weight as a function of the bending power. Vertical axis weight (t), horizontal axis integral of BdL (Tm)*

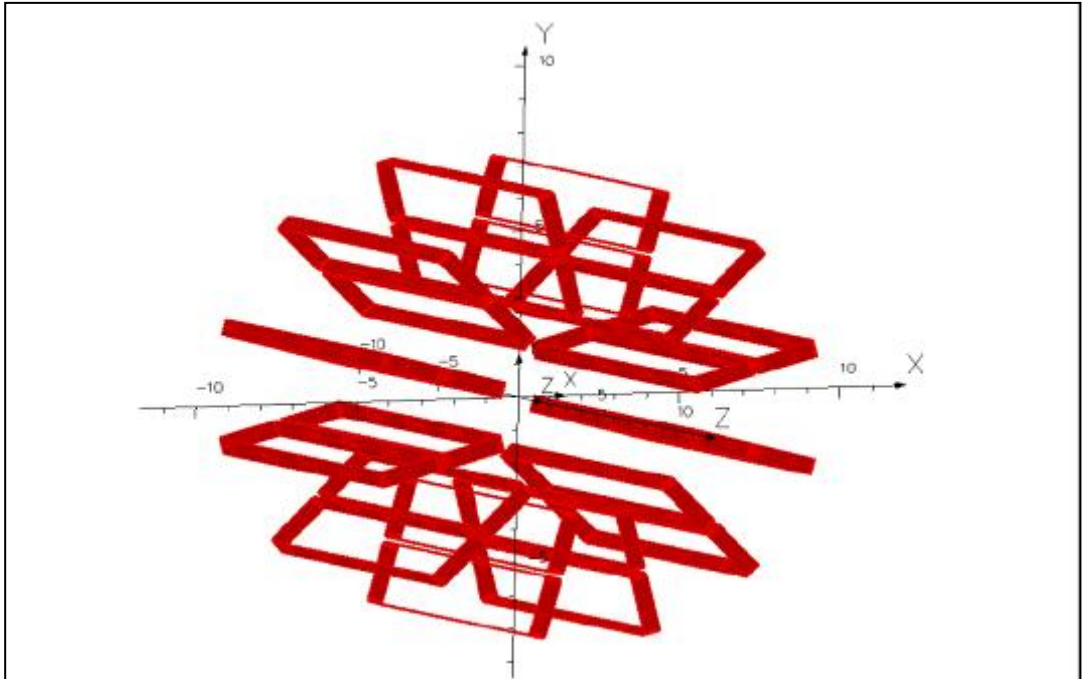

*Figure 4.4  A double racetrack toroidal magnet structure for large bending power*

## 4.3 Study of a shield based on a "Double Helix" toroid magnet

The results of the design of the the "*racetrack*" toroid magnetic shield provide a reference for understanding the potential of this configuration for future space travels. Using the state of the art *ITS* cable like *MgB$_2$*, the bare coils weight of a *5 Tm* magnet would be around *26 t* while the structural mass would add about *20 t*.

The "racetrack" configuration has the following disadvantages:

1) for a given weight, the mass is concentrated into structural beams and radially distributed coils: to make use of the shielding effect of the materials against *HZ GCR* or low energy particles from SPE, a uniform distribution around the habitable module would be more desirable. A concentrated mass would also increase the amount of secondary particle production due to nuclear interactions.
2) the "*racetrack*" coil configuration is intrinsically *"non modular"* : due to the difficulty of performing *SC* joints in space, all the *SC* joints should be made on ground, preventing a modular deployment of the coils and assembly in space.

A coil configuration able to stand the magnetic forces with less structural material and suited to modular deployment, would of course be desirable. We then considered ways to design coils which would reproduce the toroid field but using a different, modular geometry.

As a first step in this direction we have developed the concept of a novel configuration of toroid magnetic shield, based on an array of *Double Helix (DH)* coils[6], a magnet technology which allows for the generation of magnetic multipoles with unmatched field homogeneity thanks to a manufacturing process that stabilizes the conductors in precisely machines grooves. As a result, the conductors are mechanically very stable and large Lorentz forces, present in superconducting magnets can be handled effectively.

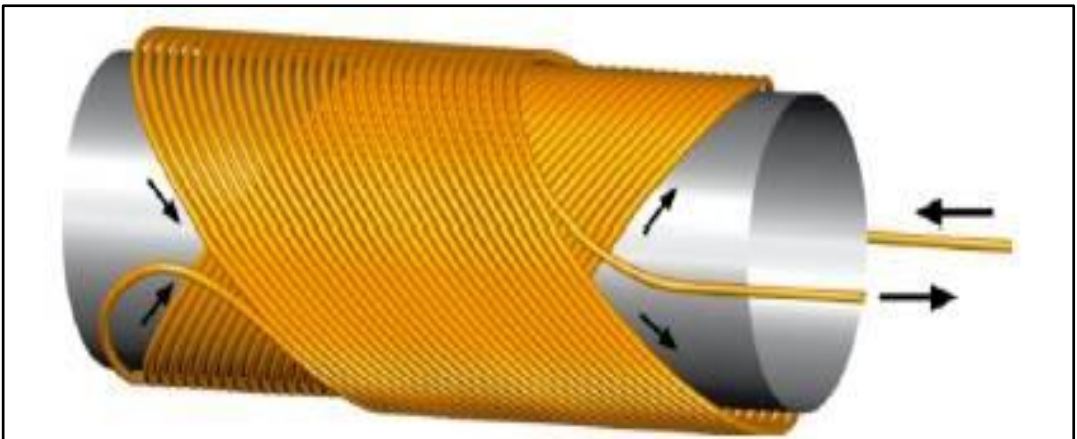

*Figure 4.5 Example of a 2-layer winding used to form a DH dipole magnet*

The double-helix coil configuration uses concentric pairs of oppositely-tilted helical windings to generate transverse magnetic fields. *Figure 4.5* shows a *2-layer* magnet generating a transverse dipole field.

The *DH* solenoid-like windings are embedded in concentric cylinders of high-strength material. Together with an over banding of high-modulus, high-strength fibers, forces can be contained easily. The minimum bend radius of the conductors in the *DH* coil configuration is significantly larger than in racetrack-shaped coils used in conventional magnets. This facilitates the use of strain sensitive (brittle) materials such as high temperature superconductors while keeping substantially smaller dimensions. In *DH* dipole magnets, each layer generates a tilted dipole field with respect to the axis as shown in *Figure 4.6*. Each layer generates a field with transverse and axial components; by combining several layers together, a purely transverse field or a purely axial field can be obtained, at the price of doubling the amount of winding. Switching from transverse field to axial field can be done by changing flow direction in the oppositely tilted layers. The *DH* geometry is not limited to dipole fields, but can be used in coils with all the advantages of *DH* windings with any multi-pole order.

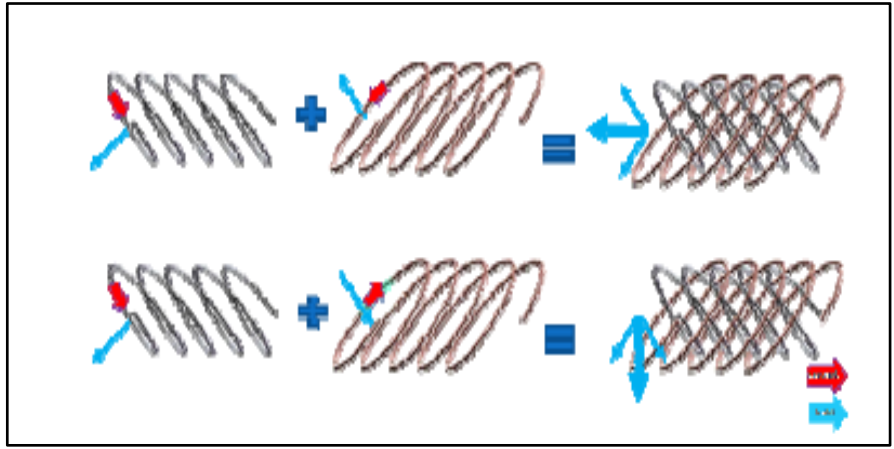

*Figure 4.6 Principle of field direction switching.*

We studied an active radiation shield based on set of *DH* coils assembled around a cylindrical habitable. The basic parameters (size, bending power, shielding efficiency), were chosen so to be compared with the classical *"racetrack"* toroid design. We developed this concept with the help of the engineers which have invented the *DH*, at *Advanced Magnet Laboratory (AML),* Melbourne, Florida (US)[7]. The results of the study suggest that this approach is very promising as solution to the radiation shielding design problem.

We considered *DH* coils with circular cross section, *2 m* diameter (*Figure 4.7*) and a dipolar field which, along the diameter, has *4 Tm* of *BdL* integral. The coil parametric models has been built in *CoilCad™*, which completely describe the coil windings on a turn by turn level. The models allow a calculation of all key parameters of the coils including field and force calculations.

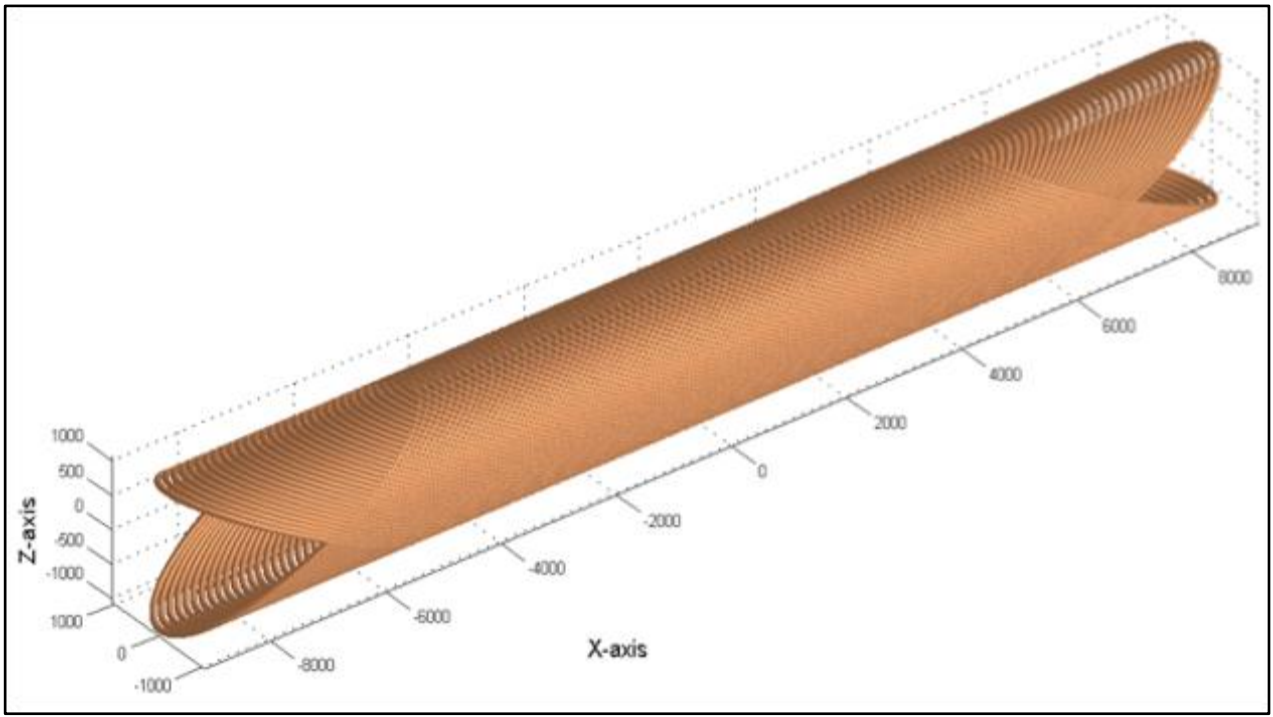

*Figure 4.7 The DH dipole cylindrical coil considered in this study. For reasons of clarity number of turns reduced and only 2 layers shown*

In this study we used the characteristics of a state of the art version of *YBCO HTS* cable, the *2G YBCO* tape (*Figure 2.18* and *2.19*), although this high performance cable still suffers from the typical issue of *YBCO* (the small value of the *"coherence length"*) which limits the manufacturability and the performances of long cable. At a temperature of *25 K,* a *50-mm* wide tape with a *2 micron YBCO* layer and a peak field of about *2.5* to *3* Tesla ⊥ to the tape would tolerate a critical current of about *10,000 A*. The detailed analysis performed by *AML* of a *8 layers* tape configuration coil suggests that the tape will not undergo large values of *B* field perpendicular to its surface. A conductor with these properties is within the reach of the next few years of *R&D* on *YBCO* cables. The thickness of available *2G* conductor with near appropriate performance is of *0,2 mm*, for a total thickness of the bare conductor of about *1,6* ⌈m. The mass/meter of *50 mm* wide tape would be *0,040 kg*.

For a *18 m* coil with a field plateau of *10 m* length at *2 T* about *25* to *30 km* of tape are required. Despite the very small thickness of the conductor (*0.2 mm*) the total weight of both coils is about *1000 kg* each (the coils extends over a length of *18 m* to allow the field to decrease smoothly at the end: this aspect should be further optimized).

A detailed estimate of the acting forces has been performed: the coil inner magnetic pressure is on average *16 atm,* with peaks at *30 atm*. A detailed force calculation has been performed, and structural reinforcement have been considered to counter the induced deformations on the coil. The axial forces in *DH* coils are from layer to layer in opposite direction and produce internal shear forces in the assembly (*Figure 4.8)*. Given a sufficiently strong coil substrate, no additional axial support like in saddle or race track coils is therefore required. This is an advantage of the *DH* coils design.

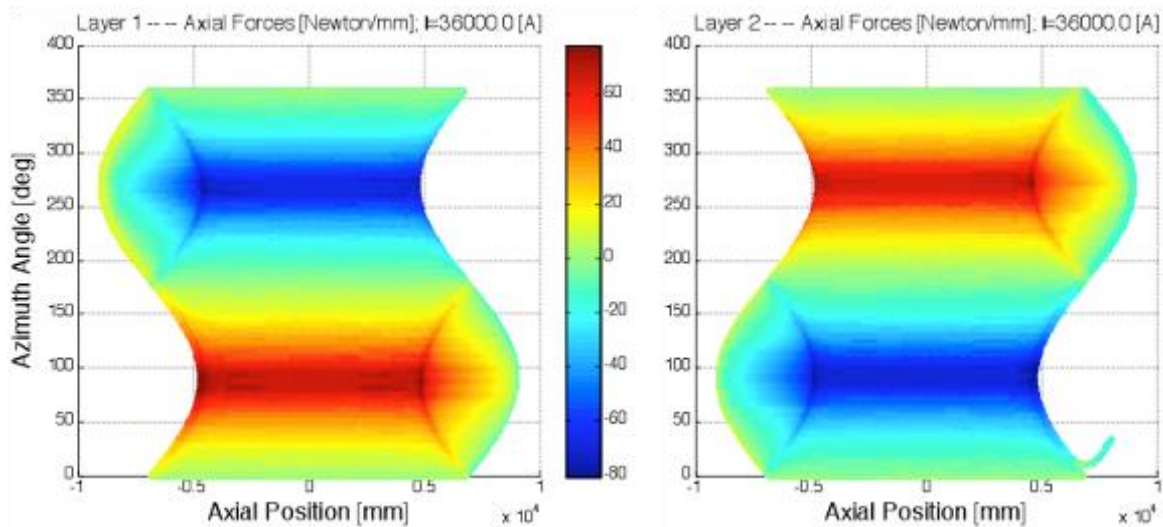

*Figure 4.8 DD map of the module of axial forces acting on the two layers of a DH dipole coil: the forces are opposite on the two layers as clearly visible in the plots.*

The arrangement of *12* coils would in addition create an inward pressure of *3,6 atm*, a reasonably small value which can be handled by a suitably reinforced plate.

A complete shield surrounding the space ship with *12* coils would require a total conductor weight of about *12 t* (*Figure 4.9* and *4.10*). A detailed calculation has been performed on the structural design of this configuration: adding the weight of the coil structural support, the support plate and the launch support structures, we obtain *30 t*, to be compared with the corresponding figure of *40 t* for a *4 Tm racetrack* configuration. Although is not easy to compare these numbers, since the two studies have been performed using different design assumptions, there is an indication that the *DH* coil design is characterized by a better weight/shielding efficiency ratio.

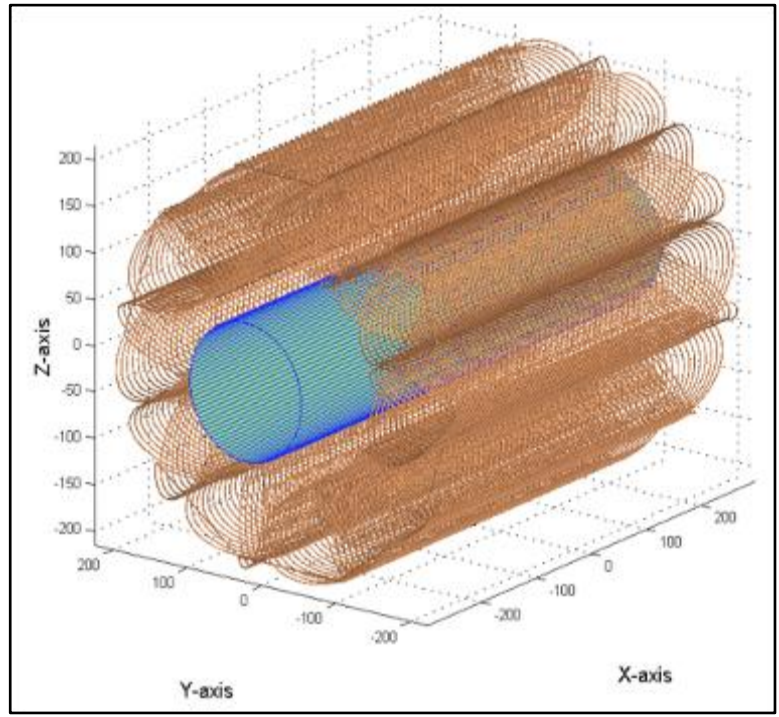

*Figure 4.9  The full configuration with  12 x 2 m DH coils located around the habitable module.*

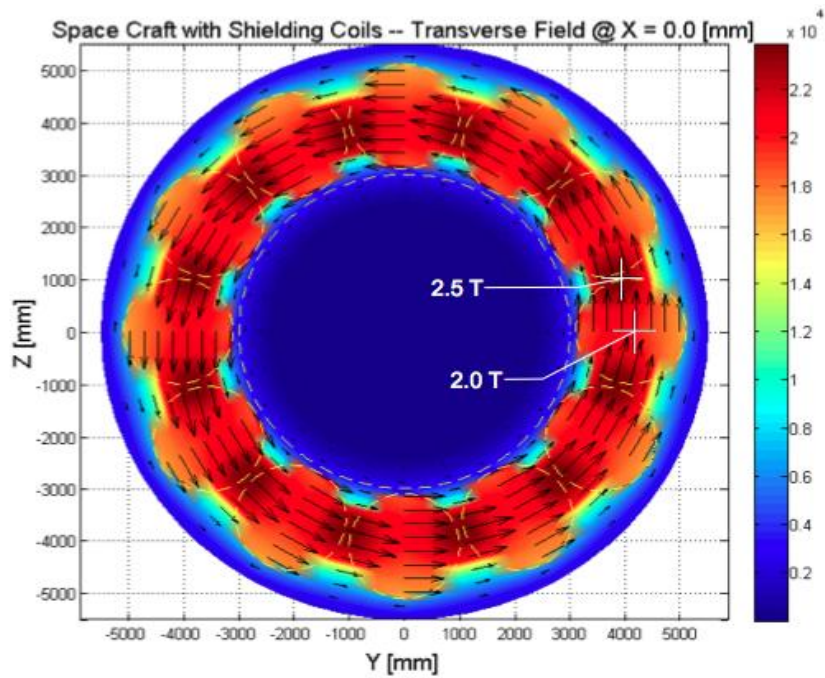

*Figure 4.10  The Halbach array of 12 DH coils providing  2.0 T field within the single coil*

## 4.4 Final remarks on the DH shield design

The *DH* radiation shield design exhibits some  potential advantages which would make it attractive  with respect to the *"racetrack"* design and should be systematically explored, namely:

- the distribution of magnetic forces is somewhat easier to handle, so the mechanical support structure will be, correspondingly, lighter:
a) axial forces between adjacent layers of tapes, are internal to the coil and tend to cancel or been taken from the local strength of the material building the supporting structure;
b) radial forces within the coil tends to modify its cylindrical shape, and can be treated using diametral wires/plates;
c) forces between adjacent coils are opposite to forces (b): they tends to cancels out, reducing the stresses on the diametral wires/plates;
d) the *Halbach DH* array exhibits its own mechanical stability due to intercoil azimuthally oriented forces;
e) the distribution of forces on adjacent tape layers allows for the use of relatively low strength, light tapes.
f) the inwards pressure acting on the habitable module is reasonably small.

- the total weight of the bare 12 coils, each having one meter of radius and *18 m* of length, has been computed to be of the order of *50 %* the corresponding $MgB_2$ bare *"racetrack"* coils. This is mostly due to the assumption made on the SC cable: improved $MgB_2$ cables, which is expected to be developed within the coming *3-4* years, would reduce substantially the bare coil difference in weight.
- for a given *BdL,* scaling of the *DH* coils to larger radius could make the system lighter (sic!): an increase in radius would imply a corresponding decrease in *B*, that is in the total current, that is in the number of *SC* tape layers which are needed. As a consequence, as a first approximation a *2 m* diameter, *4 Tm* coil would weight the same as a *6 m* diameter, *4 Tm* coil, but then only half of the coils (*6* vs *12*) will be needed in order to shield the habitable module: the weight would then be reduced *O(50%)*. Since the magnetic forces are proportional to $B^2$ operating at lower *B* would allow for simpler/lighter mechanical structures. Obviously the deployment of large diameter coils would require the development of some advanced mechanical design like the use of deployable structures.
- an important merit of the *DH* design resides in its intrinsic modularity: power supply can be provided independently to each coil and in case of a quench each (layer of each) coil will be protected independently. Obviously this might create asymmetries in the magnetic field which might induce forces or torques which should be compensated by the supporting structure. From the deployment point of view this open the way to a number of possibilities like modular transport/assembly/disassembly in space, modular transport/assembly/disassembly on the surface of a planet, and son on.
- the cylindrical coil shape represent a large volume surrounding the habitat that can be used for hosting cold materials like cold fluid/solid propellant which could also play a role as temporary passive shield
- the material traversed by the *GCR* particles is distributed all around the habitat, roughly independent from the azimuthal angle of incidence, namely the magnet structural mass act as an uniform passive at the same time as an active shield. Since at large Z values, the passive shielding provided by a thin layer of material is very effective, this is a major advantage of the DH design.

- the flexibility of the DH design of adjusting the field direction as a function of the project requirements would allow for solutions which are still to be explored.

The exploitation of these potential advantages would require a special case-by-case study and a *R&D* effort which are outside the scope of this project.

## 4.5 References

1) ASG Superconductors S.p.A. C.so F. M. Perrone, 73, 16152 Genova, Italy, http://www-as-g.it
2) Columbus Superconductors SpA, Via delle Terre Rosse, 30 - 16133 Genova, Italy, http://www.columbussuperconductors.com http://www.hexcel.com
3) Hexcel, http://www.hexcel.com
4) ATLAS Experiment at CERN, http://atlas.ch/photos/inner-detector-pixel.html
5) Y.Iwasa, " Stability and protection of Superconducting Magnets- A Discussion", IEEE Trans. Appl. Supercond. 1615, 2005
6) R.B. Meinke et al., Direct Double Helix Magnet Technology, Proceedings of PAC09 (2009) ; C.L. Goodzeit et al., The double-helix dipole a novel approach to accelerator magnet design, IEEE Trans. Appl. Supercon, 13, 2, 1365 – 1368, (2003); R. B. Meinke, Modulated double-helix quadrupole magnets, IEEE Trans. Appl. Supercon, 13, 2, 1369 – 1372, (2003)
7) Advanced Magnet Lab, 1720 Main St. Ne – Bldg. 4, Palm Bay, FL 32905, http://www.magnetlab.com

# 5  The Physics Simulation

## 5.1 Performance Assessment for various Magnetic Configurations

The complexity of the *GCR* particles magnetic transport and interactions with the spacecraft material requires the use of reliable, complete physics simulation. A major fraction of the work presented in this report is based on the study of the radiation shielding performance of several different magnetic configurations (called *Geomxx* in the following).

The spacecraft geometry is shown in Figure 5.1: in this case we show, as an example, the toroid magnet configuration of a previous study [1]. The compact geometry allows to evaluate the performance of an active screen composed of barrel and end cap magnets. The passive absorber, an aluminum plug in the simulation, represents the uninhabited part of the spacecraft, for example the propulsion system. The presence of the passive absorber plays a fundamental role for the magnet screens scenarios studied. Conversely, the presence of an end cap magnet depends on the design of the spacecraft and may be avoided. The *4 m □, 5.5 m* long cylindrical habitat is surrounded by a *1.8 cm* thick aluminum surface.

The physics simulation code allows for the determination of the yearly doses deposited by the *GCR*. For this purpose 6 *"astronauts"* are placed inside the shielded volume and the doses sampled as a function of the position and of the depth of the body: they are simulated by six water cylinders placed in the air-filled volume to provide comparative dose estimates on/off the cylinder axis and between the end cap/absorber sides.

The incident cosmic ray nuclei are generated on the surface of a *10 x 10 x 10 m³* cube containing the magnets and habitat (*Figure 5.1*). The relative position of the cube and the spacecraft results in an aluminum thickness of *1.15 m* on the absorber side.

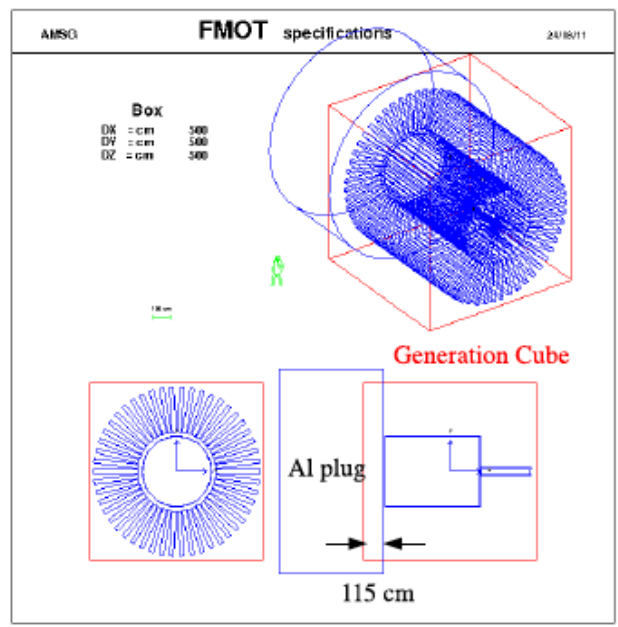

*Figure 5.1 The generation surface, a 10 x 10 x 10 m³ cube, used in the simulation. The absorber (aluminum tank) thickness, defined by the relative position of the cube with the spacecraft, is 1.15 m.*

The number of particles which penetrate the shields and deposited energy in the *"astronauts"* depends on the particle energy spectrum, on the the magnetic field strength and extension, as well as on the material to be traversed (*Figure 5.2*)

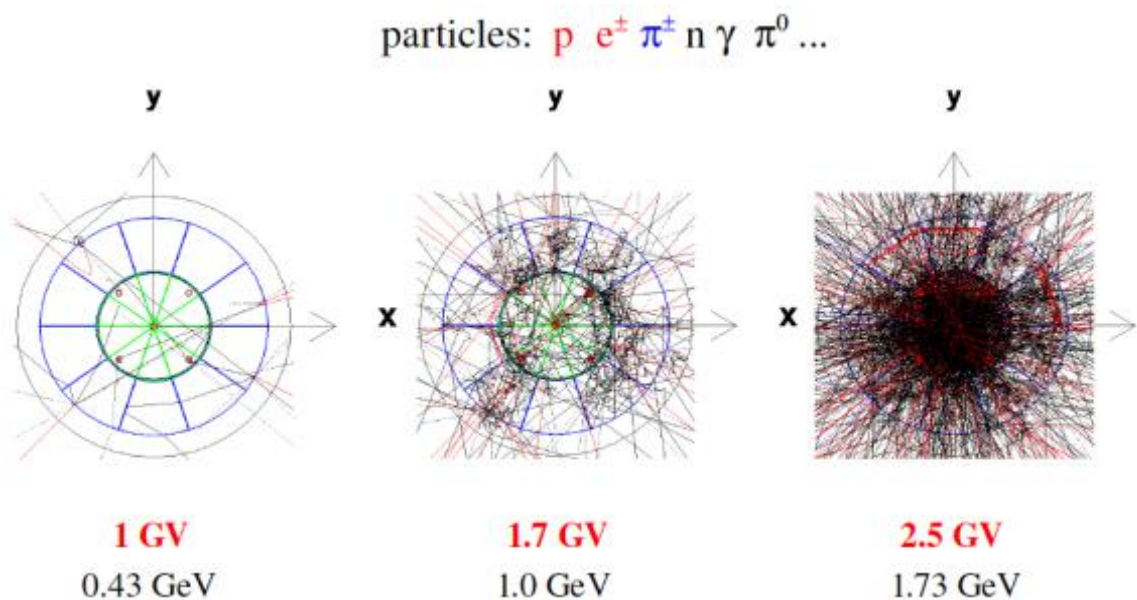

*Figure 5.2 Track Distribution vs **Proton Rigidity** / **Kinetic Energy** for particles traversing the barrel tubes in the direction of the center of the habitat (case of a toroid with 8 Tm )*

The coil structures used in these simulation are made out of aluminum. For the *DH* solenoids *2 cm Al* thickness is used for a *4 T* field; the thickness decreases proportionally with the field flux density. In this way it has been possible to study, as first order approximation, the effect of the material traversed by the particles.

As an example of these simulation, the results for the configuration studied in Ref. 1) (*Geom02*) are presented in *Table 5.1*: we see that the dose reduction for *BFO* at solar minimum is about a factor *2,5 (©44%)* with respect to free space, a relatively small reduction considered that this configuration is characterized by a large a bending power, *18,8 Tm*.

| Z | Free Space | | | Geom02 | | |
|---|---|---|---|---|---|---|
|  | skin | BFO | body | skin | BFO | body |
| 1 | 10.8 | 11.3 | 11.1 | 16.4 | 11.4 | 12.0 |
| 2 | 5.3 | 5.2 | 5.1 | 6.8 | 4.5 | 4.8 |
| 3-10 | 35.9 | 22.2 | 11.8 | 9.4 | 6.1 | 3.0 |
| 11-20 | 38.4 | 16.6 | 14.8 | 7.3 | 4.1 | 2.9 |
| 21-28 | 27.3 | 7.1 | 8.7 | 3.6 | 1.2 | 1.2 |
| total | 117.7 | 62.4 | 51.5 | 43.5 | 27.3 | 23.9 |
| fraction of free space dose | | | | 0.37 | 0.44 | 0.46 |

*Table 5.1 Annual equivalent doses for the configuration studied in (1) (Geom02) compared to the free space dose at solar minimum (units cSv/rem).*

The program also provides the various relevant quantities which can be used for the dose and rate calculations. In this way it is possible to determine the uniformity of the dose computed in various habitat location. The variations of the dose at various body depths as a function of the position within the shielded volume have been determined not to exceed *30%*.

More detailed simulations, taking into account the details of the *SC* coil structure have been done for the two schemes (*"racetrack"* and *DH*) which have been discussed in Chapter 4.

*Geom12*, *Geom13* and, in particular, *Geom14* represent the most advanced versions simulated using the *Double Helix* coils with the relevant material budget (*Figure 5.3* and *5.4*). *Geom14* with respect to *Geom13* has a more accurate simulation of the composition of the *YBCO* tape.

*Geom15* represents the most advanced version simulated using the *"racetrack"* coils with the relevant material budget (*Figure 5.5*). Particular care has been used to reproduce the material composition of the $MgB_2$ *SC* cable and of the coil structural support.

ACTIVE RADIATION SHIELD FOR SPACE EXPLORATION MISSIONS

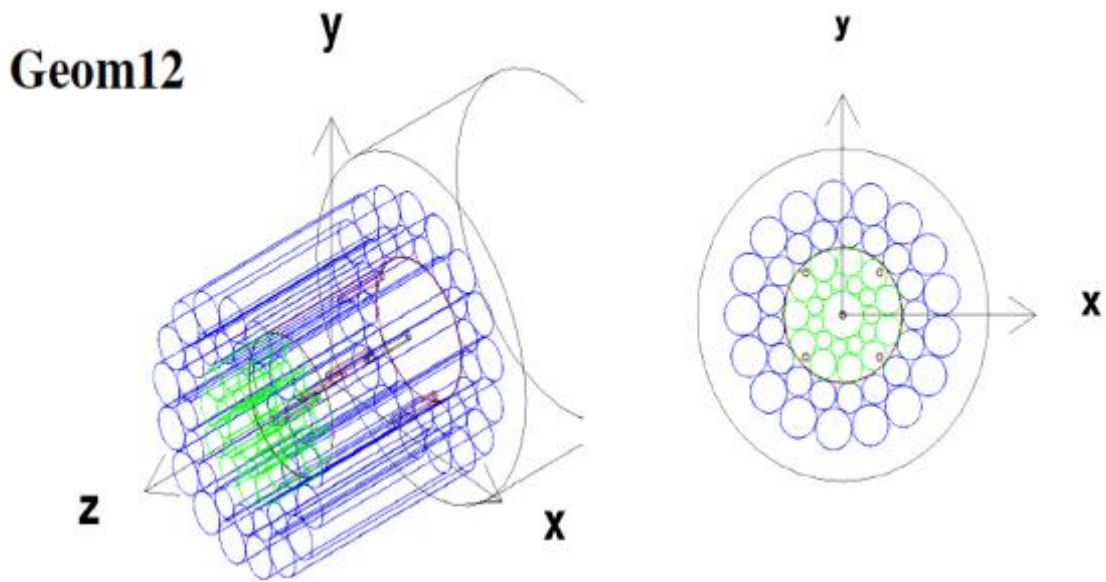

*Figure 5.3 Geom12: Barrel and Endcap composed of two layers of Double Helix coils; B- field = 2T confined to the cylindrical volumes defined by the coils.*

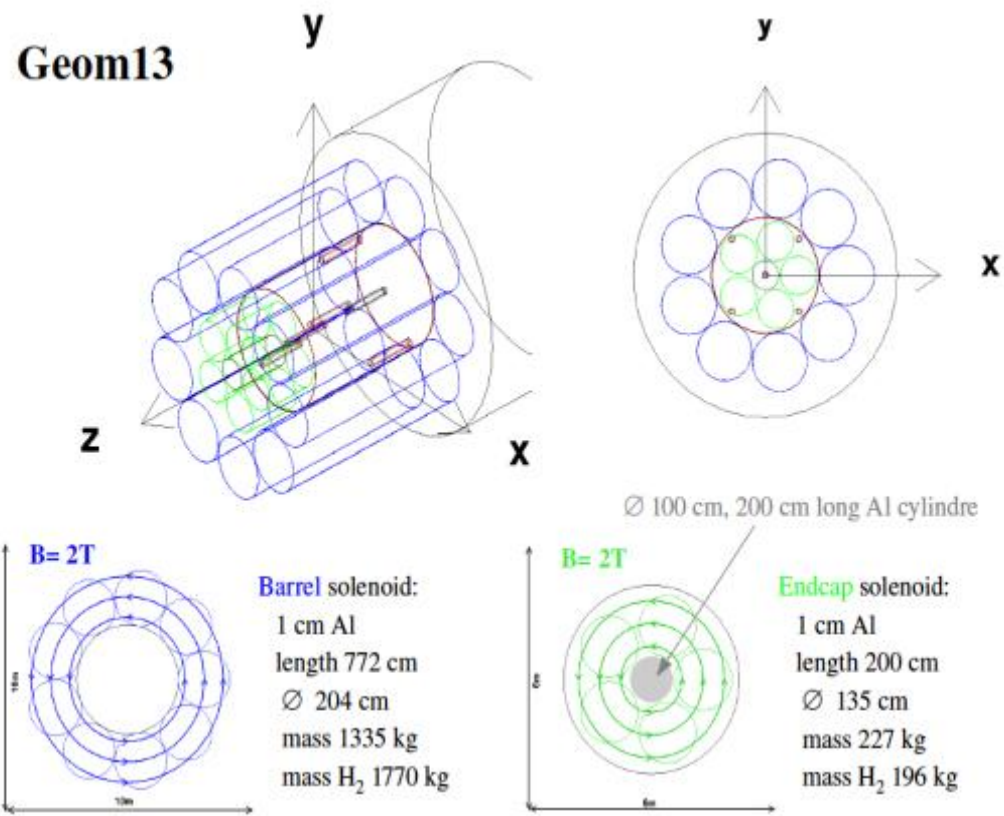

*Figure 5.4 Geom13: Barrel and Endcap composed of 2,04 m and 1,35 m diameter Double Helix coils; B-field=2T confined to the cylindrical volumes defined by the coils.*

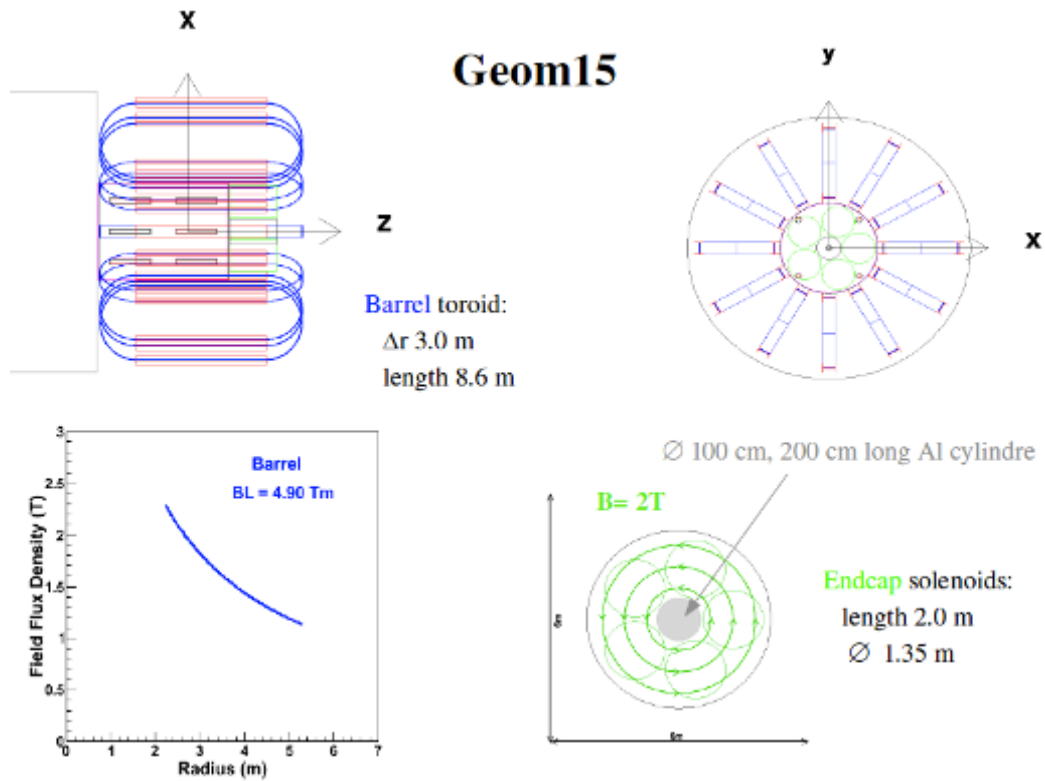

*Figure 5.5 Geom15: Barrel and Endcap composed of 12 racetrack coil and 5 DH coils, respectively. Barrel B-field=5Tm.*

While studying *Geom13* we have analyzed the role of the different parts of the shield in the case of the *4 Tm* array of *DH* coils, a relatively weak bending power:

1) the magnetic field (*4 Tm*) with/without the material of the coils
2) the material of the coils with/without the magnetic field (*4 Tm*)
3) the role of hydrogen as filler material for the *DH* coils ($<t_{LiH2}> = 11$ g/cm$^2$)

For this we compared the *"free space"* GCR yearly dose with the corresponding dose within the module shielded by the various configuration.

The analysis of these tables provides a clear evidence of the complex interaction among the various passive/active shielding element, as a function of the particle charge *Z*. The following considerations can be made (*BFO* doses):

- the presence of *DH* coils with *4 Tm* (*Geom13*) reduces by nearly a factor of 2 (©56%) the yearly dose (most effective at the highest *Z*) ;
- looking more in details, *4 Tm B* field "*without the presence of the coils*", would reduce by a factor of 2 the *HZE*, but would not affect significantly *p, He*.
- the coil material (passive) does contribute with a *10%* shielding effect, mostly on *HZ* ions ;

- filling the coils with *liquid $H_2$* (*<$t_{H2}$>= 11 g/cm²*) improves the shielding effect to an overall factor of *3* reduction (*©35%*), mostly due to the absorption of *HZ* ions by *liquid $H_2$* : the weight of the filling *10 m* of length of each coil would however triple the weight of the shield (from *1* to *3 t/coil*);
- if the coils are filled with *liquid $H_2$* (*<$t_{H2}$>= 11 g/cm²*) the contribution of the relatively weak shield bending power (*4 Tm*) becomes negligible (*10%* additional effect).

| Z | Free Space | | | Geom13 | | | Geom13sb | | |
|---|---|---|---|---|---|---|---|---|---|
| | skin | BFO | body | skin | BFO | body | skin | BFO | body |
| 1 | 10.8 | 11.3 | 11.1 | 19.4 | 13.6 | 14.4 | 17.7 | 12.6 | 13.3 |
| 2 | 5.3 | 5.2 | 5.1 | 8.2 | 5.9 | 6.0 | 7.7 | 5.2 | 5.6 |
| 3-10 | 35.9 | 22.2 | 11.8 | 12.8 | 8.8 | 4.5 | 17.6 | 11.9 | 5.8 |
| 11-20 | 38.4 | 16.6 | 14.8 | 9.5 | 5.5 | 4.2 | 14.7 | 7.6 | 6.3 |
| 21-28 | 27.3 | 7.1 | 8.7 | 4.1 | 1.4 | 1.4 | 7.5 | 3.1 | 2.7 |
| total | 117.7 | 62.4 | 51.5 | 54.0 | 35.2 | 30.5 | 65.2 | 40.4 | 33.7 |
| fraction of free space dose | | | | 0.46 | 0.56 | 0.59 | 0.55 | 0.65 | 0.65 |

*Table 5.2 Geom13: Annual Skin, BFO and Whole Body Equivalent Doses for Solar Minimum (cSv/rem). Geom13: 4 Tm, DH coils; Geom13sb, 4 Tm, no coils*

| Z | Free Space | | | Geom13 | | | Geom13h | | | Geom13sc | | |
|---|---|---|---|---|---|---|---|---|---|---|---|---|
| | skin | BFO | body | skin | BFO | body | skin | BFO | body | skin | BFO | body |
| 1 | 10.8 | 11.3 | 11.1 | 19.4 | 13.6 | 14.4 | 17.6 | 13.3 | 13.5 | 18.5 | 14.6 | 14.5 |
| 2 | 5.3 | 5.2 | 5.1 | 8.2 | 5.9 | 6.0 | 6.6 | 4.8 | 4.9 | 6.8 | 4.8 | 5.0 |
| 3-10 | 35.9 | 22.2 | 11.8 | 12.8 | 8.8 | 4.5 | 3.8 | 2.6 | 1.8 | 4.4 | 3.0 | 1.9 |
| 11-20 | 38.4 | 16.6 | 14.8 | 9.5 | 5.5 | 4.2 | 1.3 | 0.9 | 0.7 | 1.3 | 0.8 | 0.6 |
| 21-28 | 27.3 | 7.1 | 8.7 | 4.1 | 1.4 | 1.4 | 0.4 | 0.2 | 0.2 | 0.4 | 0.2 | 0.2 |
| total | 117.7 | 62.4 | 51.5 | 54.0 | 35.2 | 30.5 | 29.7 | 21.8 | 21.1 | 31.4 | 23.4 | 22.2 |
| fraction of free space dose | | | | 0.46 | 0.56 | 0.59 | 0.25 | 0.35 | 0.41 | 0.27 | 0.38 | 0.43 |

*Table 5.3 Geom13: Annual Skin, BFO and Whole Body Equivalent Doses for Solar Minimum (cSv/rem). Geom13: 4 Tm, DH coils; Geom13h, 4 Tm, DH coils filled with $LiH_2$; Geom13sc, 0 Tm, DH coils filled with liquid $H_2$.*

The study of *Geom14* as a function of *BdL* has been performed. At the (low) value of *4 Tm* bending power which are feasible with currently available technologies, a *DH* shields would provide a *40%* reduction of the absorbed dose. Doubling *BL* to *8 Tm* the values do not changes significantly while reaching *16 Tm* one begins to see the *proton/He* dose decreasing, reaching an overall *61%* (factor *2,6*) dose reduction (*Table 5.4*), a better result if compared with *Table 5.1* (*Geom02*).

| Z | Free Space | | | Geom14-2T | | | Geom14-4T | | | Geom14-8T | | |
|---|---|---|---|---|---|---|---|---|---|---|---|---|
| | skin | BFO | body | skin | BFO | body | skin | BFO | body | skin | BFO | body |
| 1 | 10.8 | 11.3 | 11.1 | 18.3 | 13.0 | 13.5 | 18.7 | 13.4 | 14.0 | 16.5 | 11.1 | 11.7 |
| 2 | 5.3 | 5.2 | 5.1 | 7.7 | 5.4 | 5.6 | 8.1 | 5.6 | 6.0 | 6.9 | 4.7 | 4.8 |
| 3-10 | 35.9 | 22.2 | 11.8 | 15.5 | 10.3 | 5.2 | 13.7 | 9.7 | 4.7 | 7.1 | 4.6 | 2.5 |
| 11-20 | 38.4 | 16.6 | 14.8 | 11.4 | 6.6 | 4.9 | 10.3 | 5.2 | 4.1 | 5.0 | 3.2 | 2.1 |
| 21-28 | 27.3 | 7.1 | 8.7 | 5.6 | 1.8 | 1.8 | 4.4 | 2.0 | 1.6 | 1.8 | 0.8 | 0.7 |
| total | 117.7 | 62.4 | 51.5 | 58.5 | 37.1 | 31.0 | 55.2 | 35.9 | 30.4 | 37.3 | 24.4 | 21.8 |
| fraction of free space dose | | | | 0.50 | 0.59 | 0.60 | 0.47 | 0.58 | 0.59 | 0.32 | 0.39 | 0.42 |

*Table 5.4 Annual equivalent doses for Geom014 with a 2, 4 and 8 T field for the 2 m □ barrel solenoids, compared to the free space dose at solar minimum (units cSv/rem).*

The comparison of *Geom14* and *Geom15* is also of interest (*Table 5.5*). At the low shielding level corresponding to *4 -5 Tm* bending power which are feasible with currently available technologies, *DH* shields provide comparable performances to the "*racetrack*" coil shield, providing a *40%* reduction of the absorbed dose with respect to *deep space*.

| Z | Free Space | | | Geom14-2T | | | Geom15 | | |
|---|---|---|---|---|---|---|---|---|---|
| | skin | BFO | body | skin | BFO | body | skin | BFO | body |
| 1 | 10.8 | 11.3 | 11.1 | 18.3 | 13.0 | 13.5 | 21.3 | 14.4 | 15.4 |
| 2 | 5.3 | 5.2 | 5.1 | 7.7 | 5.4 | 5.6 | 9.0 | 6.6 | 6.5 |
| 3-10 | 35.9 | 22.2 | 11.8 | 15.5 | 10.3 | 5.2 | 12.1 | 8.4 | 4.3 |
| 11-20 | 38.4 | 16.6 | 14.8 | 11.4 | 6.6 | 4.9 | 8.6 | 4.7 | 3.6 |
| 21-28 | 27.3 | 7.1 | 8.7 | 5.6 | 1.8 | 1.8 | 4.5 | 1.2 | 1.6 |
| total | 117.7 | 62.4 | 51.5 | 58.5 | 37.1 | 31.0 | 55.5 | 35.3 | 31.4 |
| fraction of free space dose | | | | 0.50 | 0.59 | 0.60 | 0.47 | 0.57 | 0.61 |

*Table 5.5 Geom13: Annual Skin, BFO and Whole Body Equivalent Doses for Solar Minimum (cSv/rem). Geom14-2T: 4 Tm, DH coils; Geom15, 4,9 Tm, Racetrack coils*

Increasing the bending power by keeping the same *DH* coil radius but increasing the number of *SC* tape layers would necessarily increase the number of secondaries from *p* and *He GCR* interactions. In order to reduce *p, He* dose contribution using the effect of the magnetic field it is important that the coil material budget (thickness) would be kept minimal. This is another reason of choosing *DH* coils of the largest possible radius: for the same *BdL,* larger radius means less material traversed by the *GCR*. This behavior is described in *Figure 5.6* where the dependence on Z of the dose reduction with respect to *deep space,* due to the spacecraft body and due to the *DH*, *4 Tm* active shield is shown. The effectiveness increase with increasing Z: for low Z the effect of the passive shielding is increasing the dose.

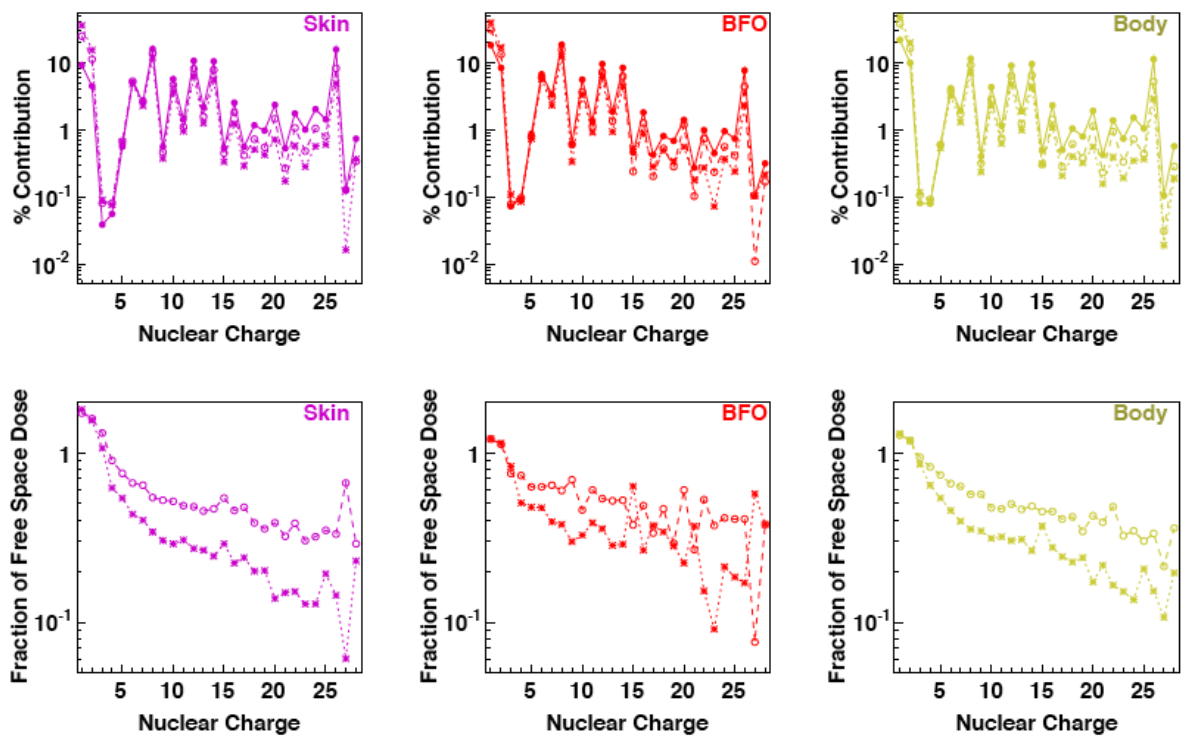

*Figure 5.6 The relative dose contributions of the charged nuclei to the annual free space (solid line), spacecraft without (dashed line) and with (dotted line) the HTSC Double Helix solenoid shield (top); the corresponding ratio of the dose of the spacecraft without (dashed line) and with (dotted line) the active magnetic shield to the free space (bottom).*

*Table 5.6* summarize the annual dose results for the **barrel region only** for the main shield configuration analyzed in the physics simulation. Note that the barrel mass quoted in the table represent only the mass used to describe the material traversed by the particles in the physics simulation and does not represent the weight for the magnetic shield in the various configurations.

| Configuration | BL (Tm) | Barrel Mass (kg) | Annual BFO Equivalent Dose (cSv/rem) |
|---|---|---|---|
| Geom02: Reference Toroid | 19 | 12 472 | 19.2 |
| Geom13sc: Liquid $H_2$ | - | 27 945 | 17.3 |
| Geom14-2T: DX Solenoid | 4 | 4 833 | 29.3 |
| Geom14-8T: DX Solenoid | 16 | 19 296 | 16.8 |
| Geom15: Toroid | 4.9 | 24 100 | 26.7 |

*Table 5.6 A comparison of the doses in the barrel region only of the different shielding configurations. The quoted mass refers to the barrel mass present in the physics simulation.*

The dose levels for the configuration *Geom4-2T* have been evaluated at *solar maximum*. Compared to *solar minimum*, the annual doses decrease by *30 %*. The *BFO* annual equivalent dose in the barrel region is 20,*6 cSv* (*29,3 cSv* in *Table 5.6*).

Finally, we evaluated the shielding effects of the main configuration with respect to the shielding provided by the structural material of the walls of a standard habitable module, namely *1,5 cm (4 g/cm$^2$)* of *Al (Geom01)*. In this way it is possible to separate the effect of the basic structural spacecraft material from the effect of the magnetic field and the magnet structural mass. The result of these comparisons is shown in *Table 5.7*.

| Z | Free Space | | | Geom01 | | | Geom14-2T | | | Geom14-8T | | |
|---|---|---|---|---|---|---|---|---|---|---|---|---|
|   | skin | BFO | body | skin | BFO | body | skin | BFO | body | skin | BFO | body |
| 1 | 10.8 | 11.3 | 11.1 | 18.4 | 13.5 | 14.1 | 18.3 | 13.0 | 13.5 | 16.5 | 11.1 | 11.7 |
| 2 | 5.3 | 5.2 | 5.1 | 8.3 | 5.9 | 5.9 | 7.7 | 5.4 | 5.6 | 6.9 | 4.7 | 4.8 |
| 3-10 | 35.9 | 22.2 | 11.8 | 19.9 | 13.2 | 6.6 | 15.5 | 10.3 | 5.2 | 7.1 | 4.6 | 2.5 |
| 11-20 | 38.4 | 16.6 | 14.8 | 17.3 | 9.4 | 7.1 | 11.4 | 6.6 | 4.9 | 5.0 | 3.2 | 2.1 |
| 21-28 | 27.3 | 7.1 | 8.7 | 10.3 | 3.2 | 3.1 | 5.6 | 1.8 | 1.8 | 1.8 | 0.8 | 0.7 |
| total | 117.7 | 62.4 | 51.5 | 74.2 | 45.2 | 36.8 | 58.5 | 37.1 | 31.0 | 37.3 | 24.4 | 21.8 |
| fraction of free space dose | | | | 0.63 | 0.72 | 0.71 | 0.50 | 0.59 | 0.60 | 0.32 | 0.39 | 0.42 |
| fraction of spacecraft dose | | | | | | | 0.79 | 0.82 | 0.84 | 0.50 | 0.54 | 0.59 |

*Table 5.7 Annual **Skin**, **BFO** and **Whole Body** Equivalent Doses for Solar Minimum (cSv/rem). Free space: no shield; Geom01: 1,5 cm of Al; Geom14-2T: 4 Tm, DH coils; Geom14-8T: 16 Tm, DH coils.*

Following the results of these physics simulation studies, we note that:

1) the role of the material traversed by *GCR* is important: since we are dealing with spacecraft component the amount of material traversed by *GCR* has to be be rather small (max *10-20 g/cm$^2$*). This thin layer of material causes two competing effects on the dose:
- for *Z<3* particles increase their flux and their dose due to secondary production
- for *Z>2* particles decrease their flux and their dose due to the large stopping power
2) the dose reduction obtained with *4 Tm DH* are comparable to the *"racetrack"* configuration with a similar value of *Tm*: the typical reduction, with respect to deep space, is around *40-43%*, with respect to a *4 g/cm$^2$* vessel is around *20%*
3) higher shielding effects could be reached by developing very light, high *BdL*, coils;
4) this level of *GCR* shielding effect would reduce the total yearly dose for a interplanetary flight during solar maximum to about *26-28 rem/y,* close to the current *ISS* values, a factor *1,8* below the maximum allowed yearly professional dose of *50 rem/y;*
During solar minimum the dose would be decreased to *37 rem/y*, *1,35* times below the maximum allowed yearly professional dose of *50 rem/y;*.
From the point of view of the dose reduction, both solutions could be considered for a Mars interplanetary mission.

5) both the active shielding configurations considered in this study have a rigidity cutoff of about *250 MeV* and would be very effective in shielding *SPE* burst.

## 5.2 References

1) V. Choutko, H. Hofer and S.C.C. Ting, "The AMS Experiment and Magnet Faraday Cage for Human Space Exploration", presented at the NASA Active Radiation Shielding Workshop, Ann Arbor, MI, August 17-18, 2004.

# 6  Conceptual definition of an exploration mission

## 6.1 Conceptual definition for a 4 crew compartment

We analyzed a possible conceptual implementation of a superconducting magnetic shield system on a space system aimed at transporting a crew of *4* from the Earth orbit to a Mars orbit and back. The shield is based on the Double-Helix concept implemented as *12* coils placed to form a cylindrical shield around the pressurized modules.

We define a reference mission based on the results of the *ESA* study *Integrated Reference Architecture for Exploration*[1], which has investigated and identified a possible mission architecture for the first manned exploration of Mars. The overall mission foresees several launches from the Earth based on the availability of a *Large Launch System* with a payload capability to *LEO* of *125 tons* and the capability to assembly modules in Earth orbit. At first, two cargo transportation missions will bring on Mars surface the habitation module and other surface elements (rovers, power plant, etc.); also the manned lander/ascent vehicle will be transported to Mars orbit, where it will wait for the *4*-people crew arrival. Then, the human mission foresee the assembly in Earth orbit of the transfer vehicle, with the propulsion stage and the pressurised elements. The crew then reaches the transfer vehicle in *LEO* and the trip to Mars starts.

The mission timing foresees a transfer phase from Earth to Mars orbit of about of *200 days,* a stay on Mars surface of about *400-500 days* and a return trip of about *200 days* (so-called *'conjunction mission'*). An alternative mission (the *'opposition mission'*), would need about *300 days* each way and a stay on Mars of about *40 days*. The whole crew is transferred to the Mars surface, leaving the transfer vehicle in dormant mode until the crew return.

## 6.2 Overall Spacecraft Conceptual Configuration

*Figure 6.1* shows a conceptual configuration for the manned transfer vehicle and refers to the elements definition given in *Ref. 1*.

The Habitation Modules are mounted on top of the large propulsion stages (shown on the left, *10 m* diameter and more than *50 m* long).

The configuration of the *Habitation Modules* given in *Figure 6.1* is somewhat different from the one identified in Ref.1 in order to be compatible with a radiation shield to be placed around it: in particular the diameter has been reduced from *7.2 m* to *5.6 m* and the length extended from *8.5 m* to about *18 m*, with the overall pressurised volume slightly increased. The mass of the pressurised module is in *Ref. 1* is *23000 kg* plus *16000 kg* of consumables items, for a total of *39000 kg*. This mass includes a not better identified passive radiation shield *'equivalent to 10 g/cm$^2$'*; it is not possible with the information available to assess how much of the said mass is specifically dedicated to radiation shielding (and not also, e.g., to structural functions), but it can be estimated to be in range of several tons: assuming a *2 cm*-thick aluminium alloy shell around a structural central core (the only shielded volume) of the habitation module would have a mass in the range of *8-10 tons*, the typical weight of *ISS* modules.

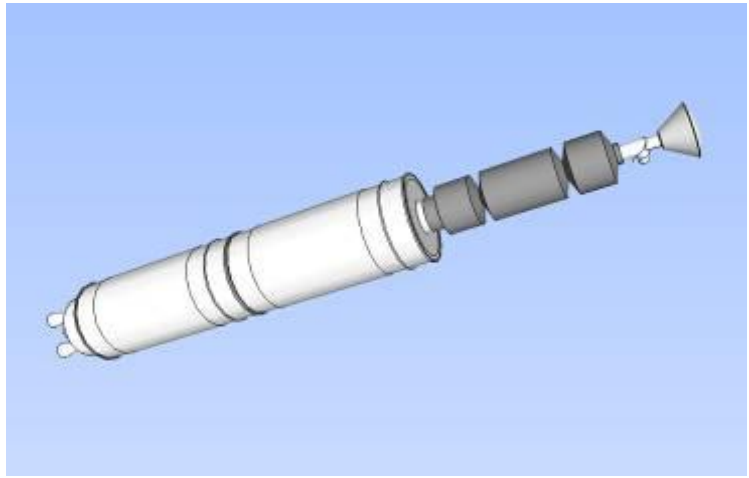

*Figure 6.1 General configuration of the transfer vehicle (w/o radiation shield)*

The *Habitation Modules* are linked via a pressurised corridor to the *Earth Re-entry Module* (right) to be used at the end of the mission. The pressurised corridor presents another docking station for connecting, once orbiting Mars, to the vehicle for transporting the crew down to Mars surface and back (this vehicle is waiting in Mars orbit for the arrival of the transfer vehicle).

The central *Habitation Module* of the configuration assumed for this study is the main area where the astronauts will live; the concept outlined here has general dimensions of *5,6 m* diameter and *10 m* long, providing about *60 $m^3$/crew member*. The other two modules are pressurised modules, containing service equipment and subsystems, which can be visited by the astronauts but where they are not supposed to stay for long times; therefore these modules do not require to be significantly shielded from radiations. The mass of the Habitation Modules has been extrapolated from in *Ref. 1* considering the mentioned *39.000 kg* minus the *9.000 kg* of the estimated mass dedicated only to the passive radiation shielding function, which in our case is covered by the superconductive system: therefore the mass of the *Habitation Modules* is assumed to be *30.000 kg*.

A superconductive shielding system based on *12 DH* coil surrounds the main Habitation Module (*Figure 6.2*). In order to protect the *Main Habitation Module*, the cylinders are longer (about *18 m*) than the module itself, enveloping also the two service module in front and behind the Main Habitation Module. The *'front' Service Module* can work, thanks to its mass, also as a *'conventional'* radiation shield on the front section of the main module. On the other side, the shielding is provided by the other *Service Module* and by the *Propulsion Stages*. Then the *Superconductive Shield* is required to envelop and protect only the lateral sides of the main module.

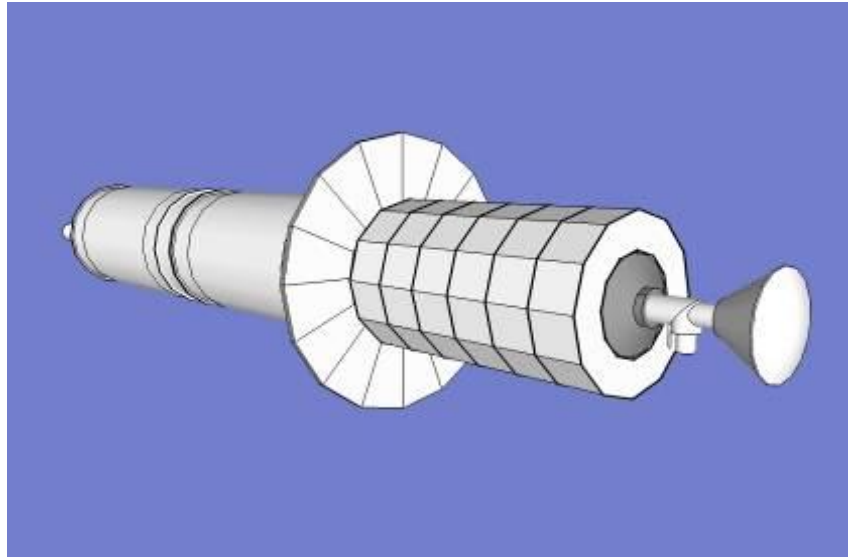

*Figure 6.2 General configuration of the transfer vehicle (with radiation shield)*

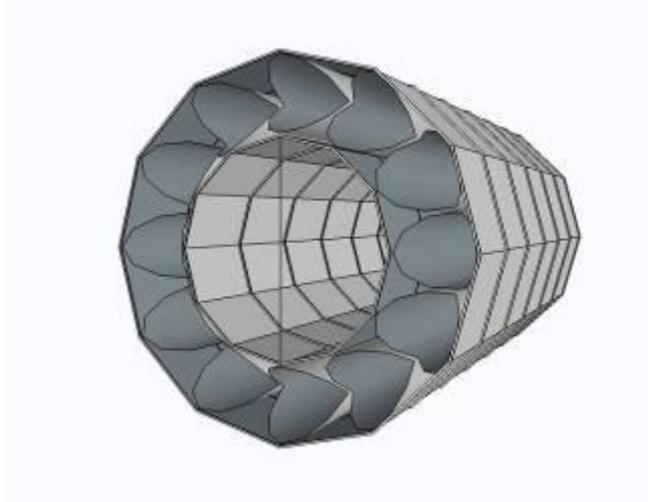

*Figure 6.3 Radiation Shield general layout*

Considering the cryogenic temperatures required for the functioning of the superconductive shield, the thermal issues represents one of the most important and critical aspects to be addressed, the feasibility of the system itself depending on them. To reach and maintain the required cryogenic temperatures, the thermal losses must be minimised.

For this, a sophisticated *Thermal Protection Subsystem* must be implemented. The main elements of the *Thermal Protection Subsystem* are:

- the *Sunshield* (the disk on the right of *Figure 6.4*);
- the *Thermal Insulation Shell* assembly (on the left of *Figure 6.4*), which envelopes the *12 cylinders* both on the internal side and on the external one.

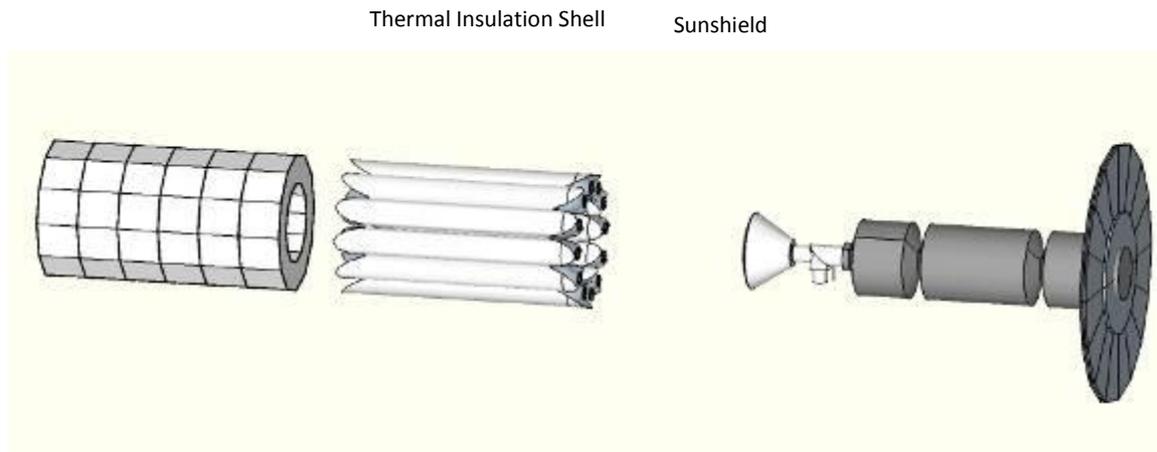

*Figure 6.4 Main elements of the Thermal Protection Subsystem*

## 6.3 Thermal Protection Subsystem concept

A first element is given by the sunshield, which is shown in *Figure 6.2* as the large circular plane to be positioned between the propulsion stages and the pressurised elements. It is a deployable element composed of multi-layer insulation (*MLI*) blankets; more than one *MLI* blanket could be used. The vehicle is maintained oriented with one side (in this case, the propulsion stage side) facing the sun; in this way the sunshield keeps under its shadows the superconducting system. A similar concept is used by the *NASA* James Webb Space Telescope (see figure below) to shield cryogenic instrumentation from sunlight (and also from spacecraft bus).

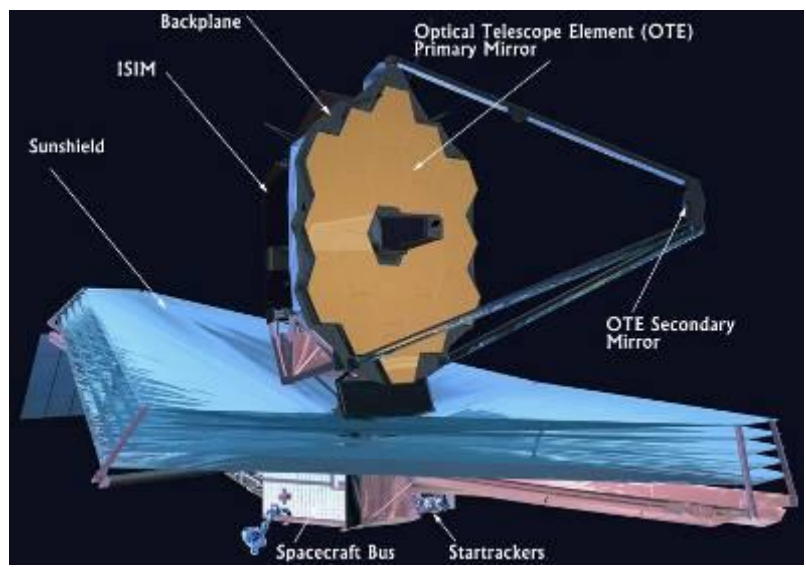

*Figure 6.5 The NASA James Webb Space Telescope*

The sunshield allows the superconducting system to be exposed only to the deep space (at *3K*), in this way easing the thermal control of the system.

When the spacecraft would use its engines, it has to orient the longitudinal axis in the required thrust direction so that it is not possible to keep the sun-pointing attitude. In any case the propelled phases are very short (max few hours) in comparison to the mission duration. Other situations may require specific attitudes, such as rendezvous and docking manoeuvres in Mars orbit. The cumulative duration of these situations can be estimated to about one day (or maximum few days) and has to be considered when dimensioning the *Thermal Insulation Shell* and cryogenic system.

Similarly, the sun-pointing concept has to be assessed during Mars orbiting conditions, where the planet albedo and *IR* emission have to be accounted for and cannot be shadowed by the sunshield.

The area of the sunshield is about *200 m$^2$*. The present concept would allow up to *9°-10°* deviation from sun-pointing direction while still keeping shadowed the system, thus reducing the need for attitude manoeuvring, especially while orbiting Mars.

An alternative solution would be to have a thermal blanket wrapping the whole cryogenic system. In this case the attitude of the vehicle would not be constrained. The overall surface would be of about *800 m$^2$*.

The advantage of the sunshield with respect to this solution is that only a limited area is exposed to the sun and can be optimised for such condition (sunshield); with the other solution, different parts of thermal protection would be exposed to the sun at different times, so that the whole surface has to meet the more stringent thermal performance requirements. For the same reason, the sunshield would also provide a more homogeneous effect on the radiation shield, with respect to a wrapping thermal protection.

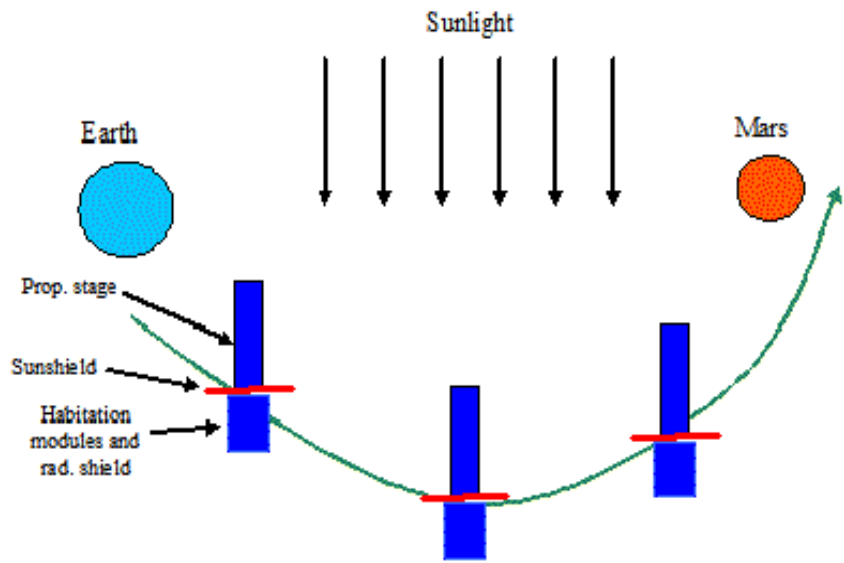

*Figure 6.6 System flight attitude during transfer flight*

## 6.4 The Thermal Insulation Shell

The *Superconductive Shielding* coils assembled together to form a cylindrical shape must be thermally insulated from the crewed elements which are at much higher temperature (about *295K*). The radiative heat flux can be controlled with *MLI* blankets; the conductive fluxes through structural connections between pressurised elements and the *Superconductive Shield* could be minimised if such structural interfaces are reduced to a minimum.

For this reason and considering that such structural connections would be sized by launch loads, a structural separation can be envisaged between these two items, where each of the two assemblies, the *Habitation Modules* and the *Radiation Shield*, is directly connected to the launch vehicle, avoiding load paths from one assembly to the other. The structural interfaces between the assemblies can then be sized to the small loads occurring during in-space flight, where accelerations and vibration are much less demanding; in this way the section areas where heat fluxes can flow is reduced to a minimum; a proper design of such structural connections can further decrease heat fluxes (e.g., high strength wire system, long paths).

The space-facing side of the *Radiation Shield* ideally should take advantage of seeing directly the deep sky at 3K. But at these very low temperatures (*20K*), the radiative flux out of the *Coil Assembly* toward deep space is very small, indeed negligible (less than *5 W* for the whole *Coil Assembly* surface) with respect to the heat fluxes from the rest of the spacecraft toward the *Coil Assembly*.

On the other side, a problem arises when the vehicle is in martian orbit, where the albedo and planet *IR* radiation would heat the cryogenic system to unacceptable levels in case no thermal insulation is used. A similar situation, with higher fluxes but shorter exposures, occurs when the spacecraft has to leave its sun-pointing attitude for performing propelled phases.

For this reasons it is deemed necessary to insulate with *MLI* also the external side of the *Coils Assembly*.

Therefore the thermal insulation represents a double shell covering the *Coil Assembly* both on the internal side (toward the pressurised modules) and on the external side (to deep space); also the annulus surfaces closing the envelope between the internal and external shells is covered with *MLI*.

The internal side of the *Thermal Insulation Shell* is facing the pressurised module at *295 K* and this represents the area of the highest thermal exchanges and therefore the insulation is more critical. *MLI* blanket performance cannot be improved beyond a certain level by increasing the number of layers, because manufacturing aspects (sewing, mechanical interfaces, ...) are limiting such performance to asymptotic figures: an equivalent emissivity for a good *MLI* is in the range of *0,03*.

Therefore, to further improve insulation, an additional blanket has been envisaged covering the pressurised modules. More complex solution could involve a third (or more) blankets, or a layer of active heat removal system, for instance based on two phase liquids as developed for the *AMS-02* silicon tracking system to remove *O(100) W*. In this case also proper supporting structures (suitable for launch environment, mainly vibrations) have to be foreseen.

## 6.5 Preliminary Thermal Analysis

In order to roughly estimate the heat fluxes to be removed from the coils and assess the criticality of the thermal issues, a thermal model has been implemented.

The system is composed by three main components: the *Habitation Module* (red in figure), the insulating shell and coils structure (blue) and a sun-shield (grey) (*Figure 6.7*).

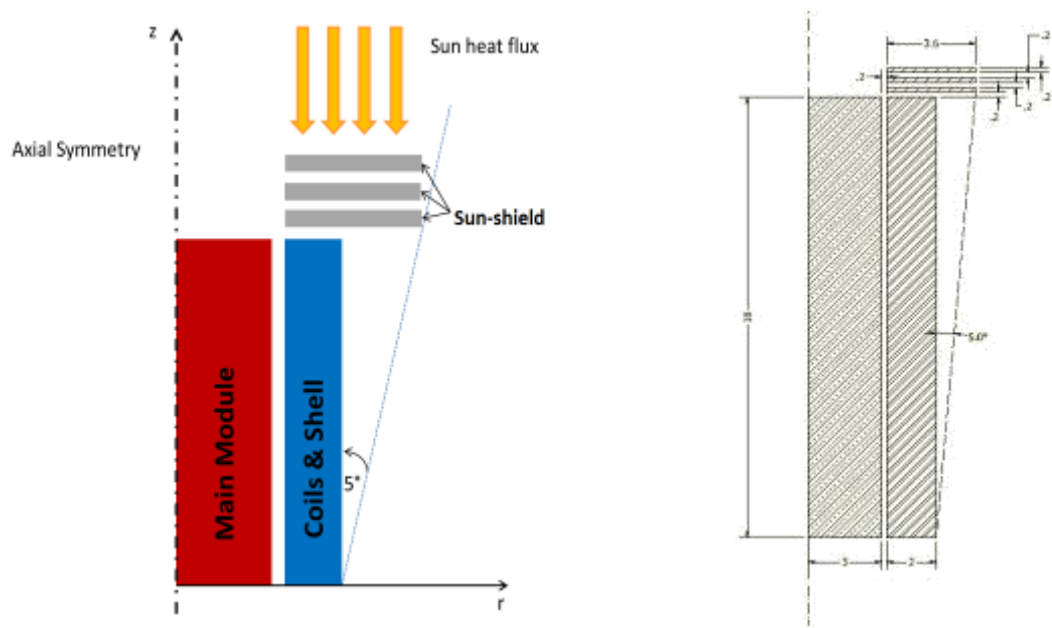

*Figure 6.7 Thermal model description.*

It is assumed that the *Main Module* is always kept at *295K*; the Sun is always aligned to the z direction. The surfaces of the insulation shell are covered with *MLI* with an effective emissivity of *0,03*; the same values are used for the different layers (*3*) of the sunshield, the only exception being the sun-looking face which is covered with *OSR* (*Optical Solar Reflector*) with high emissivity and very low absorptivity ($\Sigma = 0,86$, $\langle =0,07$). The external surface of the *Habitation Module* is covered with *MLI* as well (emissivity *0,03*).

Thermal conductivity is neglected at a first instance and the view factors are numerically computed. The following graph (*Figure 6.8*) shows the heat power to be removed from the *Coil Assembly* as a function of coils temperature.

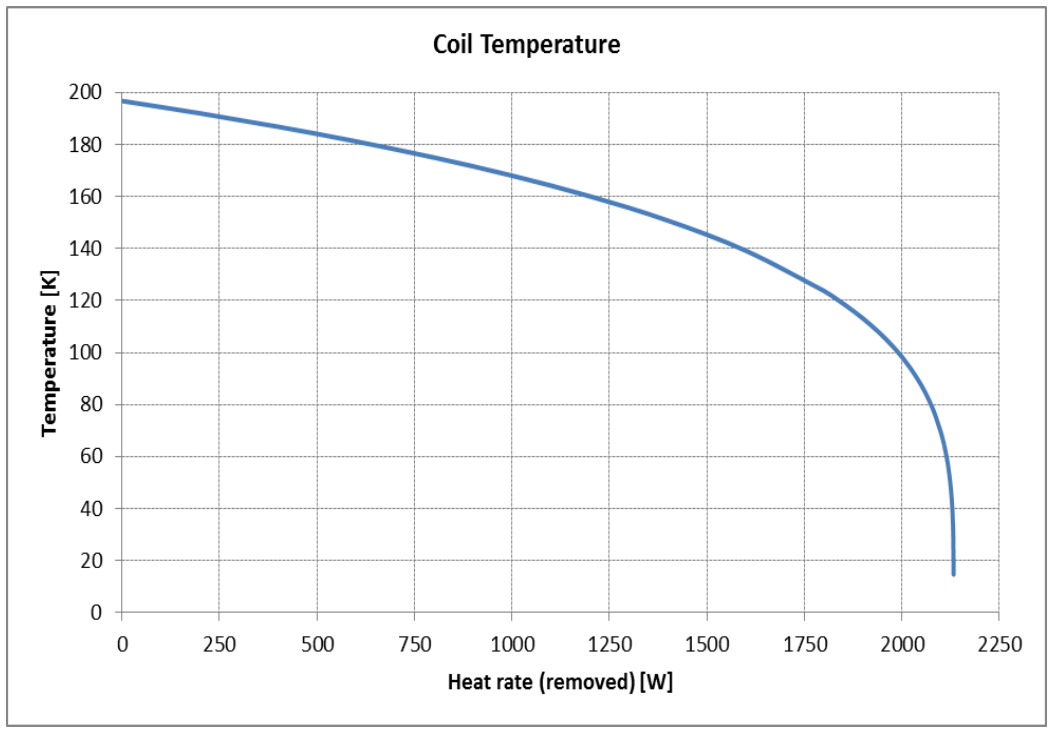

*Figure 6.8 Coil temperature as a function of removed heat rate*

It results that is necessary to remove about *2,1 kW* to keep a *20K* temperature.

The case in which an additional *MLI* blanket is used between the *Coil Assembly* and the *Habitation Module* has been assessed as well: in this case the heat to be removed by the cryogenic system is about *950 W*.

A rough estimate of conductive fluxes due to structural elements connecting the *Habitation Modules* and the *Coil Assembly/Thermal Insulation Shell* has led to figure in the order of few hundreds Watt, under the assumption of a *'weak'* structural connection as previously described.

These are preliminary results which, with their limitations, allows in any case to understand the orders of magnitude at stake.

## 6.6 The Cryogenic Subsystem

The *Cryogenic Subsystem* has to provide the cryogenic fluid to allow the superconductive functionality of the system.

As coolant fluid two possible solutions can be envisaged: the use of liquid hydrogen which has a liquefaction temperature of about *20K* compatible to the one required by the superconductive material and, as an alternative, the use of gaseous helium; a trade-off between the two solutions could not be performed in the frame of this study and should be done within future activities. As a qualitative consideration the low thermal capacity of gaseous helium could require large gas flow rates and a correspondingly large cryogenic plant.

The implementation of each of these solutions in any case poses important technological implications.

For the liquid hydrogen solution, the unavoidable thermal losses occurring in the *Radiation Shielding* system must be compensated implementing one of the these two possible approaches:

- To use the boil-off of hydrogen which is then lost;
- To re-liquefy the gaseous hydrogen to liquid phase and recycle it.

The first solution is normally used in space application, for instance in cryogenic propulsion stages, but the consumption of hydrogen is such that this solution can be implemented usually only for short mission durations (days/weeks).

The second solution has never been implemented in space applications, although is very common on ground. Specific *R&D* would then be required.

Regarding the use of gaseous helium, a cryogenic refrigerator has to be used, but again, a helium closed loop system has never been implemented in space applications. Specific *R&D* would then be required.

Considering the hydrogen-based solution, the results of the thermal analysis previously described allow to perform a further assessment.

On the base of the value of the heat of vaporization of hydrogen, using a boil-off technique would require about *400 kg* of hydrogen per day (*150 tons* per year) assuming a *2,1 kW* heat rate. This is not compatible with mission durations of *1-2 years*, then a different approach should be developed where the vaporized hydrogen is re-liquefied and re-enters into the cycle. Hydrogen liquefaction is a very energy-consuming effort on earth applications, where gaseous hydrogen at ambient temperature has to be made liquid; for the application of interest the gaseous hydrogen is already at very low temperatures, not far from *20K* so that the required energetic impact should be much lower. In any case, as said above, this is a new technology to be developed, which would be very useful more in general also for other space applications, for instance the use of cryogenic propulsion also for very long missions.

The *Cryogenic Subsystem* has also to provide the distribution and circulation of the coolant to the Coils.

The various equipment composing the *Cryogenic Subsystem* as well as the hydrogen tanks could be located within some of the Coils, near one of the ends of the coil cylinder.

## 6.7 The Coil Assembly Subsystem

The *Coil Assembly Subsystem* is formed by the *12* Coil elements mechanically joined to form a single assembly.

In *Figure 6.9* a section of the spacecraft is shown; the pressurized element is in the centre (the size of the man sketch gives the proportions); it is covered with a *ML*I blanket (yellow) with high performance. Around the pressurized module the *Radiation Shield* is composed by the *12 Coil Elements* and by the *Thermal Insulation Shell*. Each *Coil Element* is structurally connected to the nearby ones to form a very strong and stiff structure working as one element during launch and able to stand the inward radial electromagnetic forces acting on the *Coil Assembly* during its functioning.

The *Thermal Insulation Shell* is composed by the inner and the outer *MLI* blankets (yellow; the shell is then closed by two annulus *MLI* blankets at the ends) and is supported by a lighter structural assembly (shown in grey) attached to the *Coil Elements* structures.

The *Coil Element* structure is supposed to be composed by a cylindrical continuous sheet of aluminum alloy: aluminum alloys are the choice for structural materials which must provide high mechanical properties -strength, stiffness- at cryogenic temperatures.

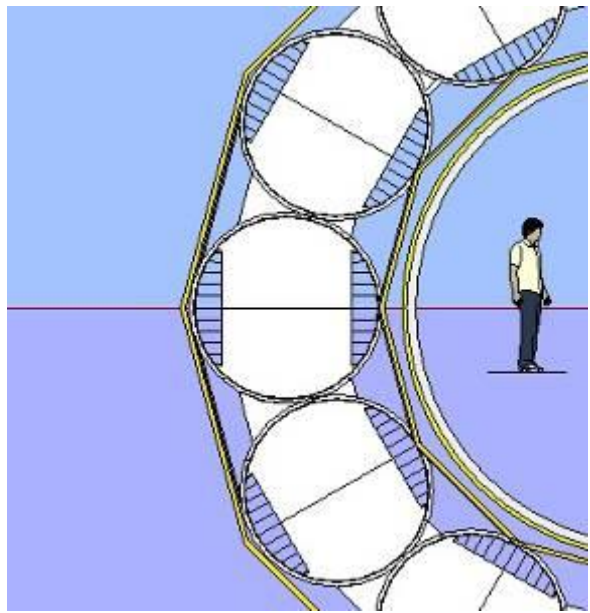

*Figure 6.9 Generic section of the Coil Assembly*

## 6.8 Single Coil Structural Configuration

The sizing loading conditions for the Coils are given by:
- the launch requirements (static and dynamic loads, minimum frequency requirements);
- the electromagnetic forces acting on the coil conductors during the Radiation Shield functioning.

The electromagnetic forces are acting between adjacent coils to attract each other with a force distributed on the surface, with higher electromagnetic pressures in the parts closer to the adjacent coil. To contain the superconductive elements, then, the Coil cylindrical structure must be placed externally with respect to the conductor. Considering the circular distributions of the *12* Coils, the net force acting on one Coil as result of the forces from the two adjacent Coils has a radial inward component, which has been estimated (in other part of the study) to be *10.3 MN* per Coil. These forces act as a *"pressure"* force which is supported by the *Coil Assembly*, where the Coils are connected one each other to form an overall cylindrical 'pressure vessel' (see blue arrows in Figure 6.10). To avoid "buckling" of the thin *DH* coil, a suitable coil mechanical supporting layer will be needed.

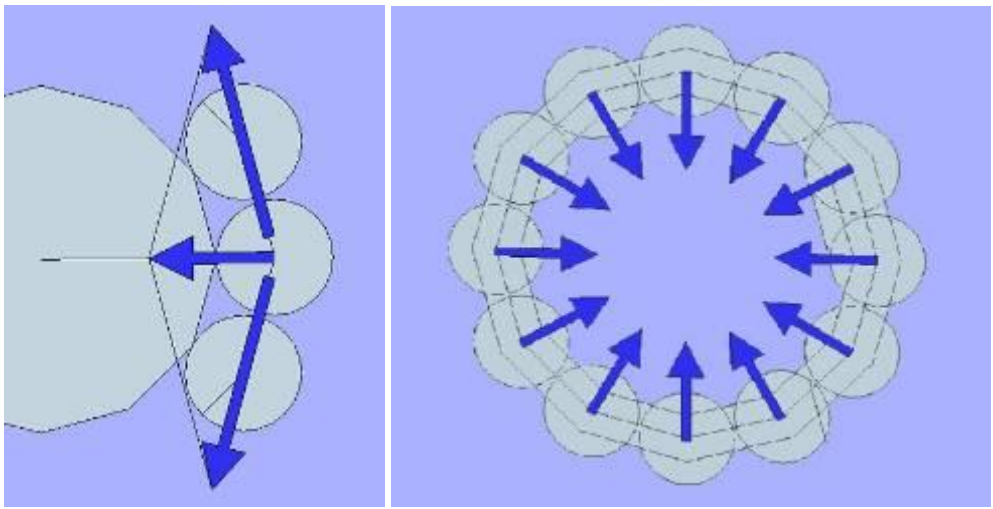

*Figure 6.10 Forces acting of the Coil Assembly*

Another effect is that the electromagnetic forces act to extend each Coil in the radial direction of the *Coil Assembly*. So, the electromagnetic forces are acting on one side to compress the coil cylinder section in one direction (blue arrows in *Figure 6.10*) tangential to the *Coil Assembly*, to withstand the 'pressure' loads and, on the other side, to extend the section in the other direction (see red arrows).

The coil structural concept is shown in the same figure. A purely cylindrical structure, without any internal element, is not suitable to bear such loads. The two *"compressed"* sectors are then connected with plates which bear the compressive force (stiffeners and/or corrugated surfaces should be implemented to avoid structural instability issues). The two *"extended"* sectors are connected with wires/thin bars which bear the tension forces.

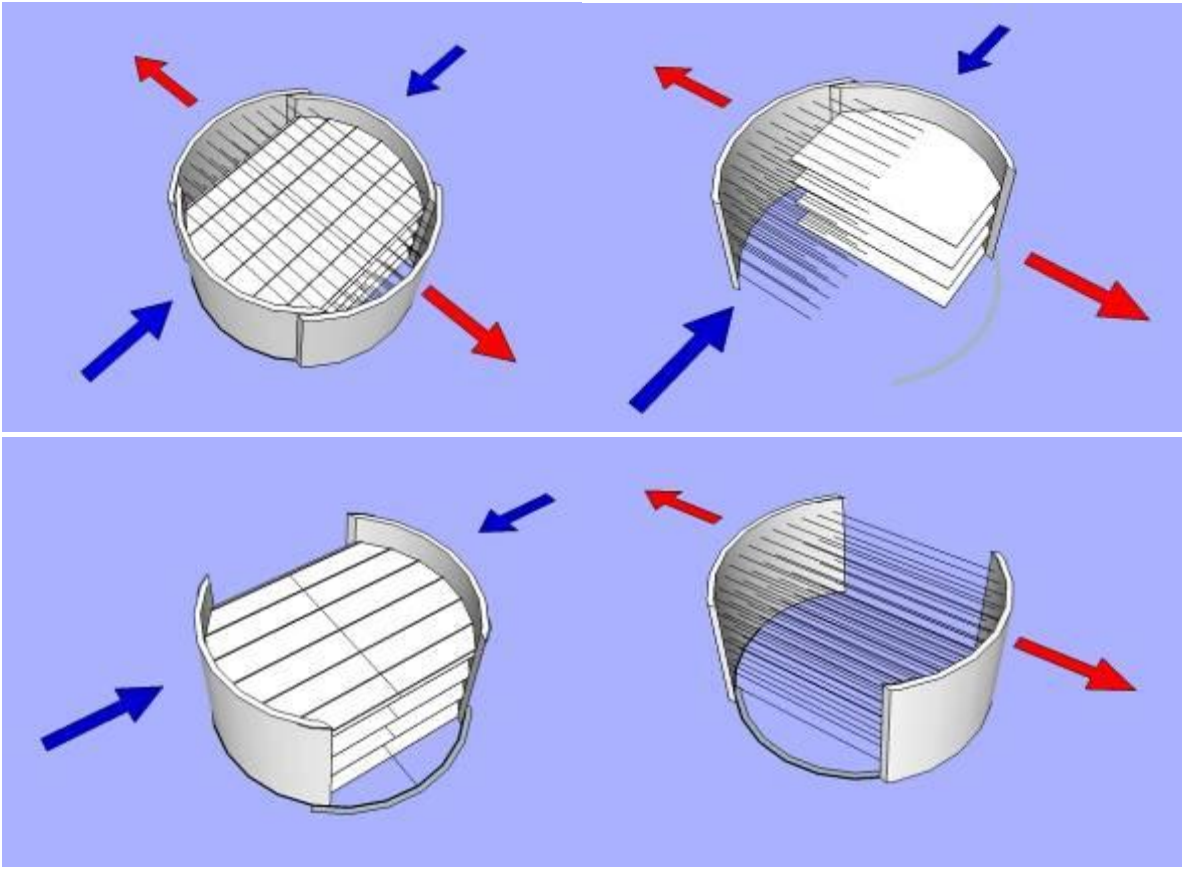

*Figure 6.11 Coil structural concept*

## 6.9 Coil Assembly Structure

It has been said that to minimise the conductive heat transfer, the *Habitation Modules* and the *Coil Assembly* should be structurally decoupled during launch, so that the interconnecting structures could be sized only for bearing the much lower in-space loads.

Assuming then that the *Radiation Shield* is brought to orbit in a single launch together with the *Habitation Modules*, the *Figure 6.12* shows the arrangement for launch. On the left is the launch adapter on top of the launcher (which is not shown). The *Habitation Modules* (orange) are connected with the green ring to the cone adapter (green) on the launcher; each of the Coil element has it own interface ring (about *1 m* diameter, shown in red) with other similar ring adapters on the launcher side (red as well) (the *Thermal Insulation Shell* is not shown in the figure for better clarity but it is supposed integrated with the *Coil Assembly*). The launch accommodation, with *12* interface rings with the launcher allows a simple launch load path along the Coils.

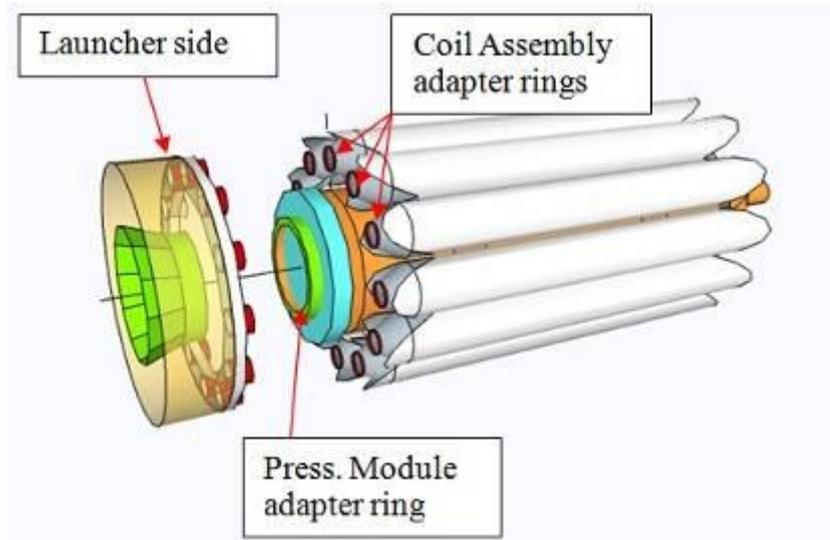

*Figure 6.12 Launch arrangement of the Coil Assembly with 12 Interface Adapters*

The system diameter, about *10 m*, could be in principle compatible with a *Large Launch System (LSS)* able to put into orbit payloads of *70-125 tons*. Such large vehicles are envisaged in the frame of manned planetary exploration to match the requirements for lifting heavy spacecraft.

The *12* Coils are structurally connected to form a single overall structure: in this way the dynamic response requirement deriving from a hypothetical launch vehicle could be much easier to met: in fact the first lateral frequency of such system has been assessed between *17* and *20 Hz* (including not also the structural mass but also the superconductive elements and accessories mounted on the coils); a typical requirement from a launch system could be envisaged at below *10 Hz*.

From the point of view of strength with respect to launch loads, the overall *Coil Assembly* working as one single structural item provides a strong system both to lateral and axial loads (static and dynamic). For instance, extensive margins exists with respect to axial loads (factor of *10*), where the resistive strength is provided mainly by the coil cylindrical part of the structure.

It has to be pointed out that this part of the coil structure has not been sized to *EM* forces, but the thickness has been preliminarily defined on the base of other considerations, such as the capability to locally transfer the *EM* forces from conductor tapes to the coil internal structural items. For this reason, it can be assumed that the mass related to such cylindrical structures is not dependant by *EM* load or launch loads.

The estimate of the mass of the *Coil Assembly* is given by

- the mass of the single Coil cylindrical structures *(954 kg)* plus the cables mass *(102 kg)*, as estimated in the previous section, plus the mass of the conductor tapes (about *1000 kg* per Coil); plus the mass of the launcher adapter (*100 kg*); such sum (*2156 kg*) multiplied by *12*;
- The mass of the structures (plates) required to withstand the 'pressure' force (*4940 kg*);

*85*

- As a first approximation a *20%* margin can be identified to include contribution from the structural connections between the Coils, of the Coil with the Habitation Module, of the Coil with the Thermal Insulation Shell; mass of ancillaries, ...

Therefore the Coil Assembly mass is estimated to be:

$$1,2 \times (2156 \times 12 + 4940) = 36.975 \ kg$$

This is a preliminary estimate to identify the order of magnitude of structural masses, and can be subject to significant changes, when considering further issues beyond the strength characteristics of the material and more realistic and complete structural load conditions

## 6.10 Power and Control Subsystem

This subsystem has to manage the functioning of the Radiation Shielding:
- controlling the electrical currents flowing in the coils during the various phases and operating modes;
- monitoring the system in all its aspects (functioning, performance, safety,...) - *Health Monitoring System*;
- interfacing with the rest of the spacecraft, in terms of electrical power supply, monitoring data, commands;
- automatic handling of emergency situations.

This subsystem would be basically composed of electronic units located inside the pressurized modules.

## 6.11 System Mass Budget

The overall mass of the *Radiation Shield* has to include the mass of all the identified subsystems:

- *Coil Assembly Subsystem*;
- *Cryogenic Subsystem*;
- *Thermal Protection Subsystem*;
- *Power* and *Control Subsystem*.

For the first one, the *Coil Assembly*, the mass has been already estimated above.

For the *Cryogenic Subsystem*, at this stage it can only be assessed a possible wide range for its mass. The required amount of liquid hydrogen would be envisaged in the range of several cubic meters; the coil cylindrical surface where the conductors are placed is about *1.300* square meters; with one centimeter thickness layer of cooling hydrogen it means *13 $m^3$* ; with a *100%* margin on such volume, the hydrogen mass is about *2.000 kg*.

The mass of the cryogenic equipment (for liquefaction, pumps, valves, ...) plus distribution piping, and including redundancies, could be guessed in the range of *2-4 tons*.

**Thermal Protection Subsystem**

The *Thermal Insulation Shell* is represented by a surface of about *1.000 $m^2$* covered by *MLI*; the mass is given mainly by the *MLI* supporting structure, which can be envisaged as made in aluminum or composite honeycomb panels; the panels could be connected directly to coils structures with some further possible stiffening structure.

Assuming a honeycomb mass of 1.*5 kg/$m^2$* (*0,2 – 0,25 mm* thick sheets), the mass the Insulation Shell is estimated at *2.200 kg* (*50%* margin).

The Sunshield is a deployable structure of about *200 $m^2$* composed of *3* or more insulating layers. The mass of this system is driven by the deployable elements and can be roughly extrapolated by similarity with other deployable systems at about *1.000 kg*.

Then the *Thermal Protection Subsystem* mass is about: *2.200 + 1.000 kg = 3.200 kg*.

**Power and Control Subsystem**: mass estimated at less than *2000 kg*.

All this considered the preliminary estimate of the overall Superconductive DH Magnet Radiation Shield mass has been estimated to be:

| Subsystem | Mass (kg) |
|---|---|
| Coil Assembly Subsystem | 37.000 |
| Cryogenic Subsystem | 4.000 - 6.000 |
| Thermal Protection Subsystem | 3.200 |
| Power and Control Subsystem | 2.000 |
| TOTAL | 46.300-48.200 |

*Table 6.1 Radiation Shield Mass Budget*

## 6.12 System Arrangements for Launch and In-orbit Assembly

*Figure 6.11* has already introduced the general concept for transportation of the *Radiation Shield System* on the launch vehicle together with the *Habitation Modules*. The overall assembly would have a mass compatible with the expectable capabilities of a *Large Launch System* which is required for human exploration (in the range of *80-100 ton* payload in *LEO*). As well a system diameter of *10 m* would be most likely compatible with the payload envelope of such a launcher.

*Figure 6.12* shows this assembly (*Habitation Modules + Radiation Shield*) installed within the fairing of the launch vehicle.

This configuration has the advantage that it is not required to assemble in orbit the Radiation Shield out of a set of independently launched modules; nor it is required to integrate in orbit the *Radiation Shield* with the *Habitation Modules*. This would greatly simplify the orbital operations and also the Radiation Shield would result in a simpler system, as no cryogenic connections or superconductive electrical connections would have to be implemented in orbit.

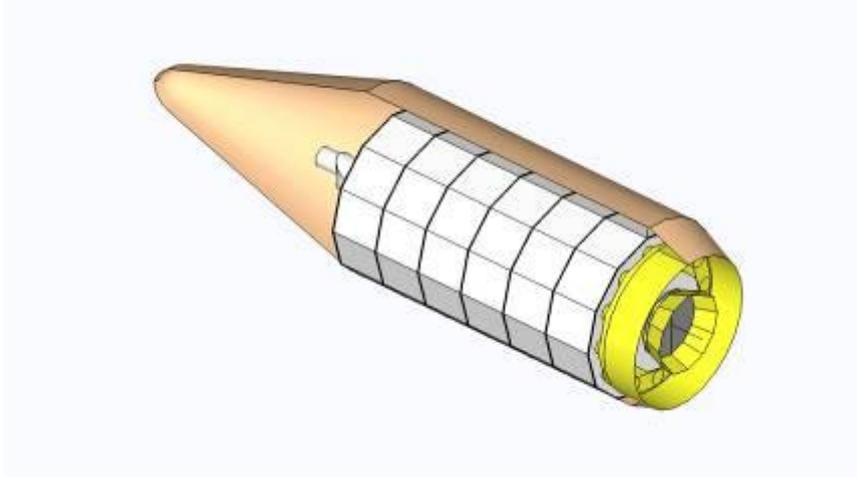

*Figure 6.12   Single launch configuration (no assembly/deployment in orbit)*

In case a *Large Launch System* of the characteristics described above would not be available, the need of in-orbit assembly would be unavoidable. Hereafter, two possible cases have been investigated.

In the first case, *Figure 6.13*, the *Radiation Shield System* is split in *6* parts, each composed by two adjacent Coils; the *6* parts would be put in Earth orbit with two separate launches (three coil pairs per launch. The launch adapter interfaces mounted on each coil structure and described in above sections could be used as well for this configuration.  The stowed *Sunshield* could be accommodated for launch in the room available among the *6* coils). Some mass penalty with respect to the single-launch configuration could be expected. The *Habitation Modules* would require an additional launch .

In this configuration the required launcher capabilities are:
- *20-25* ton payload in *LEO*;

- Payload envelope diameter: *6.2 m*;
- Payload envelope height: *18 m*.

These characteristics are not so far from the *Ariane5* capabilities.

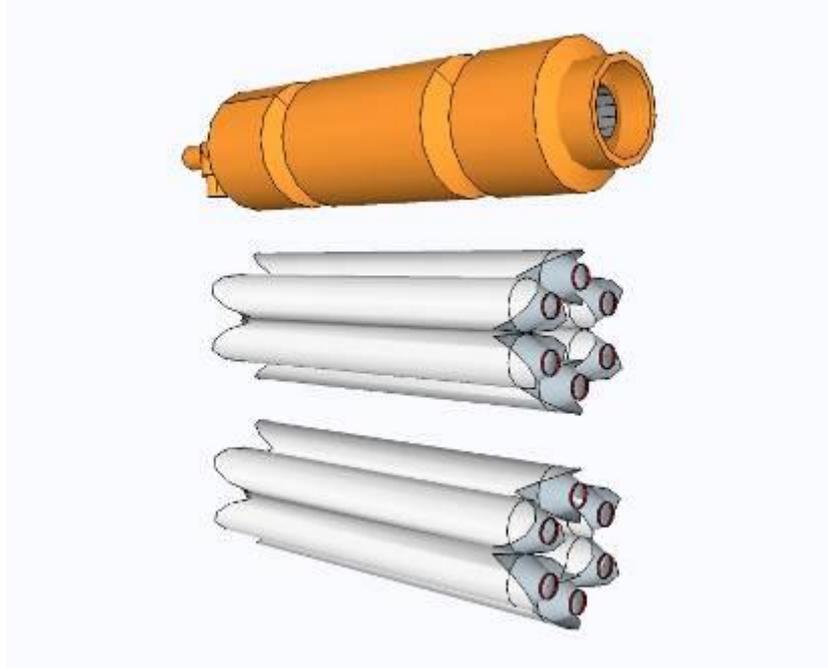

*Figure 6.13 Multiple launch configuration - 1*

Another possible configuration when multiple launches are required is shown in *Figure 6.14*.

In this case the Radiation Shield is divided in *4* parts, each composed of *3* adjacent Coils. With respect to the other solution, the in-orbit assembly would be easier as less components are involved.

The requirement for the launch system are the same as for the first multiple-launch configuration, but for a wider payload envelope diameter: now *7.6 m* are required.

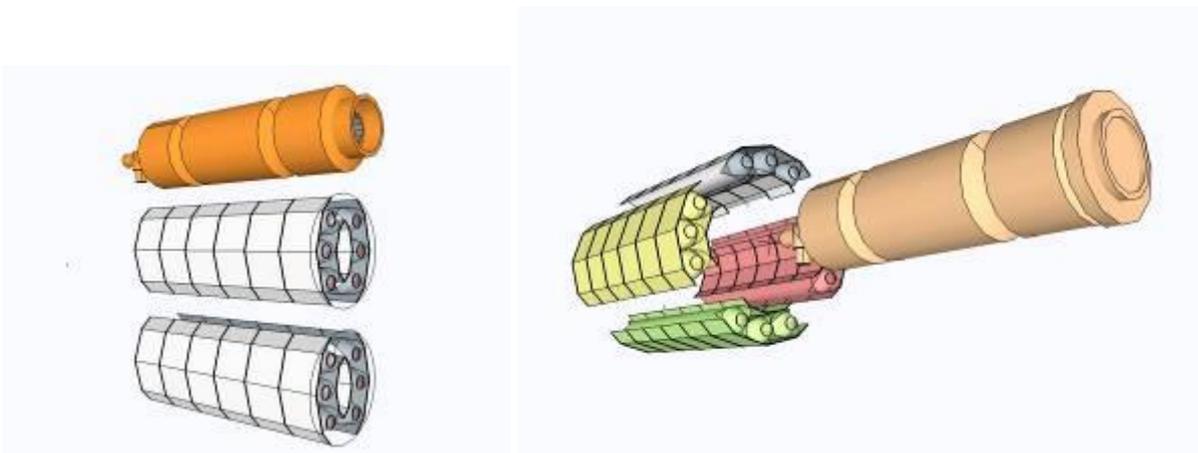

*Figure 6.14 – Multiple launch configuration – 2*

## 6.13 System conceptual definition for a 8 crew compartment

The definition of a *Radiation Shield* configuration relevant to *8*-crew mission (with respect to the 4-crew considered in above sections) starts on the assumptions that a similar *Large Launch System* as the one considered before: payload capability in the range of *80-100* ton in *LEO* and a payload envelope diameter of *10 m.*

On such base, the simplest configuration is to double the *4*-crew system (*Figure 6.15*).

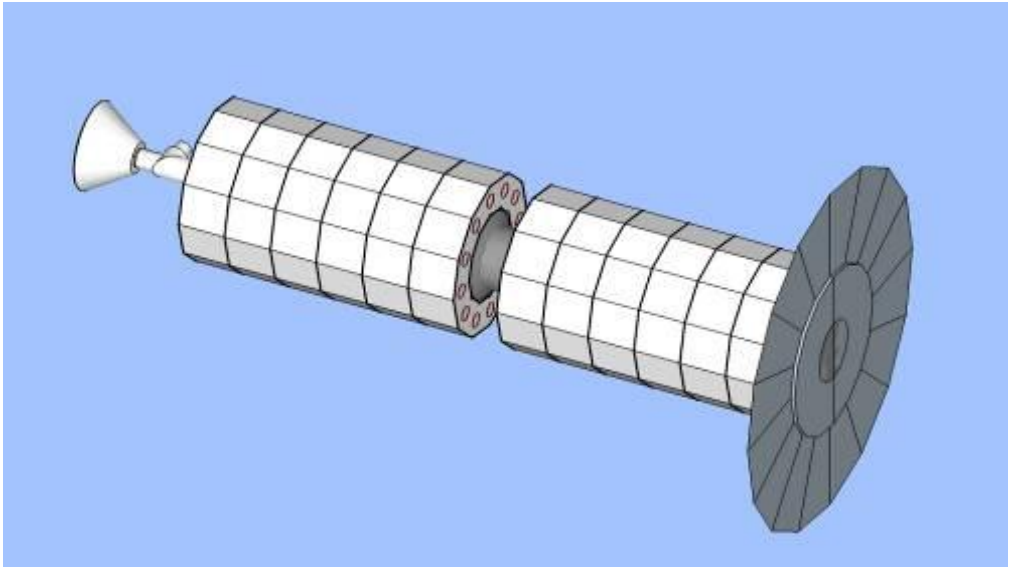

*Figure 6.15  Radiation Shield configuration for a 8-crew mission*

Each of the two modules, including the *Radiation Shield* and the *Habitation Module* for a crew of *4*, would be independent from the other. They would have to be launched with two separate launches and assembled in orbit; the assembly would be relatively easy involving only mechanical and electrical interfaces.

Only one *Sunshield* is required and probably would be somewhat enlarged to allow the same shadowed cone as in the *4*-crew configuration.

Such an overall configuration would provide more safety as the two *Radiation Shields* are independent. In case of a complete failure of one of the Shields, the crew accommodations could be changed so that all the *8* people are living most of their time (including sleeping and free time) in the still shielded pressurized modules. The other ones would still be habitable but the stay there would have to be reduced to a limited percentage of the time for each crew member (*15-25%).*

## 6.14 References

1) Integrated Reference Architecture for Exploration, HME-HS/STU/TN/OM/2008-XX

# 7 Active Radiation Protection System Roadmap

## 7.1 Critical Technologies Identification

The heritage of AMS-02 based technologies represent good starting ground for a *SC Active Radiation Shield* development. However, since a radiation shield would require a strong magnetic bending capability, which require high field, large lever arm or both, it would benefit from a different coil configuration than the racetrack toroid and from the use of higher temperature *SC* wires like $MgB_2$ or *YBCO* with an higher superconductor current density.

We have identified the following *10 Critical Technologies* which would need significant *R&D* to meet the requirements of an active shield for Space Exploration.

- Critical Technology #1 *ITS* and *HTS* wires of better suitable quality (*MgB$_2$, YBCCO*)
- Critical Technology #2 Double helix design and assembly
- Critical Technology #3 Cryogenically stable, light mechanics
- Critical Technology #4 Gas based recirculating cooling systems
- Critical Technology #5 Cryocoolers operating a low temperature
- Critical Technology #6 Magnetic field flux charging devices
- Critical Technology #7 Quench protection for *HTS* coils
- Critical Technology #8 Large cryogenic cases
- Critical Technology #9 Superinsulation, Radiation Shielding, Heat Removal
- Critical Technology #10 Deployable *SC* Coils

For same of these technologies, breakthrough are expected/needed. In particular:

- Critical Technology #1 *ITS* and *HTS* wires of better suitable quality (*MgB$_2$, YBCCO*)
- Critical Technology #4 Gas based recirculating cooling systems
- Critical Technology #5 Cryocoolers operating a low temperature
- Critical Technology #10 Deployable *SC* Coils

## 7.2 Critical Technologies Development Tree

The technology development tree of the critical technologies which have been identified, that is the interconnection between the various technologies and their development and verification

*91*

path at subsystem and at system level (ground based and space demonstrators) is shown in *Figure 7.1*. The corresponding *TRL* matrix is shown in *Figure 7.2*.

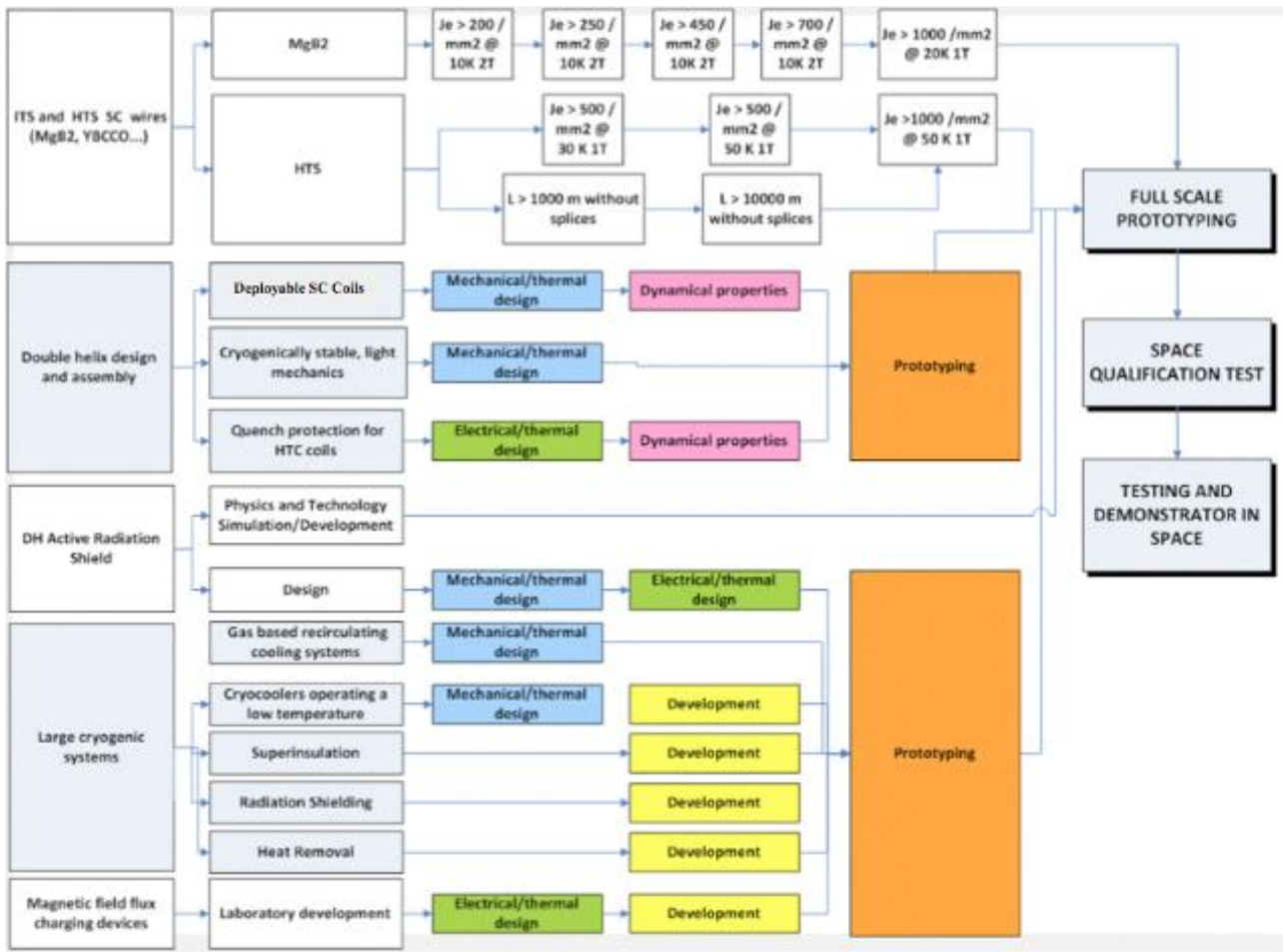

*Figure 7.1 Critical technologies development tree*

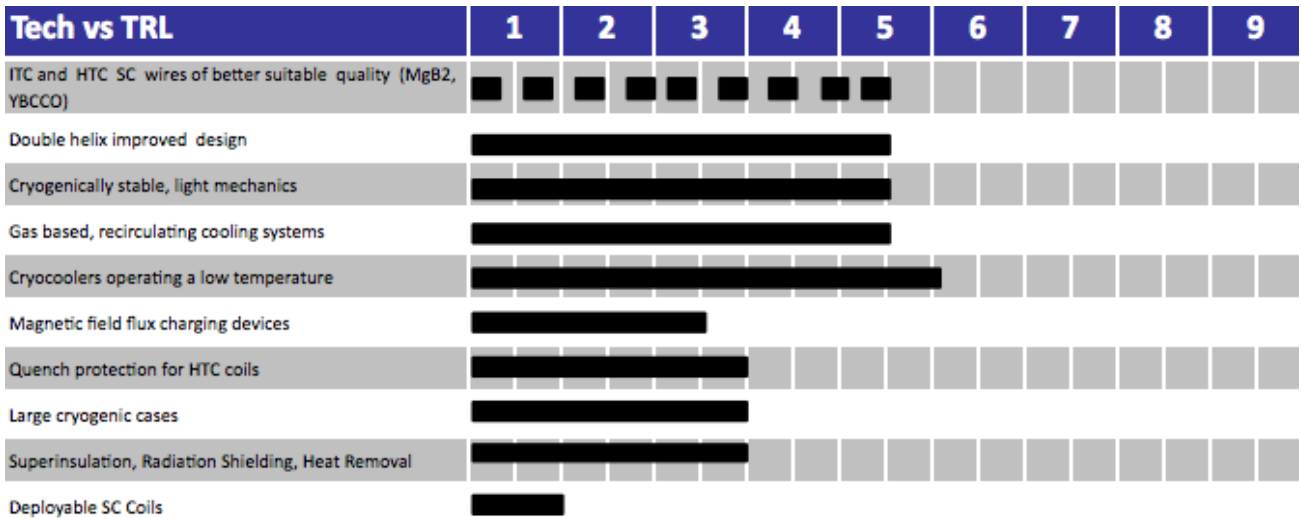

*Figure 7.2  Critical Technologies TRL matrix*

## 7.3  Demonstrators Development

*R&D* activities should progress in parallel on the various topics. These progress would allow for the development of a set of demonstrators:

### 7.3.1 Ground Demonstrators

A ground demonstrator could consist, initially, in the design and construction of a *DH* coil based on $MgB_2$ or *YBCO* cables/tapes. The first coil could be a scaled down version (*1 m* diameter, *5 m* long), operating in a cryostat. A second demonstrator would be a *1:1* scale coils. The goal of these demonstrators, would be to test the design technology developed by *AML* from the magnetic and *SC* point of view (stability, quench control system etc.). These demonstrators could operated using qualified hardware (power supply, quench detection system, gas cooling system, cryocoolers) from the *AMS-02 SC* magnet.

Once the first *DH* coil demonstrator will be tested successfully, one would use this demonstrator as test bed for the various technologies needed to develop an active shield like: flux charging, superinsulation, setting and controlling persistent mode operation, low temperature cryocoolers….

### 7.3.2 Flight Demonstrators

Light radiation shields based on DH technology should be designed to effectively take advantage from the low temperature, low pressure condition which can be found in space. After a suitable amount of ground based *R&D* and demonstrator development, a small scale (*1/10*) array of *DH* coils should be developed for operation in space, likely as free flyer payload operating in *LEO*. The purpose of this demonstrator would be to test the sun shielding and heat removal technologies which are necessary for operating light *SC DH* coils in space. Given the limited size of the flight demonstrator, extensive test in *TV* chambers on ground would be performed before attempting deployment in space.

## 7.4 Facilities and resources definition

For the purpose of technology and demonstrator development and coordination of the *R&D* effort a dedicated laboratory should be conveniently set up at a suitable research institute. This laboratory should operate for an initial period of a minimum of five years to reach some well defined milestones which should be the starting point for defining the specification of a series of flight demonstrators.

Such a laboratory could be similar to the *AMS-02* cryogenic and integration laboratory which has been set up at *CERN* in the years *2007-2010*. *CERN* would also be a suitable institution where to develop such a laboratory due to the extensive experience of its engineers in the field of superconductivity and cryogenics. At *CERN* are also available all the flight units as well as *GSE* form the *AMS-02 SC* magnet development: they would be available as starting ground for the development of new, advanced space magnet system.

Extensive exposure of materials and subsystems of Thermo Vacuum conditions would require the continuous availability of dedicated large Space Simulator, like the one available at the *SERMS* Laboratory in Terni (I) which is equipped with a *2.2 m ☐x 2.2 m* space simulator located in a clean room environment, as well as with additional material testing and thermal and vibration facilities fully devoted to research and development.

Larger prototypes/models could be tested at the world class facilities located at the *ESA-ESTEC* Center of Noordwiik (*NL*): access to these facilities, however, would be affordable only for limited amount of time, due to the large operation cost of these infrastructures.

## 7.5 Radiation Protection System Phasing and Planning

A *Development Plan for the Radiation Protection System* investigated in the frame of this study has to be phased and harmonized with the development of the overall vehicle system on which it will be used.

The reference mission which has been assumed for the definition of the system has been a human mission to Mars. The mission timing foresees a transfer phase from Earth to Mars orbit of about of *200* days, a stay on Mars surface of about *400-500* days and a return trip of about *200* days (so-called *'conjunction mission'*). An alternative mission (the *'opposition mission'*), would need about *300* days each way and a stay on Mars of about *40* days. The whole crew is transferred to the Mars surface, leaving the transfer vehicle in dormant mode until the crew return. These reference information derives from the *ESA* exploration architectural studies summarized in *Ref. 1*.

According to the results presented in '*The Global Exploration Roadmap'*[(2)] as result of the activities of the *ISECG (International Space Exploration Coordination Group)*, a first step toward a final objective of human Mars exploration could a mission to *NEA (Near Earth Asteroids)*, the so-called 'deep space first' as an alternative to a 'moon-first' approach. Such a mission would require as well a R*adiation Protection System* for the crew as the mission duration could be in the range of *400* days, so that radiation protection becomes an issue.

The scenario identified in the mentioned Roadmap foresees the development of capabilities necessary to demonstrate crewed missions in space for longer durations at increased distances from Earth. In this framework, crew radiation protections is identified as one of the critical capabilities to be developed.

An important step in this process has been identified by *ISECG* with the early deployment of a **'Deep Space Habitat', DSH**, to Earth- Moon Lagrange point 1 (*EML 1*), allowing demonstration of habitation and other critical systems in a deep space environment progressively longer demonstrations of the ability to live without a regular supply chain from Earth. The *DSH* launch is placed in 2023.

The scenario foresees that the *Deep Space Habitat* would serve in its first *5* to *6* years of operations at *EML-1* as a platform for advancing enabling research for human exploration in the deep space environment and for demonstrating advanced habitation systems. The *Deep Space Habitat* is then sent on a human mission to a *Near Earth Asteroid*. A second human mission to another *Near Earth Asteroid* may follow by the mid *2030s* following a similar approach. After two deep space missions, the preparation for a human mission to the Mars system (either to the Mars orbit or to Phobos/Deimos), can in principle commence by advancing further the capabilities developed.

The above described framework is here assumed as a reference for defining a plan for the development of the *Superconductive Radiation Shielding System* which would be part of the *DSH*.

The development plan is in principle divided in 4 main sequential phases:

1. **Technology development**, in which the critical and/or not yet mature technologies are brought to *TRL5*. *Sect.7.1* has identified these critical technologies for a *Radiation Shield* based on a *Double-Helix* coil concept.

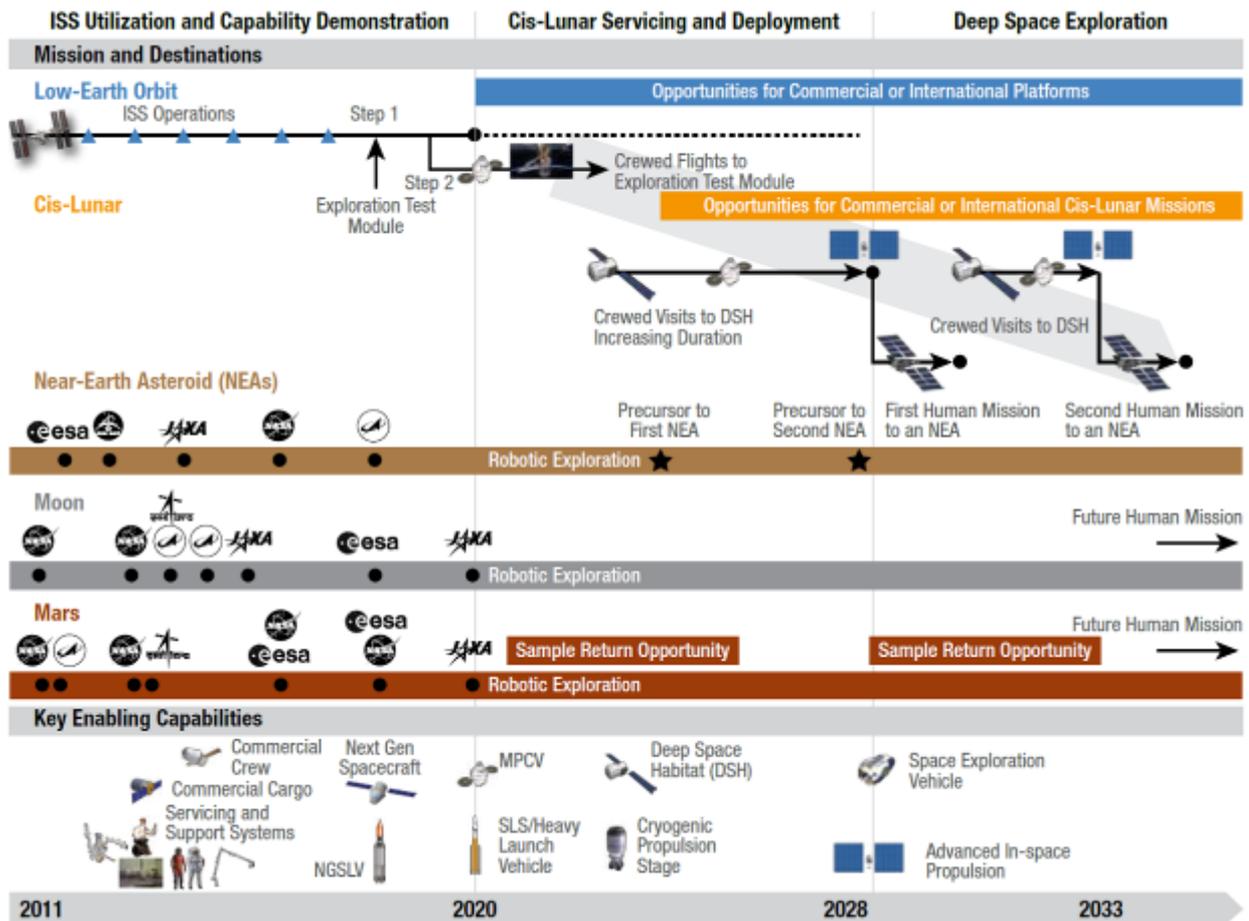

*Figure 7.3  The 'Deep Space First' scenario (from RD 33) Ref*

2. **Ground Demonstrators**, aiming at integrating, with different steps, the required different technologies in one system, allowing to validate the capability of the system in providing the expected functions and performance. The following figure shows the integration flux of the different technologies within these Ground Demonstrators. The results obtained from these Demonstrator shall give the go-ahead for the viability of the in-space system (demonstrator). This phase requires the realization of a proper laboratory facility where the different Demonstrators have to be integrated and operated

3. **In-space Demonstration**, where a representative *Radiation Shield* model is flown in *LEO* (maybe as a free-flying item in proximity of the *ISS*). The Demonstrator could be a scaled version of the final system or can be limited to one or few coils, with the objective of demonstrating reliability and key performance parameters of the superconductive radiation shield under a representative operational environment. The fact that the *LEO* is not representative of the deep space radiation environment is not considered an issue for the purpose of this Demonstrator. In case such representativity is deemed necessary a *MEO/GEO* orbit could be considered as target orbit.

4. **DSH Radiation Shielding System**; this is the operational system for the *DSH*.

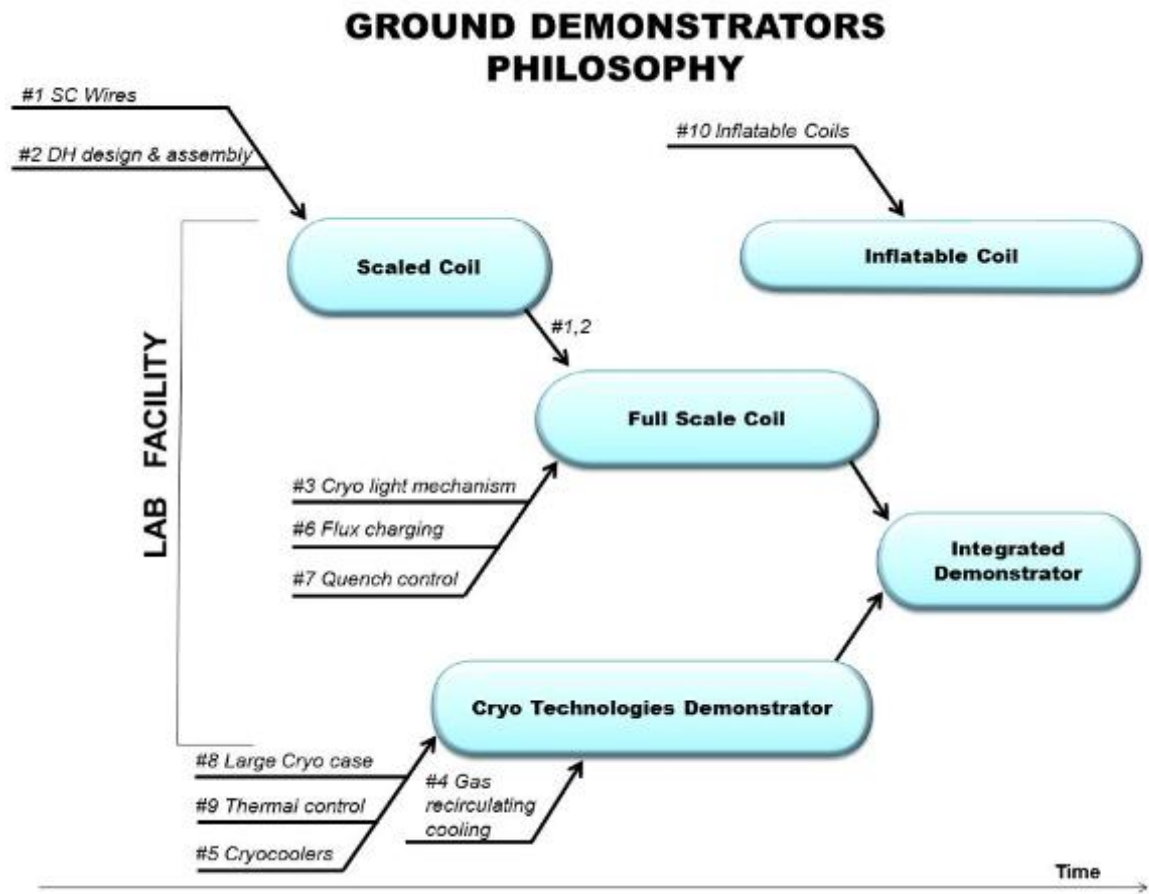

*Figure 7.5 Ground Demonstrator Philosophy*

For the first three areas, parallel activities have to be performed at system level to coordinate the activities with the higher level exploration effort (e.g., technology/capabilities demonstrations in-orbit/ISS; *DSH* pre-development) to define and keep a Radiation System baseline as reference for the preparatory development activities.

For the development of the *Radiation Shielding System* for the *DSH*, the effort must be part of the *DSH* development, fully integrated in it.

The general timeline for the implementation of above phases is given in the following *Figure 7.6*.

The technology developments are starting in *2012* and will proceed in parallel for each technology; an overall coordination effort will guarantee that the work on technologies is driven by a common vision at system level.

A time period of *3* years is deemed adequate to reach *TRL 5* for all technologies: at this stage component and/or breadboard are used for validation in relevant environment; the basic technological elements are integrated to some extent into realistic supporting elements; some level of integration between different technologies could also be reached.

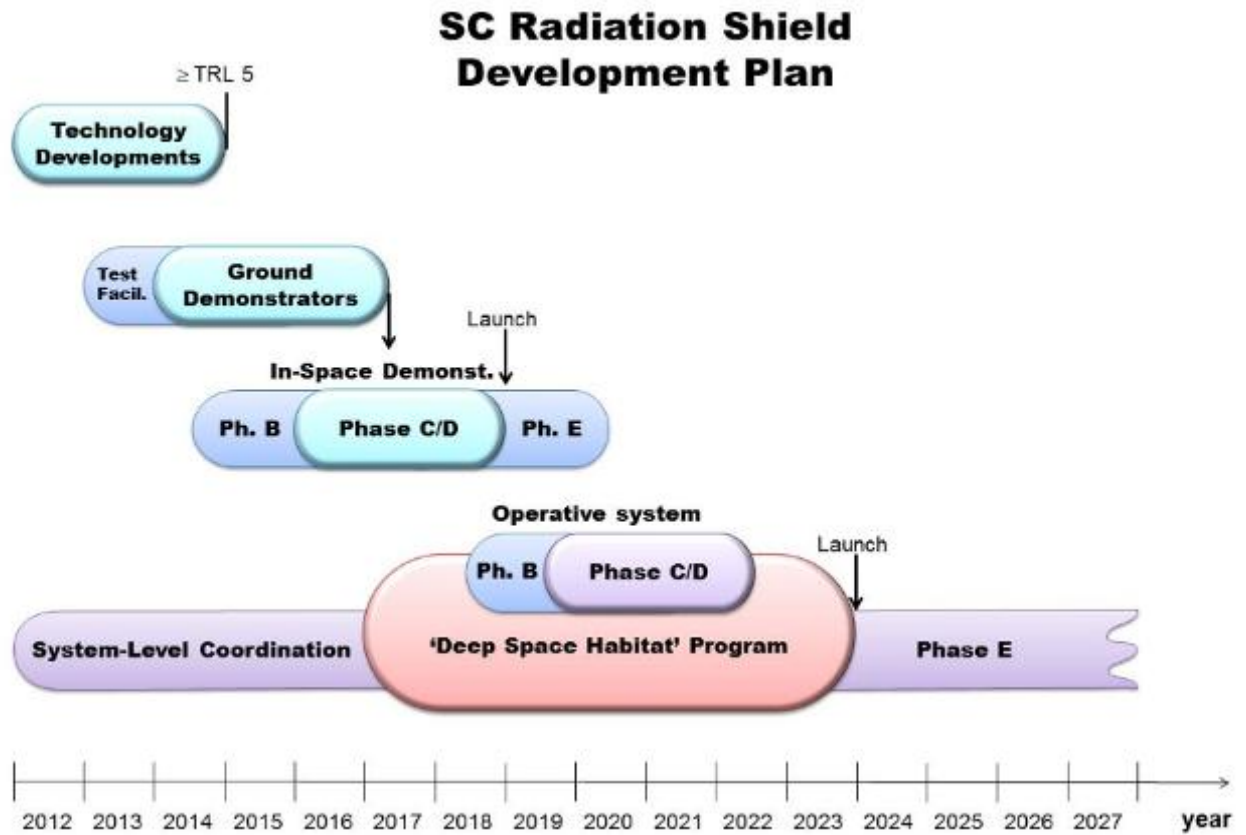

*Figure 7.6 Development plan timeline*

The In-space Demonstrator phase *B* will start in *2015* (phase A feasibility studies are considered included in the system level coordination bar in the figure). A *3*-years phase C/D lead to a launch in *2019* with an in-orbit activity (Phase E) of at least in order to demonstrate the durability capabilities of the developed technologies. at this stage it can be said that *TRL 8+* is reached: the technologies are *'flight proven'*.

The development of the *Radiation Shielding System* for the *DSH* will be an integral part of the *DSH* program; anyway, as far as the *Radiation Shield* is concerned, in order to reach a launch date in *2023*, a Phase B must be completed within *2018*, with full development (Phase C/D) launched in *2019*.

According to the *ISECG* scenario, the *DSH* and its *Radiation Shielding System* will operate in *EML-1* for *5-6* years before being used for the first mission to *NEA*, after integration with a propulsion stage module.

## 7.6 References

1) Integrated Reference Architecture for Exploration, HME-HS/STU/TN/OM/2008-XX
2) The Global Exploration Roadmap, September 2011, International Space Exploration Coordination Group